\documentclass[a4paper,11pt]{article}
\pdfoutput=1 
\usepackage{jheppub} 
\usepackage[T1]{fontenc} 
\usepackage{amsmath,amssymb,amsfonts}
\usepackage{graphicx}
\usepackage{enumitem}
\usepackage[usenames,dvipsnames]{color}
\usepackage{verbatim}
\usepackage[percent]{overpic}
\usepackage{rotating}
\usepackage{hyperref}
\usepackage{xcolor}
\usepackage{array}
\usepackage{mathtools}
\usepackage{lmodern}
\usepackage[normalem]{ulem}
\usepackage{braket}
\usepackage{tensor}
\usepackage{dsfont}
\usepackage{bm}
\usepackage{tikz}
\usetikzlibrary{decorations.pathmorphing}

\usepackage{CJKutf8}

\newcommand{\bx}{\bm{x}}

\newcommand{\R}{\mathbb{R}}
\newcommand{\dd}{\text{d}}
\newcommand{\bk}{{\bm{k}}}
\newcommand{\rr}[1]{\left(#1\right)}
\renewcommand{\H}{\hat{H}}
\newcommand{\id}{\mathds{1}}
\newcommand{\sx}{\mathsf{x}}

\newcommand{\ii}{\mathsf{i}}
\newcommand{\pd}{\partial}
\newcommand{\loc}{\text{loc}}
\DeclareMathOperator{\tr}{Tr}
\renewcommand{\bar}{\overline}
\title{Harvesting correlations in Schwarzschild and collapsing shell spacetimes}

\author[a,b,1]{Erickson Tjoa,\note{Corresponding author.}}
\author[a,c]{Robert B. Mann}

\affiliation[a]{Department of Physics and Astronomy,
\\University of Waterloo, Waterloo, Ontario, N2L 3G1, Canada}
\affiliation[b]{Institute for Quantum Computing, \\University of Waterloo, Waterloo, Ontario, N2L 3G1, Canada}
\affiliation[c]{Perimeter Institute for Theoretical Physics,
\\Waterloo, Ontario, N2L 2Y5, Canada}

\emailAdd{e2tjoa@uwaterloo.ca}
\emailAdd{rbmann@uwaterloo.ca}

\abstract{We study the harvesting of correlations by two Unruh-DeWitt static detectors from the vacuum state of a massless scalar field in a background Vaidya spacetime consisting of a collapsing null shell that forms a Schwarzschild black hole (hereafter Vaidya spacetime for brevity), and we compare the results with those associated with the three preferred vacua (Boulware, Unruh, Hartle-Hawking-Israel vacua) of the eternal Schwarzschild black hole spacetime. To do this we make use of the explicit Wightman functions for a massless scalar field available in (1+1)-dimensional models of the collapsing spacetime and Schwarzschild spacetimes, and the detectors couple to the proper time derivative of the field. First we find that, with respect to the harvesting protocol, the Unruh vacuum agrees very well with the Vaidya vacuum near the horizon even for finite-time interactions. Second, all four vacua have different capacities for creating correlations between the detectors, with the Vaidya vacuum interpolating between the Unruh vacuum near the horizon and the Boulware vacuum far from the horizon. Third,  we show that the black hole horizon inhibits \textit{any} correlations, not just entanglement. Finally, we  show that the efficiency of the harvesting protocol depend strongly on the signalling ability of the detectors,  which is highly non-trivial in presence of curvature. We provide an asymptotic analysis of the Vaidya vacuum to clarify the relationship between the Boulware/Unruh interpolation and the near/far from horizon and early/late-time limits. We demonstrate a straightforward implementation of numerical contour integration to perform all the calculations.}

\begin{document} 
    \maketitle
    \flushbottom

\section{Introduction}

In recent years fruitful progress in our understanding of quantum field theory (QFT) and fundamental physics has been made by applying insights from quantum information theory. In particular, a great deal of attention has been focused on  entanglement in quantum field theory. For example, different regions of QFT vacua contain both classical correlations and entanglement even if the regions are causally disconnected \cite{summers1985bell,summers1987bell}. Entanglement plays a crucial role in many fundamental phenomena such as the black hole information problem  \cite{hawking1975particle,almheiri2013black,marolf2017black, haco2018black}. Viewed from a quantum information perspective, entanglement is a resource that can be used to perform information-processing tasks \cite{nielsen2000quantum,chitambar2019quantum,Brandao2015reversibleQRT}.

The pioneering work of Valentini \cite{Valentini1991nonlocalcorr} and later by Reznik et al. \cite{reznik2003entanglement,reznik2005violating} showed that one can indeed extract entanglement from the QFT vacuum using a pair of initially uncorrelated quantum systems: in recent years this protocol became known as {\it entanglement harvesting}. Entanglement harvesting has been shown to be sensitive to accelerations \cite{salton2015acceleration}, time dependence of the interaction and the number of spacetime dimensions \cite{pozas2015harvesting}, spacetime curvature \cite{Steeg2009,kukita2017harvesting,henderson2018harvesting,ng2018AdS}, spacetime topology \cite{smith2016topology}, boundary conditions \cite{henderson2019entangling,cong2019entanglement}, as well as indefinite causal ordering \cite{Henderson2020temporal,Foo2020:2003.12774v3,Foo2020:2005.03914v1}. Entanglement harvesting has also been investigated in more practical, experimental settings \cite{olson2011entanglement,olson2012extraction,sabin2010dynamics,sabin2012extracting,EMM2013farming,ardenghi2018entanglement,beny2018energy}. In such studies, it is common to employ a form of particle-detector model such as the Unruh-DeWitt (UDW) model \cite{Unruh1979evaporation,DeWitt1979}, which is a simplified scalar model derived from realistic light-matter interaction. This model captures the essential physics if no angular momentum is exchanged \cite{pozas2016entanglement}. Furthermore, the model can be easily modified into more complicated variants (`UDW'-like models), such as
non-linear couplings \cite{Sachs:2018trp},
coupling to a vector field (`dipole coupling') \cite{pozas2016entanglement} or to the conjugate momentum of the scalar field (`derivative coupling') \cite{Aubry2014derivative}.

Virtually all studies of entanglement harvesting -- and indeed of detector response -- are for spacetimes that are static. Rather little is known beyond this case, with two notable exceptions being a study of detector response for a rotating black hole in (2+1) dimensions \cite{Hodgkinson2012corotatingBTZ}, and a new study of entanglement harvesting in the presence of a gravitational wave \cite{Xu:2020pbj}. Very recently, the transition rate of a UDW detector coupled to a test scalar field in a $(1+1)$-dimensional Vaidya spacetime background describing collapse of a null shell into a black hole was investigated \cite{Aubry2018Vaidya}.  The objective was to see whether the traditional expectation that the vacuum of a dynamical collapsing star spacetime is well-described by  the so-called Unruh vacuum associated with  a test field on a  \textit{static} Schwarzschild background  \cite{Unruh1979evaporation}. The conclusion was that in the causal future of the shell this is indeed the case at appropriate limits, and in particular in the long time (`late time') limit. Furthermore, it was also found that the Unruh vacuum 
gives an upper bound for radiation flux at future null infinity, and a stationary observer carrying a detector at fixed orbit in the exterior black hole region measures a Planckian spectrum in the late time limit \cite{Aubry2018Vaidya}.

We consider here for the first time the correlation properties of the quantum vacuum in a spacetime containing gravitational collapse. Specifically, we investigate the harvesting protocol in (1+1)-dimensional Vaidya spacetime consisting of a collapsing null shell that forms a black hole (hereafter simply called \textit{Vaidya spacetime} for brevity), and compare it with the eternal Schwarzschild black hole spacetime.  Our study is motivated from two considerations. First, it has been known for quite some time \cite{Steeg2009}  that a single inertial detector cannot distinguish a Minkowski thermal bath from the conformal vacuum of de Sitter expanding universe, but in some regimes two detectors can do so via their entanglement dynamics.  Further studies have indicated that the entanglement harvesting properties of a pair of detectors yields novel behaviour that cannot be discerned from that of a single detector \cite{henderson2018harvesting,ng2018AdS,smith2016topology,henderson2019entangling,cong2019entanglement}. It is therefore natural to ask if the Unruh vacuum is still a good approximation of the vacuum describing gravitational collapse (henceforth called the \textit{Vaidya vacuum}) when two detectors are involved. Second, the detector transition rate 
was found to be well-approximated  by the Unruh vacuum near the horizon and also `at late times'  \cite{Aubry2018Vaidya}. While this calculation demonstrated clearly the Planckian behaviour seen in the Unruh vacuum, the detector is turned on both sharply and for infinitely long times after the black hole has formed, thus the notion of `late time' also subsumes `long time' limit. We would like to explore this for both single and  double detector scenarios for more realistic, strictly finite-time interactions to see how the timescales of the interaction affect the detector dynamics. 

We present four main results regarding the dynamics of two detectors interacting with a quantum massless scalar field in a (1+1) Vaidya spacetime in comparison with  corresponding  results  from the three preferred vacua -- Boulware, Unruh, Hartle-Hawking-Israel -- of a Schwarzschild spacetime. First we find that, with respect to harvesting protocol,  the Unruh vacuum agrees very well with the Vaidya vacuum even for finite-time interactions near the horizon,  complementing recent results for long-interaction times  \cite{Aubry2014derivative}. Second, all four vacua have different capacities for creating correlations between the detectors: the Vaidya vacuum interpolates between the Unruh vacuum near the horizon and the Boulware (and hence \textit{Minkowski} vacuum) far from the horizon. Third, by investigating the mutual information of the detectors, we show that the black hole horizon inhibits \textit{any} correlations, not just entanglement, complementing   results found in \cite{henderson2018harvesting}. Finally, we also show that the efficiency of the harvesting protocol depends strongly on the ability of two detectors to signal, which is highly non-trivial in presence of curvature.  Our study also includes an asymptotic analysis of the Vaidya vacuum, which clarifies how its respective
approximations by Unruh and Boulware/Minkowski vacua are related to the early/late time and near/far from horizon limits of the detector-field interaction.

We shall employ derivative coupling for the detector-field interaction in order to remove the problematic infrared (IR) divergence associated with massless scalar fields in two-dimensional Schwarzschild spacetimes \cite{hodgkinson2013particle}.  In addition to  reproducing some aspects of the short-distance Hadamard property of the Wightman distribution in $(3+1)$ dimensions, this also allows us to make a fairer comparison with Vaidya spacetime (which has no IR divergence).  It turns out that we shall require a rather careful numerical treatment after we obtain the general expression for the joint density matrix elements of the two detectors for three reasons. First, the problem lacks a lot of symmetry exploited in other investigations for performing simplifications using Fourier transforms or stationarity; thus numerical treatment becomes essential (especially for the Vaidya vacuum). Second, the Wightman distribution for the Vaidya spacetime has very a complicated pole structure, making the standard $\ii\epsilon$ procedure numerically unstable\footnote{Actually the same situation holds for the Boulware, Unruh and Hartle-Hawking states.  This is because even though we can use stationarity property of the Wightman distribution for computing transition probabilities, the matrix elements that depend on spacetime events at two different radial coordinates do not share this symmetry due to different gravitational redshifts.}. Finally,  transition rate calculations \cite{Aubry2014derivative,Aubry2018Vaidya} involve one-dimensional integrals that allow for better control of the numerics\footnote{For example, in \textit{Mathematica} there are a lot more integration schemes and settings available for computing one-dimensional integrals than for multi-dimensional ones.}. We take this opportunity to present a straightforward way of circumventing these issues in Appendix~\ref{appendix: numerical-contour} without resorting to bottom-up numerical schemes, which makes use of the smoothness and strong support property of Gaussian switching function and basic complex analysis. This method should be applicable to other studies in quantum field theory and beyond where integrals over Green's functions, propagators, kernels, etc. are involved.

Our paper is organized as follows. We first review the geometrical aspects of Klein-Gordon QFT in Schwarzschild spacetime in Section~\ref{sec: geometry}, and then in  Vaidya spacetime in section \ref{sec: Vaidya}. In Section~\ref{sec: UDW-model} we outline the setup for Unruh-DeWitt detector model coupled to $(1+1)$-dimensional background spacetime.  In Section~\ref{sec: results} we present our main results. 

In this paper we employ the natural units $c = \hbar = 1$ throughout. We take the metric $g$ to be such that $g(\mathsf{V},\mathsf{V})=g_{\mu\nu}V^\mu V^\nu <0$ if $\mathsf{V} = V^\mu\partial_\mu$ is a timelike vector. We also write $\sx\equiv x^\mu$ to denote the spacetime coordinates.

\section{Klein-Gordon field in Schwarzschild spacetime}
\label{sec: geometry}

In the next two sections we first review the geometrical and quantum field-theoretic aspects of a quantum massless scalar field on Schwarzschild and Vaidya background spacetimes\footnote{We believe \cite{Aubry2018Vaidya} is one of the most concise yet clearest and most illuminating exposition about QFT in a black hole background suitable for our purposes; thus we find it appropriate that we review their descriptions and notation as closely as we can.}, as done in \cite{Aubry2018Vaidya}.

\subsection{Schwarzschild spacetime: geometry}


We will start from the maximal extension of Schwarzschild spacetime, also known as Kruskal-Szekeres extension $(\mathcal{M}_K,g_K)$, where $\mathcal{M}_K=\R^2\times S^2$. In terms of Kruskal-Szekeres coordinates $(U,V,\theta,\phi)$, the metric reads
\begin{equation}
    g_K = -\frac{32M^3 e^{-r/(2M)}}{r} \dd U \dd V + r^2(\dd \theta^2+\sin^2\theta \dd\phi^2)\,,
\end{equation}
where $U,V\in \R$ are dimensionless, $\theta\in [0,\pi],$ and $\phi\in [0,2\pi)$. For convenience we define $M\coloneqq G\mathsf{M}$, where $\mathsf{M}>0$ is the ADM mass of the black hole, and $r>0$ can be written in terms of $U,V$, i.e.
\begin{equation}
    \rr{\frac{r(U,V)}{2M}-1}e^{r(U,V)/(2M)} = -UV 
    \Longrightarrow r(U,V) = 2M \left(1+ \mathsf{W}\left(-\frac{U V}{e}\right)\right)
    \label{eq: r(U,V)}
\end{equation}
where $\mathsf{W}(z)$ is the Lambert W-function ($\mathsf{W}(z) e^{\mathsf{W}(z)} = z$) \cite{NIST:DLMF}. This spacetime is static, spherically symmetric, asymptotically flat and globally hyperbolic. The point $r=0$ is a curvature singularity. 

The Schwarzschild spacetime $(\mathcal{M}_K,g_K)$ admits four Killing vector fields. Three of these vector fields, which we denote by $\zeta_1,\zeta_2,\zeta_3$, are globally spacelike and generate spherical symmetry:
\begin{subequations}
\begin{align}
    \zeta_1 &= \partial_\phi\,,\\
    \zeta_2 &= \sin\phi\partial_\theta+\cot\theta\cos\phi\partial_\phi\\
    \zeta_3 &= \cos\phi\partial_\theta-\cot\theta\sin\phi\partial_\phi\,,
\end{align}
\end{subequations}
while the fourth Killing field, denoted $\xi$, is given by
\begin{align}
    \xi = \frac{1}{4M}\rr{-U\partial_U+V\partial_V}\,.
\end{align}
This Killing field $\xi$ is timelike for $r>2M$, spacelike for $0<r<2M$ and null at the hypersurface $r=2M$. The null hypersurface $r=2M$ thus defines a bifurcate Killing horizon. This bifurcate Killing horizon separates $M_K$ into four regions, conventionally labelled Region I, II, III, and IV as shown in Figure~\ref{fig: Penrose1}.

\begin{figure}
    \centering
    \scalebox{0.75}{
    \begin{tikzpicture}
    \node (I)    at ( 4,0)   {I};
    \node (II)   at (-4,0)   {III};
    \node (III)  at (0, 2.5) {II};
    \node (IV)   at (0,-2.5) {IV};

\path  
  (II) +(90:4)  coordinate[label=90:$i^+$] (IItop) 
       +(-90:4) coordinate[label=-90:$i^-$] (IIbot)
       +(0:4)   coordinate (IIright)
       +(180:4) coordinate[label=180:$i^0$] (IIleft)
       ;
\draw (IIleft) -- 
          node[midway, above left]    {$\cal{I}^+$}
      (IItop) --
      (IIright) -- 
      (IIbot) --
          node[midway, below left]    {$\cal{I}^-$}    
      (IIleft) -- cycle;

\path 
   (I) +(90:4)  coordinate[label=90:$i^+$]  (Itop)
       +(-90:4) coordinate[label=-90:$i^-$] (Ibot)
       +(180:4) coordinate                  (Ileft)
       +(0:4)   coordinate[label=0:$i^0$] (Iright)
       ;
\draw   (Ileft)     -- 
            node[midway, below right] {$\mathcal{H}^+$}
            node[midway, above, sloped] {$r=2M$}
        (Itop)      -- 
            node[midway, above right] {$\mathcal{I}^+$}
        (Iright)    -- 
            node[midway, below right] {$\mathcal{I}^-$}
        (Ibot)      -- 
            node[midway, above right] {$\mathcal{H}^-$}
            node[midway, below, sloped] {$r=2M$}
        (Ileft)     -- 
        cycle;

\draw[decorate,decoration=zigzag] (IItop) -- (Itop)
      node[midway, above, inner sep=2mm] {$r=0$};

\draw[decorate,decoration=zigzag] (IIbot) -- (Ibot)
      node[midway, below, inner sep=2mm] {$r=0$};

\end{tikzpicture}
}    
    \caption{Conformal diagram for Schwarzschild spacetime.}
    \label{fig: Penrose1}
\end{figure}

Regions I and II are defined by $V>0$ and part of the Killing horizon  $\mathcal H^+$ that separates the two regions. In this region, one can use  ingoing Eddington-Finkelstein coordinates $(U,v,\theta,\phi)$ defined by $V=e^{v/(4M)}$, with $v\in \R$. In this coordinate system, we can view regions I and II as an asymptotically flat and globally hyperbolic spacetime in itself, denoted $(\mathcal{M}_{E},g_E)$, where $M_E$ is a submanifold of $M_K$ and $g_E$ is the induced metric obtained from inclusion map $i:\mathcal{M}_E\hookrightarrow \mathcal{M}_K$ by pullback, i.e. $g_E = i^*g_K$. In this coordinate system, the metric reads
\begin{align}
    g_E = -\frac{8M^2}{r}e^{-\frac{r}{2M}+\frac{v}{4M}}\dd U \dd v + r^2\rr{\dd \theta^2+\sin^2\theta\dd\phi^2}\,.
\end{align}
The Killing vectors for $\mathcal{M}_E$ are $\zeta_1,\zeta_2,\zeta_3$, and also $\xi$ under the restriction $V>0$, which can now be written as $\xi = -\frac{1}{4M}U\partial_U+\partial_v$.

Finally, Region I of Schwarzschild spacetime describing the exterior of a static spherically symmetric star or eternal black hole is defined by $U<0$ and $V>0$. Thus in addition to the ingoing Eddington-Finkelstein coordinates $v$, one can also introduce the outgoing Eddington-Finkelstein coordinates $u$ defined by $U = -e^{-u/(4M)}$. Now using the so-called tortoise radial coordinate $r_*$ we can construct two null coordinates
\begin{align}
    u &= t-r_*\,,\hspace{0.5cm}v=t+r_*\,,\hspace{0.5cm}
    r_* = r+2M\log\rr{\frac{r}{2M}-1}\,.
\end{align}
We can thus regard Region I as a standalone asymptotically flat and globally hyperbolic spacetime, denoted $(\mathcal{M}_S,g_S)$, where $\mathcal{M}_S$ is a submanifold of $\mathcal{M}_K$ and $g_S$ is the induced metric obtained from the inclusion map $i:M_S\hookrightarrow M_K$ by pullback, i.e. $g_S = i^*g_K$. In this coordinate system, the metric reads 
\begin{align}
    g_S &= -f(r)\dd t^2 + \frac{\dd r^2}{f(r)}+r^2\rr{\dd\theta^2+\sin^2\theta\dd\phi^2}\,,\hspace{0.5cm} f(r) = 1-\frac{2M}{r}\,.
\end{align}
The Killing vectors for $\mathcal{M}_S$ are $\zeta_1,\zeta_2,\zeta_3$, and also $\xi$ under the restriction $U<0$ and $V>0$, which can now be written as $\xi = \partial_t$.

\subsection{Schwarzschild spacetime: Klein-Gordon field}

A real massless Klein-Gordon field $\phi: \mathcal{M}\to \R$ conformally coupled to gravity in $(n+1)$-dimensional spacetime $\mathcal{M}$ satisfies the Klein-Gordon equation
\begin{align}
    \frac{1}{\sqrt{-g}}\partial_\mu \rr{\sqrt{-g}g^{\mu\nu}\partial_\nu}\phi - \frac{n-1}{4n}R\phi=0\,,
    \label{eq: KGE}
\end{align}
where $R$ is the Ricci scalar curvature. In order to construct an appropriate vacuum state of the theory, we will solve for the classical mode solutions. The general solution takes the form
\begin{align}
    \hat \phi(\sx) = \int \dd^{n} \bk\, \rr{\hat a_\bk^{\phantom{\dagger}}u^{\phantom{*}}_\bk(\sx) +\hat a_\bk^{\dagger}u_\bk^*(\sx) }\,.
\end{align}
The mode  (eigen)functions $\{u_\bk(\sx)\}$ satisfy the orthogonality conditions
\begin{align}
    (u_{\bk},u_{\bk'})=\delta^{(n)}(\bk-\bk')\,,\hspace{0.5cm}(u_{\bk}^*,u_{\bk'}^*)=-\delta^{(n)}(\bk-\bk')\,,\hspace{0.5cm}(u_\bk^{\phantom{*}},u^*_{\bk'})=0\,,
\end{align}
where $(f,g)$ is the Klein-Gordon inner product of $f,g$ given by
\begin{align}
(f,g) = -\ii\int_{\Sigma}\dd \Sigma^\mu \sqrt{-g}\rr{f\nabla_\mu g^* - g^*\nabla_\mu f } 
\end{align}
with respect to the Cauchy surface $\Sigma$.

The definition of a vacuum state of the field depends on the choice of timelike Killing vector field: given a timelike Killing vector $\xi$, the mode function $u_\bk$ is said to be positive frequency with respect to $\xi$ if $u_\bk(\sx)$ solves the eigenvalue equation
\begin{align}
    \ii\mathcal{L}_\xi u_\bk = \omega_\bk u_\bk\,,
    \label{eq: general-fourier-mode}
\end{align}
where $\mathcal{L}_\xi$ is the Lie derivative with respect to $\xi$ and $\omega_\bk = |\bk|>0$. Similarly, $u_\bk(\sx)$ is negative frequency if $\ii\mathcal{L}_\xi u_\bk = -\omega_\bk u_\bk$.

In our problem, there are three distinguished vacuum states that are invariant under the Killing vector $\xi$:
\begin{enumerate}[label=(\alph*)]
    \item Boulware vacuum $\ket{0_B}$: this state is defined in Region I and has modes that are positive and negative frequency with respect to Schwarzschild timelike Killing field $\partial_t$ (restriction of $\xi$ to Region I). It is considered unphysical as it is not regular on both future and past horizons $\mathcal{H}^\pm$. However, this state will be useful for the discussion of the vacuum in the Vaidya background later.

    \item Unruh vacuum $\ket{0_U}$: this state is defined in Region I and II and has modes that are positive frequency on the Cauchy surface $\Sigma=\mathcal{I}^-\cup \mathcal{H}^-$, the union of past null infinity and past horizon. The positive frequency modes on the past horizon $\mathcal{H}^-$ are obtained with respect to the null generator $\partial_U$ of $\mathcal{H}^-$  ($U$ being the null affine parameter along $\mathcal{H}^-$); the positive frequency modes on the past null infinity $\mathcal{I}^-$ are obtained with respect to the null generator $\partial_v$ of $\mathcal{I}^-$. 
    
    \item Hartle-Hawking-Israel (HHI) vacuum $\ket{0_H}$: this state is defined on the full Kruskal-Szekeres extension and has modes that are positive frequency with respect to both past and future horizon generators $\partial_U$ and $\partial_V$. This is a state representing a black hole in thermal equilibrium with a radiation bath, such that the restriction of the state to Region I is KMS at the Hawking temperature $T_H = (8\pi M)^{-1}$. Note that $T_H$ is the temperature measured by an observer at infinity (see Section~\ref{sec: thermality}).
\end{enumerate}

We restrict our attention to the dimensionally reduced (1+1)-dimensional Schwarzschild spacetime by removing the angular part of the metric in (3+1) dimensions. This allows us to obtain the closed-form expression of the vacuum states in terms of the Wightman two-point distributions by invoking conformal invariance of the Klein-Gordon equation in (1+1) dimensions. The positive frequency modes associated with each vacuum read \cite{birrell1984quantum}
\begin{subequations}
\begin{align}
    \text{Hartle-Hawking-Israel}:\hspace{0.5cm} &  e^{-\ii\omega \bar{U}}\,,e^{-\ii\omega \bar{V}}\,,\\
    \text{Unruh}:\hspace{0.5cm} &  e^{-\ii\omega \bar{U}}\,,e^{-\ii\omega v}\,,\\
    \text{Boulware}:\hspace{0.5cm} &  e^{-\ii\omega u}\,,e^{-\ii\omega v}\,,
\end{align}
\end{subequations}
where $\Bar{U} = -(4M)^{-1}U$ and $\Bar{V} = (4M)^{-1}V$. 

We are interested in constructing the Wightman two-point distribution for each vacuum state (denoted $\ket{0_\alpha}$), defined by
\begin{align}
    W_\alpha(\sx,\sx') \coloneqq \tr\rr{\hat\phi(\sx)\hat\phi(\sx')\ket{0_\alpha}\!\bra{0_\alpha}}\,,
    \label{eq: wightman-general}
\end{align}
so that for each vacuum state (here $\alpha=B,U,H$) we have\footnote{Note that in \cite{birrell1984quantum} the IR cut-off has been removed by hand.}
\begin{subequations}
\begin{align}
    W_B(\sx,\sx') &= -\frac{1}{4\pi}\log\left[-\Lambda^2(\Delta u-\ii\epsilon)(\Delta v-\ii\epsilon)\right]\,,
    \label{eq: wightman-boulware}\\
    W_U(\sx,\sx') &= -\frac{1}{4\pi}\log\left[-\Lambda^2(\Delta \bar U-\ii\epsilon)(\Delta v-\ii\epsilon)\right]\,,
    \label{eq: wightman-eddington}\\
    W_H(\sx,\sx') &= -\frac{1}{4\pi}\log\left[-\Lambda^2(\Delta \bar U-\ii\epsilon)(\Delta \bar V-\ii\epsilon)\right]\,,
    \label{eq: wightman-kruskal}
\end{align}
\end{subequations}
where $\Lambda>0$ is an IR cutoff inherent in (1+1) massless scalar field theory.

\subsection{Comment on IR ambiguity and derivative coupling}

We pause here to comment briefly on the IR divergence in two-dimensional massless field theory that will motivate the choice of detector-field interaction in this paper.

It is well-known that a  two-dimensional massless scalar field in Minkowski space exhibits an infrared (IR) ambiguity. More specifically, from Eq.~\eqref{eq: KGE} one can show that in (1+1)-dimensional Minkowski space, a massless scalar field quantized in Minkowski coordinates $(t,x)$ corresponding to inertial laboratory frame has Fourier mode decomposition given by
\begin{align}
    \hat\phi(\sx) = \int\frac{\dd k}{\sqrt{2(2\pi)|k|}}\rr{\hat a_k^{\phantom{\dagger}}e^{-\ii|k|t+\ii kx}+\hat a_k^{\dagger}e^{\ii|k|t-\ii kx}}\,.
    \label{eq: 1D-fourier-decomposition}
\end{align}
This decomposition allows us to define Minkowski vacuum $\ket{0_M}$: the Wightman distribution associated to Minkowski vacuum $W_{M}(\sx,\sx')$ can then be shown to have a logarithmic divergence:
\begin{align}
    W_{M}(\sx,\sx') &= -\frac{1}{4\pi}\log\rr{\Lambda^2(\epsilon+\ii\Delta u)(\epsilon+\ii\Delta v)}\,,
    \label{eq: Wightman-flat-0}
\end{align}
where $\epsilon>0$ is a ultraviolet (UV) regulator and $\Lambda>0$ is an infrared (IR) regulator. This IR divergence can also be seen from the Fourier mode decomposition, where the integral in Eq.~\eqref{eq: 1D-fourier-decomposition} is divergent for $k=0$. We can choose the principal branch of the logarithm so that
\begin{align}
    W_{M}(\sx,\sx') &= -\frac{1}{4\pi}\log\rr{(\Delta u-\ii\epsilon)(\Delta v-\ii\epsilon)} - \frac{\log(-\Lambda^2)}{4\pi}\,,
    \label{eq: wightman-flat}
\end{align}
where $u = t-x$ and $v=t+x$. The second term is formally divergent\footnote{Taking the principal branch, the real part diverges as $-\frac{1}{2\pi}\log\Lambda$ and the imaginary part is exactly $-\frac{\ii}{4}$.} as $\Lambda\to 0$. This IR divergence will also appear for the Schwarzschild spacetime as all two-dimensional spacetimes are conformally flat and the Klein-Gordon equation \eqref{eq: KGE} is conformally invariant for $n=1$; hence the same IR divergence appears in Eqs.~\eqref{eq: wightman-boulware}-\eqref{eq: wightman-kruskal}.

\textit{A priori}, this IR divergence is problematic for detector dynamics as the density matrix for the detector(s) would depend on the IR cut-off chosen (see e.g. \cite{pozas2015harvesting, Aubry2014derivative, EMM2014zeromode}). Typically, one either chooses $\Lambda$ based on a characteristic length scale of the system under consideration, or removes it by hand via other arguments. For instance the additive constant that appears in Eq.~\eqref{eq: wightman-flat} can be dropped using the argument that entanglement measures such as concurrence and negativity are by definition infrared-safe  \cite{cong2019entanglement}: they involve subtraction of two matrix elements that contain the same IR-divergent additive constant. Therefore the formally infinite additive constant drops out of the entanglement calculation. This is analogous to how entanglement entropy in QFT in general contains state-dependent divergences, but a quantity such as relative entropy is finite due to cancellation of divergences (see e.g. \cite{Marolf2016}). 

In this paper, we shall follow \cite{Aubry2014derivative} instead and consider in Section~\ref{sec: UDW-model}  a particular model of detector-field interaction known as \textit{derivative coupling}\footnote{This is sometimes also called momentum coupling since $u^\mu\nabla_\mu\hat\phi(\sx(\tau)) = \hat\pi(\sx(\tau))$, where $\hat\pi$ is conjugate momentum operator to $\hat\phi$.}. It is a modification of the UDW model where the pullback of the vacuum Wightman distribution along the detector's trajectory is given by the \textit{proper time derivative} of the field,
\begin{align}
    \label{WightA}
    \mathcal{A}_\alpha(\sx(\tau),\sx'(\tau')) &= \tr_\phi\rr{\pd_\tau\hat\phi(\sx(\tau))\pd_{\tau'}\hat\phi(\sx'(\tau'))\ket{0_\alpha}\!\bra{0_\alpha}}\,.
\end{align}
The proper time derivatives will remove the IR ambiguity from the (1+1)-dimensional Wightman function and the detectors' joint density matrix, and it leads to the same short-distance behaviour of the Wightman distribution in (3+1) dimensions \cite{Aubry2014derivative,Aubry2018Vaidya}. Furthermore, the derivative Wightman distribution for each vacuum state is indeed invariant under time translation generated by the timelike Killing field $\xi$, in contrast to the usual Wightman function where the Unruh vacuum is strictly speaking not invariant due to the IR cut-off \cite{Aubry2018Vaidya}. Last but not least, analysis of entanglement harvesting for a derivative-coupled UDW-like model where no IR ambiguity occurs has been shown to give similar qualitative results in flat space and (1+1)-dimensional spacetimes with moving mirror \cite{pozas2016entanglement, cong2019entanglement}.



\section{Klein-Gordon field in Vaidya spacetime}
\label{sec: Vaidya}

\begin{figure}
    \centering
    \scalebox{0.75}{
    \begin{tikzpicture}
    \node (IV)    at (2, 0)   {IV};
    \node (III)   at (1.25, 3.75)   {III};
    \node (I)  at (4, 3.25) {I};
    \node (II)   at (2.15, 4.35) {II};

\path  %
  (IV)  +( 105:5) coordinate     (Itop1) 
       +( 75:5)  coordinate[label=90:$i^+$]     (Itop2) 
       +( 30:5)  coordinate[label=0:$i^0$]      (Iright) 
       +(-105:5) coordinate[label=-90:$i^-$]      (Ibot)
       +(3.45,1.32) coordinate  (Imid)
       +(120:2.625)  coordinate   (Imid2)
       ; 
        
\draw   [decorate,decoration=zigzag] (Itop1) -- (Itop2)
            node[midway, above, inner sep=2mm] {$r=0$};
\draw   (Itop2) -- 
            node[midway, above right] {$\cal{I}^+$}
        (Iright) --
            node[midway, below right] {$\cal{I}^-$}
        (Ibot) --
        (Itop1);
\draw   (Itop2) -- (Imid2)
            node[pos=0.2, below right, inner sep=0mm] {$\cal{H}^+$};
\draw   (Itop1) -- (Imid)
            node[pos=0.7, below, sloped] {$v = 0$};
        


\end{tikzpicture}
}    

    \caption{Conformal diagram for Vaidya spacetime.}
    \label{fig: Penrose2}
\end{figure}
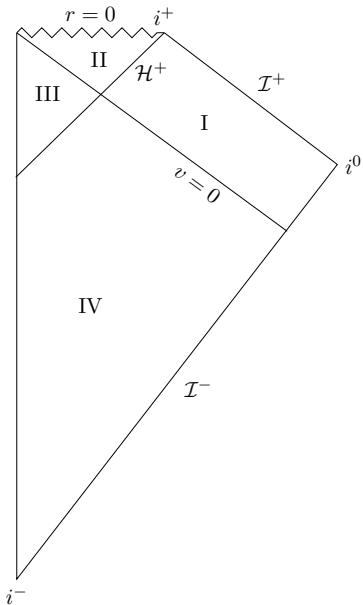

The geometry of Vaidya spacetime is given by the Lorentzian manifold $(M_V,g_V)$ with topology $M_V = \R^2\times S^2$, with the metric written in terms of the ingoing Eddington-Finkelstein-type coordinates $(v,r,\theta,\phi)$:
\begin{align}
    g_V  &= -\rr{1-\frac{2M(v)}{r}}\dd v^2 + 2\dd v\dd r+r^2(\dd \theta^2+\sin^2\theta\dd\phi^2)\,,
    \label{eq: metric-Vaidya}
\end{align}
where $v\in \R$ is a null coordinate, $r>0, \theta\in[0,\pi]$, $\phi\in[0,2\pi)$. The Vaidya metric allows for a general class of mass function $M(v)$, and a particularly simple model for null collapse is prescribed by the mass function
\begin{align}
    M(v) =\begin{cases}
    0\hspace{0.5cm}& v<0\,,\\
    M\hspace{0.5cm}& v\geq 0\,.
    \end{cases}
    \label{eq: mass-function}
\end{align}
where $\Theta(v)$ is the Heaviside step function and $M \geq 0$ is a mass parameter corresponding to ADM mass of the black hole when it is formed by the null shell. The spacetime is isometric to Minkowski space for $v<0$ and isometric to Schwarzschild spacetime for $v>0$. The conformal diagram is shown in Figure~\ref{fig: Penrose2}. For convenience, we shall simply call this particular class of Vaidya metric with mass function \eqref{eq: mass-function} as Vaidya spacetime.

Similar to the Schwarzschild case, the (1+1)-dimensional model for null collapse is obtained by removing the angular coordinates in Eq.~\eqref{eq: metric-Vaidya}. This will enable us to find the vacuum Wightman distribution with respect to the vacuum state of the theory, which we will call the \textit{Vaidya vacuum state} $\ket{0_V}$, analytically. Solving for the Klein-Gordon equation subjected to Dirichlet boundary conditions at $r=0$, the Wightman function for Vaidya spacetime is given by \cite{Aubry2018Vaidya}
\begin{align}
    W(\sx,\sx') &= -\frac{1}{4\pi}\log\frac{(\bar u-\bar u'-\ii\epsilon)(  v- v'-\ii\epsilon)}{(\bar u- v'-\ii\epsilon)( v-\bar u'-\ii\epsilon)}\,,
    \label{eq: wightman-Vaidya}
\end{align}
where $\bar{u}$ is related to the Kruskal (dimensionless) null coordinate $U$ by
\begin{align}
    \bar{u}(U) = -4M\rr{1+\mathsf{W}(-U/e)}\,,
    \label{eq: Vaidya-U}
\end{align}
with $\mathsf{W}(z)$  the Lambert $\mathsf{W}$-function. The function $\bar{u}(U)$ can be obtained by matching modes along the null shockwave $v=0$ \cite{Aubry2018Vaidya}, making use in particular the expression for $r(U,V)$ in Eq.~\eqref{eq: r(U,V)} at the junction.

\section{Unruh-DeWitt model and entanglement harvesting}
\label{sec: UDW-model}

In this section we review the basics of the Unruh-DeWitt detector model for the description of entanglement harvesting protocol. Although not the original Unruh-DeWitt model (hence sometimes said to be `UDW-like'), we will call the derivative-coupling version an Unruh-DeWitt model as well for convenience.

\subsection{Derivative coupling Unruh-DeWitt model}

Let $M$ be a $(n+1)$-dimensional spacetime manifold and consider two observers Alice and Bob each carrying a pointlike Unruh-DeWitt detector along their respective timelike trajectories\footnote{This is a common abuse of notation, since formally one considers timelike curves $\gamma_j: \R\to M$ for $j\in\{A,B\}$. In local coordinate chart $(U,\sx)$ where $\sx: U\to \R^{n+1}$, we have $\sx(\gamma_j(\tau))\in \R^{n+1}$. One then defines a shorthand $\sx_j(\tau)\coloneqq \sx(\gamma_j(\tau))$.} $\sx_j: \R\to M$, where $j\in \{A,B\}$. We consider both detectors to be non-inertial, static detectors at fixed radii $R_j>2M$ outside the black hole horizon with $R_A\leq R_B$. Consequently, the detectors experience different   gravitational redshifts at their respective locations.

Each detector is a two-level quantum system with local interaction Hamiltonian $\H_j$ of the detector-field system given by derivative-coupling Hamiltonian 
\begin{align}
    \H^t_I(t) &= \hat H^t_A(t) + \hat H^t_B(t)\,,
\end{align}
where $t$ is a time coordinate for the spacetime. The superscript $t$ is to make clear that the Hamiltonian generates time translation with respect to $t$. The local interaction between each detector and the field $\hat H^t_j(t)$ is simpler when it is written as Hamiltonian that generate time translation with respect to the proper time $\tau$, i.e. 
\begin{align}
    \H^\tau_j(\tau) &= \lambda_j \chi_j(\tau) \hat{\mu}_j(\tau)\otimes u^\mu\nabla_\mu\hat\phi(\sx_j(\tau))\,, \hspace{0.5cm} j=\{A,B\}\,.
\end{align}
Here $u^\mu$ is the 4-velocity of the detector parametrized by proper time $\tau$, and the two Hamiltonians that generate time translations with respect to $t$ and $\tau$ are related by time-reparametrization \cite{Tales2020GRQO}
\begin{align}
    \hat H^t_j(t) \equiv \hat H^t_j(\tau(t)) = \frac{\dd\tau}{\dd t}\hat H_j^\tau(\tau(t))\,.
    \label{eq: time-reparametrization}
\end{align}
$\lambda_j$ denotes the coupling strength and $\hat\mu_j$ is the monopole moment, given in terms of the detector proper time:
\begin{align}
    \hat\mu_j(\tau) = \hat\sigma^+ e^{\ii\Omega_j \tau} + \hat\sigma^- e^{-\ii\Omega_j \tau}\,,\hspace{0.5cm}j\in \{A,B\}
\end{align}
with $\Omega_j$ the detector gap, $\hat\sigma_j^\pm$ the ladder operators of $\mathfrak{su}(2)$ algebra, and $\chi_j(\tau)$ is the switching function that controls the duration of interaction.  For simplicity we will consider both detectors to be identical, i.e. $\lambda_j=\lambda$, $\Omega_j=\Omega$, with the same Gaussian switching functions 
\begin{align}
    \chi_j(\tau)=\chi(\tau)=e^{-\frac{(\tau-\tau_0)^2}{\sigma^2}}\,,
    \label{eq: switching}
\end{align}
where $\sigma$ prescribes the duration of interaction and $\tau_0$ defines the origin of the proper time. 

Due to gravitational redshift, the time evolution is best described using a common time provided by the background coordinate system, i.e. using $\hat H^t_I(t)$. Since we are interested in detector dynamics in Region I, we can use the Schwarzschild time $t$ as a common time coordinate. The time evolution operator is then given by
\begin{align}
    \hat U = \mathcal{T}\exp\rr{-\ii\int \dd t \H_I^t(t)} 
    = \mathcal{T}\exp\rr{-\ii\int \dd t\left[\frac{\dd \tau_A}{\dd t}\hat H_A^{^{\tau_A}}(\tau_A)+ \frac{\dd \tau_B}{\dd t}\hat H_B^{^{\tau_B}}(\tau_B)\right] }\,,
\end{align}
where we have used Eq.~\eqref{eq: time-reparametrization} and $\tau_A$ and $\tau_B$ are proper times parametrizing different timelike trajectories $\sx_A$ and $\sx_B$ respectively\footnote{This does not mean that there are two different definitions of proper time: the proper time of any observer is the time measured in the observer's rest frame, which is unique. However, given two timelike trajectories $\sx_A,\sx_B$, they are parametrized by two different affine parameters $\tau_A,\tau_B$ that are \textit{a priori} unrelated without further information (e.g. Alice synchronizing with Bob by sending light rays).}. We also fix the proper times of each detector $\tau_A,\tau_B$ such that $\tau_A=\tau_B=0$ when the Schwarzschild time $t=0$, which is possible because the spacetime admits a Cauchy surface given by constant-$t$ slices. We note that due to pointlike nature of the detectors, the derivative coupling particle detector model adopted here is fully covariant \cite{Tales2020GRQO,Bruno2020time-ordering}.

For weak coupling, we can perform a Dyson series expansion 
\begin{align}\label{Dyson}
    \hat U = \id + \hat U^{(1)}+ \hat U^{(2)} + O(\lambda^3)\,,
\end{align}
whose first two terms are
\begin{subequations}
\begin{align}
    \hat U^{(1)} &= -\ii\int_{-\infty}^\infty \dd t \,\H_I^t(t)\,,\\
    \hat U^{(2)} &= -\int_{-\infty}^\infty\dd t\int^t_{-\infty}\dd t'\,\hat H_I^t(t)\hat H_I^{t'}(t')
\end{align}
\end{subequations}
and where $\hat U^{(k)}$ is of order $\lambda^k$. Note that the second order correction $\hat U^{(2)}$ is time-ordered with respect to coordinate time $t$.

Our interest is in vacuum entanglement harvesting, so the initial state is taken to be the uncorrelated state 
\begin{align}
    \rho_0 = \ket{g_A}\!\bra{g_A}\otimes\ket{g_B}\!\bra{g_B}\otimes\ket{0_\alpha}\!\bra{0_\alpha}\,, \hspace{0.5cm}\alpha=B,U,K,V\,,
    \label{eq: initial-state}
\end{align}
where $\ket{g_j}$ is the  ground state of detector $j$ that satisfies $\hat \sigma_j^+\ket{g_j} = \ket{e_j}$, $\hat \sigma_j^-\ket{e_j} = \ket{g_j}$; $\ket{e_j}$ is the excited state of detector $j$, and $\ket{0_\alpha}$ is a vacuum state of the field described in Section~\ref{sec: geometry} and \ref{sec: Vaidya}. The time evolved density matrix is given by $\rho = \hat U\rho_0\hat U^\dagger$, and using the Dyson series expansion 
\eqref{Dyson} we obtain
\begin{align}
    \rho &= \rho_0 + \rho^{(1)} + \rho^{(2)} + O(\lambda^3)\,,
\end{align}
where $\rho^{(k)}$ is of order $\lambda^k$:
\begin{subequations}
\begin{align}
    \rho^{(1)} &= \hat U^{(1)}\rho_0 + \rho_0 U^{(1)\dagger}\,,\\
    \rho^{(2)} &= \hat U^{(1)}\rho_0\hat U^{(1)\dagger} +  \hat U^{(2)}\rho_0 + \rho_0 U^{(2)\dagger}\,.
\end{align}
\end{subequations}
The choice of initial state in Eq.~\eqref{eq: initial-state} implies that $\tr_\phi\rho^{(1)} = 0$ since the one-point function $\braket{0|\hat\phi(\sx)|0} = 0$ for all $\sx$. Therefore, the leading order contribution in perturbation theory is $\rho^{(2)}$.

In order to compute the entanglement between the two detectors, we find the joint reduced density matrix of the detectors by tracing out the field's degrees of freedom:
\begin{align}
    \rho_{AB}\coloneqq \tr_\phi\rr{\hat U\rho_0\hat U^\dagger}\,.
\end{align}
Using the ordered basis $\{\ket{g_A}\ket{g_B}, \ket{g_A}\ket{e_B},\ket{e_A}\ket{g_B},\ket{e_A}\ket{e_B}\}$, the matrix representation of $\rho_{AB}$ to leading order reads
\begin{align}
    \rho_{AB} &= 
    \begin{pmatrix}
    1 - \mathcal{L}_{AA} - \mathcal{L}_{BB}   & 0     & 0     & \mathcal{M}^*   \\
    0  & \mathcal{L}_{BB}   & \mathcal{L}_{BA}     & 0     \\
    0  & \mathcal{L}_{AB}   & \mathcal{L}_{AA}   & 0     \\
    \mathcal{M}    & 0     & 0     & 0
    \end{pmatrix} + O(\lambda^4)\,,
    \label{eq: density-matrix-elements}
\end{align}
where the matrix elements are given by
\begin{align}
    \mathcal{L}_{ij} &= \lambda^2\int \dd\tau_i \int \dd\tau_j'\, \chi(\tau_i)\chi(\tau_j')e^{-\ii\Omega(\tau_i-\tau_j')}\mathcal{A}_\alpha(\sx_i(\tau_i),\sx_j(\tau_j'))\,,
    \label{eq: local-noise}\\
    \mathcal{M} 
    &= 
    -\lambda^2\int_{-\infty}^\infty \dd\tau_A\int_{-\infty}^{\gamma_{\textsc{ba}}\tau_A}\dd\tau_B\, \chi(\tau_A)\chi(\tau_B)e^{\ii \Omega(\tau_A+\tau_B)}\mathcal{A}_\alpha(\sx_A(\tau_A),\sx_B(\tau_B))\, + \notag\\
    &\hspace{0.4cm} 
    -\lambda^2\int_{-\infty}^\infty \dd\tau_B\int_{-\infty}^{\gamma_{\textsc{ab}}\tau_B}\dd\tau_A\, \chi(\tau_B)\chi(\tau_A)e^{\ii \Omega(\tau_B+\tau_A)}\mathcal{A}_\alpha(\sx_B(\tau_B),\sx_A(\tau_A)) 
    \label{eq: non-local-term}
 \end{align}
 where   $\mathcal{A}_\alpha$ is given by \eqref{WightA}. The local `noise' terms $\mathcal{L}_{ii}$ correspond to the transition probability of detector $j$, so sometimes we will write this as $\Pr_j(\Omega,\sigma)\coloneqq \mathcal{L}_{jj}$. The non-local term $\mathcal{M}$ depends on the trajectories of both detectors. In the expression for $\mathcal{M}$, we have defined
 \begin{align}
     \gamma_{{ij}} \coloneqq \sqrt{\frac{f(r_i)}{f(r_j)}}\,,\hspace{0.5cm} i,j\in\{A,B\}\,.
 \end{align}
 In particular we have $\gamma^{\phantom{-1}}_{\textsc{ba}} = \gamma_{\textsc{ab}}^{-1}$. For convenience we choose the convention that $r_B\geq r_A$ (detector $B$ is at larger radial coordinate than detector $A$). The constant $\gamma_{\textsc{AB}}$ in the upper limit of $\mathcal{M}$ appears because the time-ordering in $\hat U^{(2)}$ needs to account for the redshift factor $\tau_j(t) = \sqrt{f(r_j)}t$. More explicitly, if $t=t(\tau_A)$ and $t'=t'(\tau_B)$, it follows from the time-ordering of $\hat U^{(2)}\rho_0$ (and also $\rho_0\hat U^{(2)\dagger}$) that
\begin{subequations}
 \begin{align}
     t-t'> 0 &\Longrightarrow \frac{\tau_A}{\sqrt{f(r_A)}} - \frac{\tau_B}{\sqrt{f(r_B)}} > 0 \Longrightarrow \gamma_{\textsc{ba}}\tau_A> \tau_B\,,\\
     t-t'< 0 &\Longrightarrow \frac{\tau_A}{\sqrt{f(r_A)}} - \frac{\tau_B}{\sqrt{f(r_B)}} < 0 \Longrightarrow \gamma_{\textsc{ab}}\tau_B> \tau_A\,,
 \end{align}
\end{subequations}
hence the upper limit in the expression for $\mathcal{M}$ in Eq.~\eqref{eq: non-local-term}.

Finally, in order to measure the amount of entanglement between the two qubit detectors, there are several faithful entanglement measures we can use. For simplicity, we will use \textit{concurrence} $C[\rho_{AB}]$ \cite{Wotters1998entanglementmeasure}. For the time-evolved density matrix in our scenario, this has the form  \cite{henderson2018harvesting,smith2016topology}
\begin{align}
    C[\rho_{AB}] = 2\max\{0, |\mathcal{M}| - \sqrt{\mathcal{L}_{AA}\mathcal{L}_{BB}}\}+O(\lambda^4)
\end{align}
to leading order in the coupling.
 One could also consider entanglement negativity \cite{Vidal2002negativity} (see e.g. \cite{pozas2015harvesting,simidzija2018harvesting} for its use in the harvesting setup), but we choose concurrence because it cleanly separates the effect of the  non-local term $\mathcal{M}$ and local noise $\mathcal{L}_{ii}$ on   bipartite entanglement. Furthermore, in the special case when the bipartite qubit state is pure, which is the case in this paper, concurrence is a faithful and monotone entanglement measure \cite{Wotters1998entanglementmeasure}.

We will also be interested in another type of correlation called \textit{mutual information}, which quantifies the total amount of classical and quantum correlations between the two detectors. The mutual information between the two detectors is defined by
\begin{align}
    I[\rho_{AB}]\coloneqq S[\rho_{A}]+S[\rho_{B}]-S[\rho_{AB}]\,,
\end{align}
where $S[\rho] = -\tr\rho\log\rho$ is the von Neumann entropy, and $\rho_i \coloneqq \tr_{j}\rho_{ij}$ is the reduced state for detector $i$ after tracing out the detector $j$'s internal degree of freedom. For the joint density matrix in our scenario \cite{simidzija2018harvesting}
\begin{align}
    I[\rho_{AB}] &=  \mathcal{L}_+\log  \mathcal{L}_+ +  \mathcal{L}_-\log \mathcal{L}_-  -  \mathcal{L}_{AA}\log  \mathcal{L}_{AA}   -   \mathcal{L}_{BB}\log  \mathcal{L}_{BB} + O(\lambda^4)\,,
\end{align}
to leading order in perturbation theory, where 
\begin{align}
     \mathcal{L}_{\pm} &\coloneqq \frac{1}{2}\rr{ \mathcal{L}_{AA}+ \mathcal{L}_{BB} \pm \sqrt{( \mathcal{L}_{AA}- \mathcal{L}_{BB})^2+4| \mathcal{L}_{AB}|^2}}\,.
\end{align}
This quantity is useful because if there is no entanglement between the two detectors, then we know that any correlation between them must be either classical correlation or non-entanglement quantum correlations known as quantum discord \cite{Zurek2001discord,Henderson2001correlations}.

We make a passing remark regarding the choice of switching function peaks: in practice, we can demand the detectors to be switched on such that the peak agrees in two ways: (1) at the same constant $t_0$ slice (which means the peaks are at different values of \textit{individual proper times}, or (2) at the same constant $\tau_0$ (which means the peaks are at different values of \textit{coordinate time} $t$). So long as the coordinate time and the proper times are aligned beforehand (e.g. Alice and Bob agree that their own $\tau=0$ corresponds to some fixed $t=t_0$), these two choices will lead to different protocols in the sense that the causal relationships between the detectors may be different. Our choice in this paper corresponds to (2); one could easily consider (1), which is done in the same spirit as \cite{henderson2019entangling}.

\subsection{The two-point Wightman distributions for derivative coupling}

The remaining task is to calculate the derivative-coupling Wightman function for each of  the four vacua $\ket{0_\alpha}$, $\alpha=B,U,H,V$. Let us use the shorthand $\mathcal{A}_\alpha(\tau,\tau')\equiv \mathcal{A}_\alpha(\sx(\tau),\sx'(\tau'))$, $\dot{y}\equiv \pd_\tau [y(\tau)]$, and $\dot{y}'\equiv \pd_{\tau'}[y(\tau')]$. Taking a proper-time derivative of Eqs.~\eqref{eq: wightman-boulware}-\eqref{eq: wightman-kruskal} and Eq.~\eqref{eq: wightman-Vaidya}, we obtain for Schwarzschild vacua
\begin{subequations}
\begin{align}
    \mathcal{A}_B(\tau,\tau') 
    &= -\frac{1}{4\pi}\left[\frac{\dot{u}\dot{u}'}{(u-u'-\ii\epsilon)^2}+\frac{\dot{v}\dot{v}'}{(v-v'-\ii\epsilon)^2}\right]\,,
    \label{eq: derivative-boulware}\\
    \mathcal{A}_U(\tau,\tau') 
    &= -\frac{1}{4\pi}\left[\frac{\dot{U}\dot{U}'}{(U-U'-\ii\epsilon)^2}+\frac{\dot{v}\dot{v}'}{(v-v'-\ii\epsilon)^2}\right]\,,
    \label{eq: derivative-unruh}\\
    \mathcal{A}_H(\tau,\tau') 
    &= -\frac{1}{4\pi}\left[\frac{\dot{U}\dot{U}'}{(U-U'-\ii\epsilon)^2}+\frac{\dot{V}\dot{V}'}{(V-V'-\ii\epsilon)^2}\right] \,.
    \label{eq: derivative-HHI}
\end{align}
\end{subequations}
Note that for simplicity we have written $\mathcal{A}_U$ and $\mathcal{A}_H$ in terms of  $U$ instead of $\bar U$. For the Vaidya vacuum the two-point function has two additional terms 
\begin{align}
    &\mathcal{A}_V(\tau,\tau') \notag\\
    &= -\frac{1}{4\pi}\left[\frac{\dot{\bar u}\dot{\bar u}'}{(\bar u-\bar u'-\ii\epsilon)^2}+\frac{\dot{v}\dot{v}'}{(v-v'-\ii\epsilon)^2}
    -\frac{\dot{\bar u}\dot{v}'}{(\bar u-v'-\ii\epsilon)^2}
    -\frac{\dot{v}\dot{\bar u}'}{(v-\bar u'-\ii\epsilon)^2}\right] 
    \label{eq: derivative-Vaidya}
\end{align}
due to the boundary condition at $r=0$.

\subsection{Comments on switching time and computation of joint density matrix}

Now we have all the ingredients to study the correlations between two detectors after interacting with the field. We pause here to make several comments on the procedure of computing the time-evolved density matrix $\rho_{AB}$ to leading order in perturbation theory.

First, note that in our construction the collapsing null shell occurs at $v=0$ (this could be generalized to arbitrary $v=v_0$ but we do not do this here). In terms of the Eddington-Finkelstein coordinates, this means that $t+r_*=0$. Due to the matching condition at $v=0$, it is imperative that for detectors in Region I, the switching time $\tau=\tau_0$ is chosen such that it respects $v=0$. In particular, if Alice's detector is located at $r= k r_H$ for $k > 1$ and $r_H=2M$, then inverting the null coordinate $v$ we get the constraint
\begin{align}
\label{v-constraint}
    v > 0 \Longrightarrow t > -2 (k M+M \log (k-1))\,.
\end{align}
 Accounting for redshift, this constraint can be written in terms of detector's proper time:
\begin{align}
     \tau > -\frac{2 (k M+M \log (k-1))}{1-1/k}\,.
\end{align}
Therefore, if we demand that the Gaussian strong support to be $b\sigma$ ($b>0$), the requirement that this support is contained entirely in Region I imposes the  constraint that 
\begin{align}
    \tau_0 > b\sigma  - \frac{2 (k M+M \log (k-1))}{1-1/k}\,.
    \label{eq: shell-crossing-limit}
\end{align}
In this paper we consider $5\sigma$ (analogous to ``five-sigma standard deviation'' in particle physics) to be appropriate and useful
for ensuring \eqref{v-constraint}, so we set $b=5$, though this standard is mathematically somewhat arbitrary\footnote{In principle, we could simply consider a compactly supported function from the outset, but we choose this function for convenience since it is commonly used in the literature. Furthermore, in practice we will integrate numerically only over the strong support so it is effectively compactly supported; see e.g. \cite{Cong2020horizon} for the most recent work for harvesting with compact switching. We have checked that the essential physics is unchanged whether we use strong support or compactly supported switching functions.}.

Second, we know that the three standard Schwarzschild vacua have time-translation invariant Wightman functions with respect to the Killing time $\xi$. Therefore, the excitation probability $\Pr_j(\Omega,\sigma)$ is invariant under a constant shift of the switching time $\tau_0$. However, this is not the case for the non-local terms, as the two detectors at two different radii experience different gravitational redshift. Therefore the pullback of the Wightman functions to each detector's trajectory $\mathsf{W}(\sx_A(\tau),\sx_B(\tau'))$ will not be stationary, i.e. it is not a function of $\tau-\tau'$.

Third, to our knowledge most UDW model literature to date involves   sufficiently simple settings in which numerical integration can be performed relatively straightforwardly, and in some nice cases  closed-form expressions can be obtained (see e.g. remarkable calculations in \cite{pozas2015harvesting,pozas2016entanglement} for harvesting scenario, or \cite{Aubry2014derivative,Aubry2018Vaidya} for transition rate calculations). In these cases, often the symmetry of the problem allow exact expressions, and in the case of Unruh effect calculations, transition rate is simpler because it is a one-dimensional integral obtained using stationarity of the Wightman distributions. In other contexts such as \cite{henderson2018harvesting,henderson2019entangling,ng2018AdS}, the nice properties of AdS$_3$ spacetime allow analytic computation of both the Wightman functions and reduction of numerical integrals to one-dimensional integrals. The most formidable calculations of the two-point functions of this kind are done e.g. in \cite{Ng2014Schwarzschild,Jonsson2020:2002.05482v2}, though the objectives are different.

Here we are working with (1) a derivative coupling Wightman distribution, and also (2) a time-dependent collapsing spacetime, which renders the density matrix elements intractable analytically. Therefore a numerical approach is required to make progress. However, it is not hard to check by direct computation that the usual $\ii\epsilon$ prescription easily leads to numerical instabilities, and for Vaidya spacetime where the Wightman function has a very complicated pole structure, this is practically impossible without very careful and deliberate control of the integration schemes around the poles. In certain cases, such as the flat space Minkowski vacuum, it may be possible to deal with this by a suitable rewriting of the response function (see e.g. \cite{Satz2006howoften,Satz2007transitionrate,Aubry2014derivative}) in such a way that the $\ii\epsilon$ prescription is completely eliminated. However this is an exception to the rule; for example, a spacetime with a  static mirror at the origin cannot be dealt with this way as the mirror introduces new poles \cite{cong2019entanglement}. 

In view of the above difficulties, we will compute the joint detector density matrix elements $\rho_{AB}$   using \textit{numerical contour integration}. Formally, this is equivalent to the $\ii\epsilon$ prescription but instead of `shifting the poles' and taking $\epsilon\to0$ (which is numerically unstable in general), we will perform numerical integration that involves a contour in the complex plane. By making a suitable choice of contour that takes into account the exponential suppression of the Gaussian switching functions, we will be able to simplify the numerical integration considerably so that no complicated scheme is required. Furthermore, this also serves as a simple demonstration of how contour integration can be useful in a multi-dimensional integral settings that is relatively straightforward to implement as compared to bottom-up numerical schemes\footnote{Furthermore, in the Vaidya case most computations do not require us to evaluate the $\bar{u}v'$ and $v\bar{u}'$ contribution to $\mathcal{A}_V$ in Eq.~\eqref{eq: derivative-Vaidya} as they turn out to be subleading compared to the $\bar{u}\,\bar{u}'$ and $vv'$ contributions, thus cutting down some computation time.}. We describe this procedure in Appendix~\ref{appendix: numerical-contour}.

\section{Results}
\label{sec: results}
In this section we will calculate the amount of correlations that can be extracted by the two qubit detectors and compare the differences between the three preferred states for Schwarzschild background, namely Boulware vacuum $\ket{0_B}$, Unruh vacuum $\ket{0_U}$ and Hartle-Hawking-Israel (HHI) state $\ket{0_H}$. We will then compare this to the case where the two detectors are in the black hole exterior Region I of Vaidya spacetime, corresponding to detectors interacting after the black hole collapse has occurred. We will consider both concurrence and mutual correlations as measures of classical and quantum correlations between the two detectors. We close by commenting on the KMS property and detailed balance condition associated with these four vacua.

For numerical computations, we need to choose the nearest distance to the horizon for illustrating the physics very close to the horizon. Let $d_{ij}\coloneqq d(r_i,r_j)$ be the proper distance between two radial coordinates $r_i,r_j$. We will impose the condition that the nearest Alice's detector can get to the horizon is given by the proper distance
\begin{align}
    d_A\coloneqq d(r_A,r_H) \geq 0.1\sigma\,,
\end{align}
where $r_H=2M$ is the Schwarzschild radius and $\sigma$ is the switching timescale. We can therefore effectively think of $d_A=0.1\sigma$ as having Alice's detector to be just above the horizon. We will measure distances in units of $\sigma$ except in Section~\ref{sec: thermality} where we need to vary $\sigma$. Since by convention we take Bob's detector to be farther from the horizon than Alice's detector, we always have $r_B\geq r_A$. In principle, we could go nearer to, say, $d_A\lesssim 0.01\sigma$ (since we cannot numerically evaluate the density matrix at $r=r_H$); however this would take much more optimization and computational time to work with whilst not providing new insights. Our choice is simply a matter of (practical) convenience and simplicity that still includes the relevant physics.

As a side note, we remark that for derivative coupling UDW model, the coupling constant $\lambda$ has units of $[\text{Length}]^{\frac{n-1}{2}}$ while for amplitude coupling $\lambda$ has unit $[\text{Length}]^{\frac{n-3}{2}}$. Consequently, our results will be in terms of dimensionless coupling constant $\tilde\lambda \coloneqq \lambda \sigma^{\frac{1-n}{2}}$, where $n$ denotes the number of spatial dimensions. It happens that for derivative coupling in (1+1) dimensions, we have $\tilde\lambda=\lambda$ and we will write $\Tilde{\lambda}$ throughout to remind ourselves that in general coupling constant of UDW model has dimension-dependent units.

\subsection{Harvesting entanglement}

\begin{figure}[tp]
    \centering
    \includegraphics[scale=0.57]{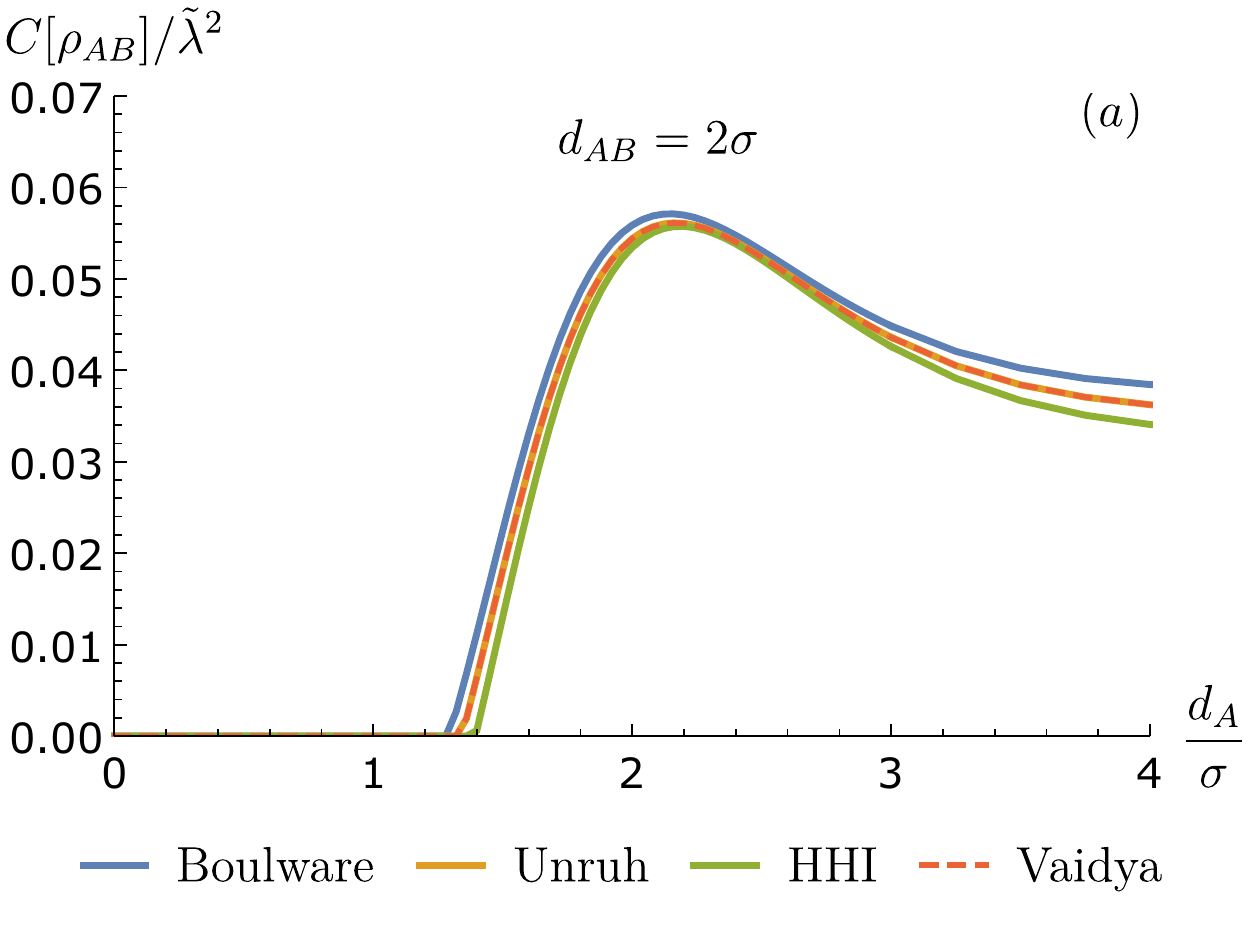}
    \includegraphics[scale=0.57]{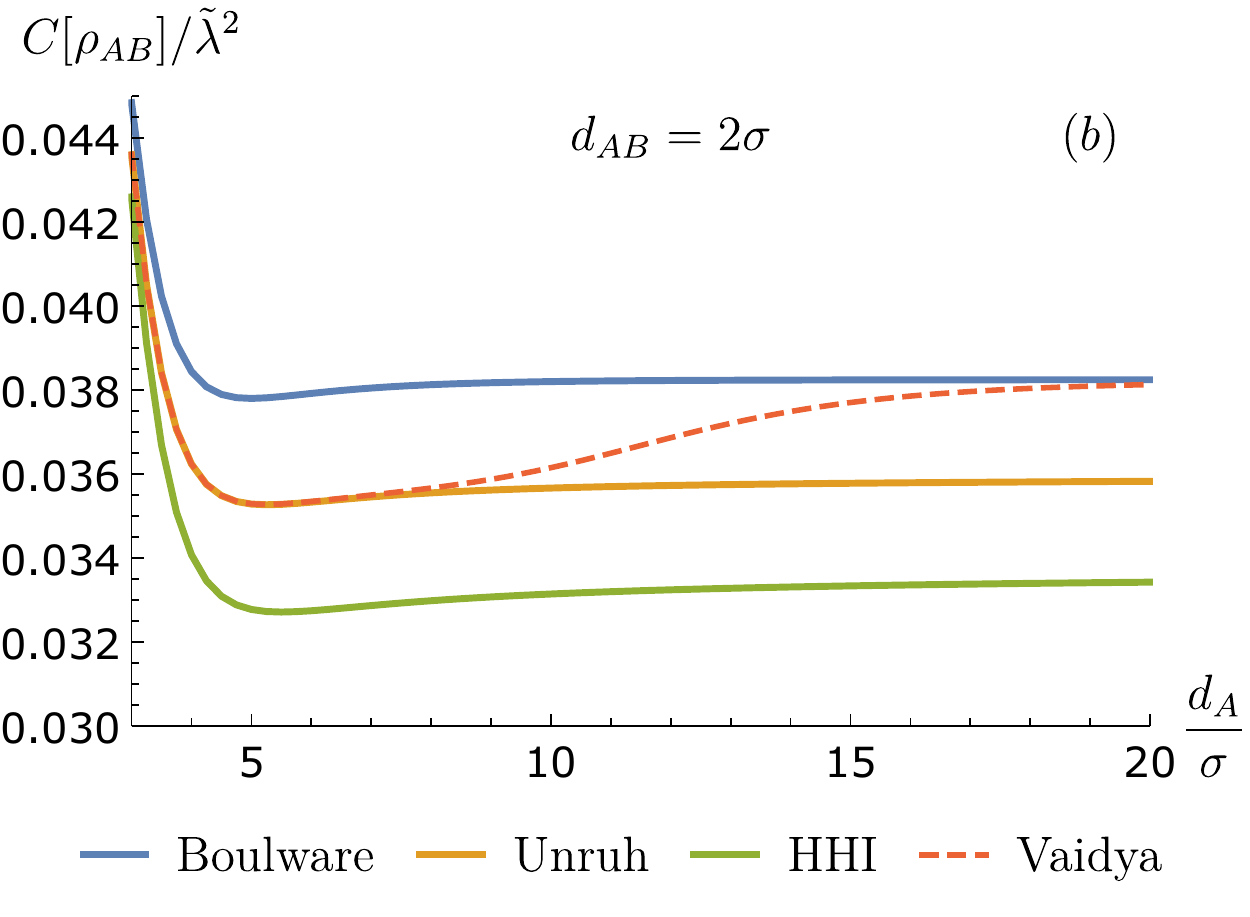}
    \includegraphics[scale=0.54]{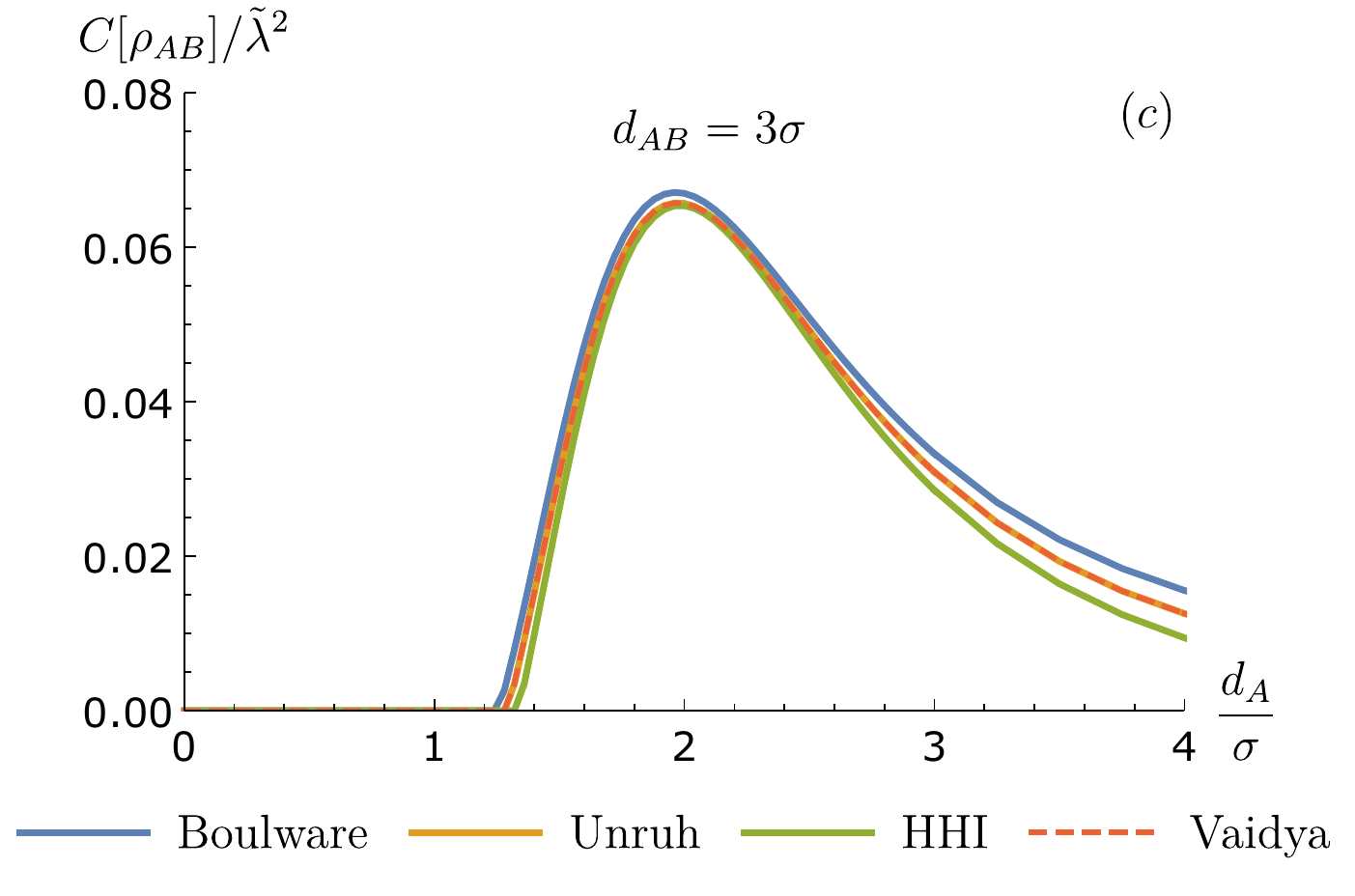}
    \includegraphics[scale=0.54]{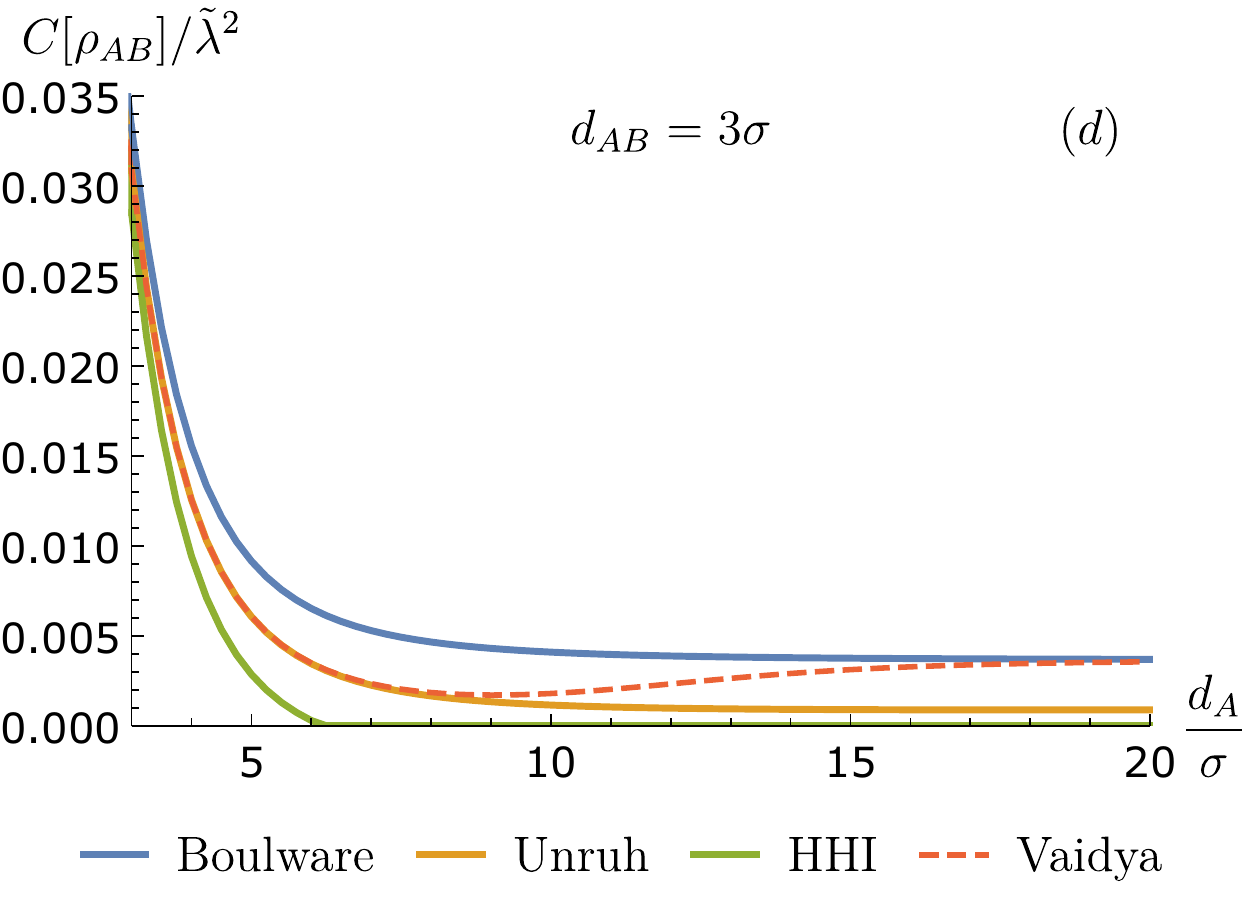}
    \caption{The concurrence as a function of proper distance of Alice's detector away from the horizon (in units of $\sigma$) for various choice of vacua. Here $\tilde\lambda = \lambda\sigma^{\frac{1-n}{2}}=\lambda$ is dimensionless coupling constant. We set $\Omega\sigma = 2$, $M/ \sigma = \frac{1}{2}$. The detectors are turned on at $\tau_0=12 \sigma$ so that the Gaussian switching peak is very far from the shell. \textbf{(a)} $d(r_A,r_B)=2\sigma$, near the horizon. \textbf{(b)} $d(r_A,r_B)=2\sigma$, far from the horizon. \textbf{(c)} $d(r_A,r_B)=3\sigma$, near the horizon. \textbf{(d)} $d(r_A,r_B)=3\sigma$, far from the horizon. }
    \label{fig: concurrence1}
\end{figure}

In Figure~\ref{fig: concurrence1} we show the concurrence as a function of proper distance of Alice's detector from the horizon $d_A$. Both Alice and Bob are static, non-inertial observers at fixed Schwarzschild radii $r_A,r_B$ respectively, separated by a fixed proper distance $d(r_A,r_B) = 2\sigma$ in \ref{fig: concurrence1}(a,b) and $d(r_A,r_B) = 3\sigma$ in in Figure~\ref{fig: concurrence1}(c,d). We first compare how the four vacua can entangle the two qubits after finite-time interaction.

First, observe that from Figure~~\ref{fig: concurrence1}(a) that there is an inhibition of entanglement extraction close to the horizon for all states, a result conjectured to hold in general   \cite{henderson2018harvesting} based on a study of this scenario for (2+1) BTZ black holes. Our results support the claim that this is a generic feature of a black hole background, since our choice of detector-field coupling and the choice of states are vastly different, and our example includes the Vaidya vacuum, which is not time-translation invariant in both the state and the black hole background. We also note that the region where concurrence is zero is slightly smaller for the Boulware vacuum and slightly larger for HHI vacuum, which is an indication that all black hole vacua do not have equal `entangling power' (to use the phrase in \cite{Steeg2009}). Furthermore, we see that the Unruh vacuum approximates the Vaidya vacuum very well near the horizon \textit{even for finite interactions}: this result therefore extends the utility of the Unruh vacuum in modelling the vacuum state for collapsing spacetime. For completeness, we note as well that with larger proper separation between the two detectors, entanglement harvesting is diminished and since each vacuum entangles differently, it is possible for some vacua to not be able entangle at some distance but other vacua could, as shown in Figure~\ref{fig: concurrence1}(d). 
This result is the black hole equivalent of that found both for  accelerating detectors and for comoving detectors in an expanding universe  \cite{Steeg2009, salton2015acceleration}.

\begin{figure}[tp]
    \centering
    \includegraphics[scale=0.52]{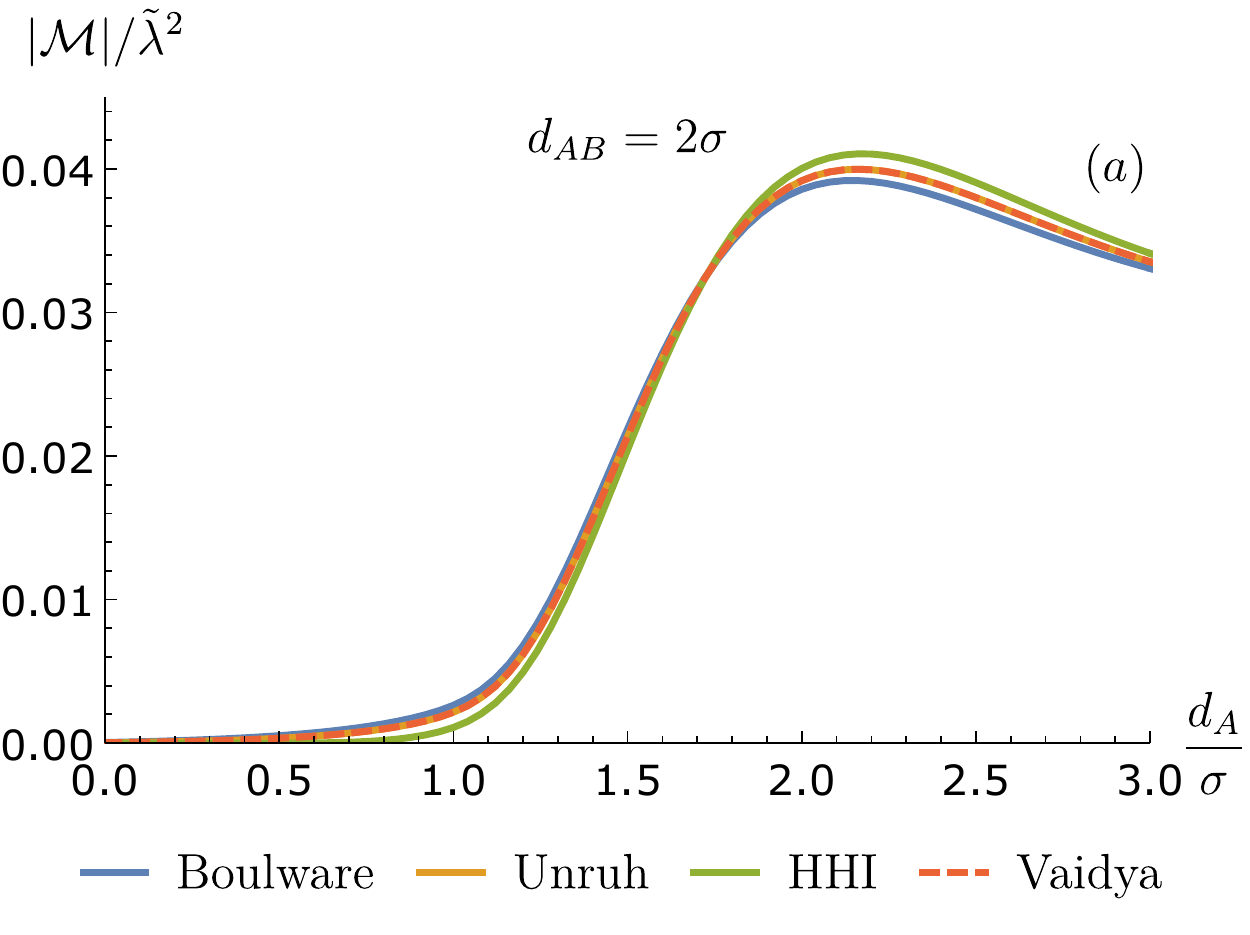}
    \includegraphics[scale=0.52]{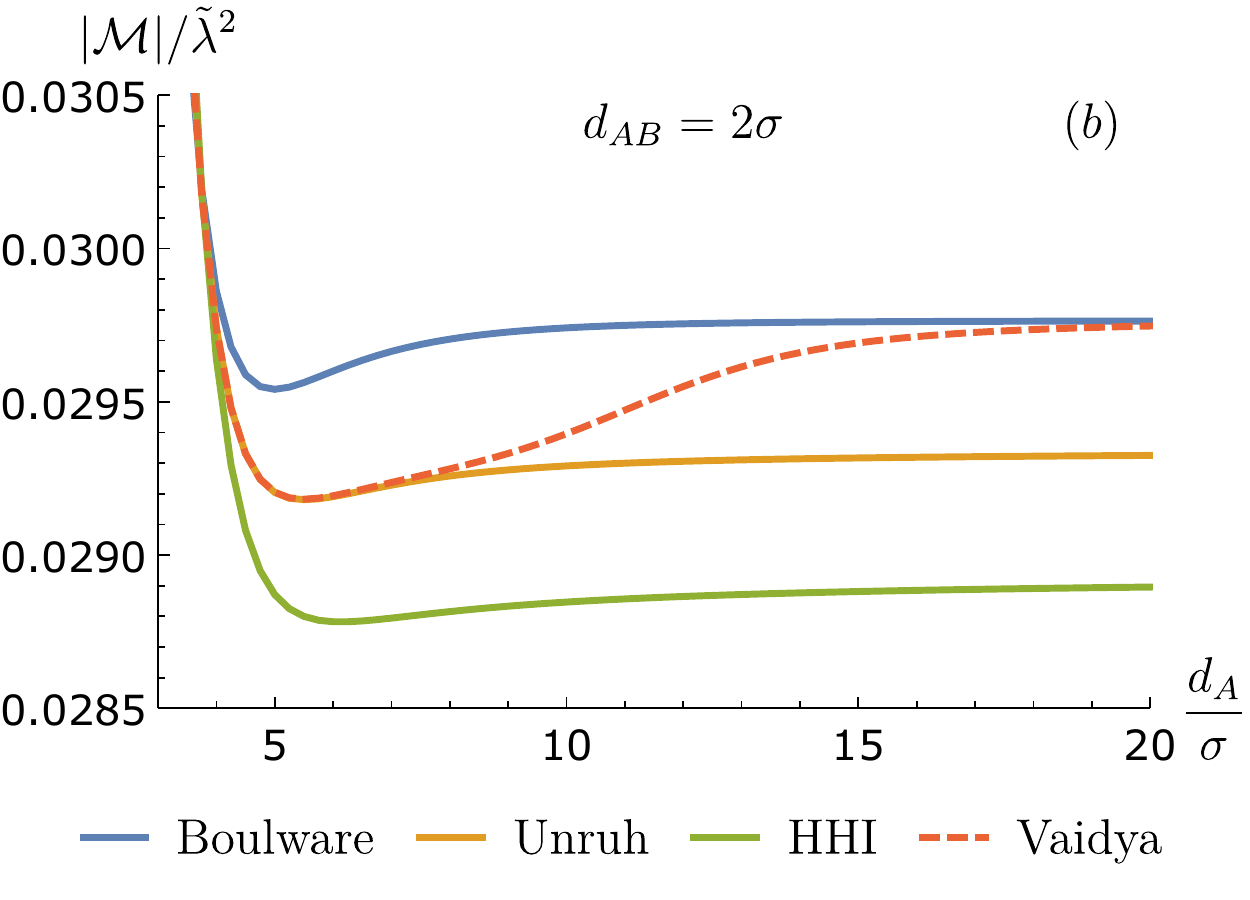}
    \includegraphics[scale=0.52]{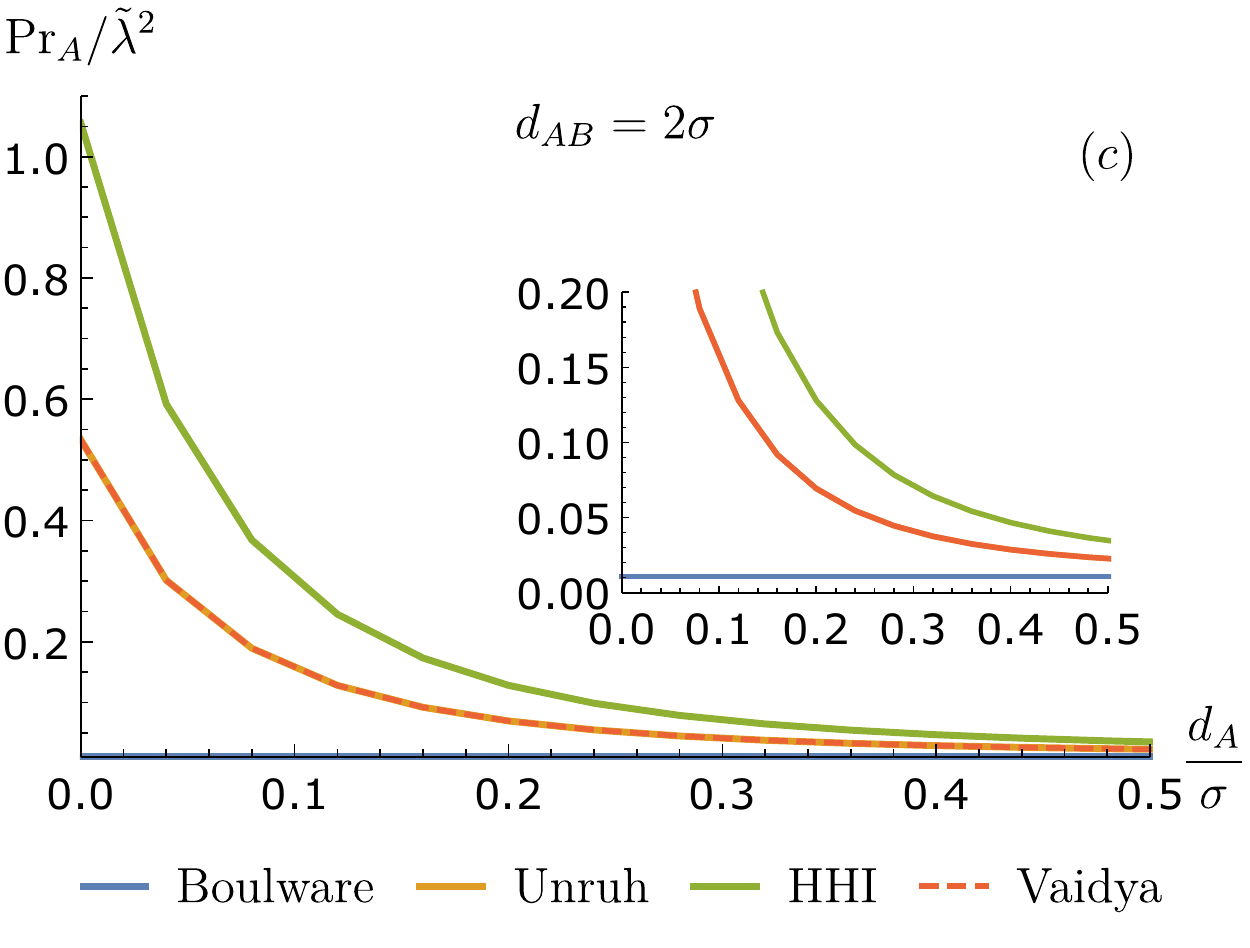}
    \includegraphics[scale=0.52]{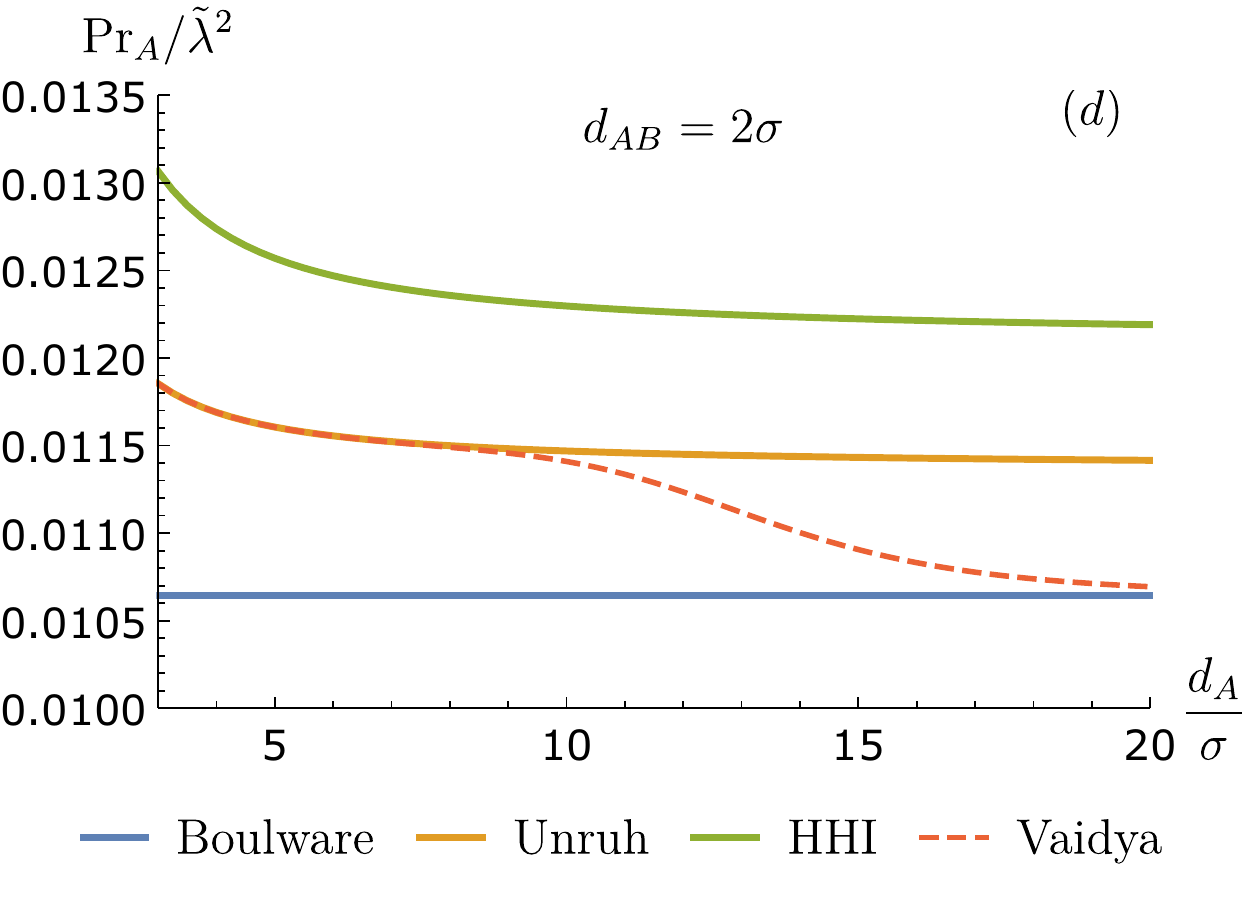}    
    \caption{The nonlocal contribution $\mathcal{M}$ and excitation probability of detector A, $\Pr_A\equiv \mathcal{L}_{AA}$, as a function of proper distance of detector A away from the horizon (in units of $\sigma$) for various choice of vacua. Here $\tilde\lambda = \lambda\sigma^{\frac{1-n}{2}}=\lambda$ is dimensionless coupling constant. We set $\Omega\sigma = 2$, $M/ \sigma = \frac{1}{2}$ and $d(r_A,r_B)=2\sigma$. The detectors are turned on at $\tau_0=12 \sigma$ so that the Gaussian switching peak is very far from the shell. \textbf{(a)} the nonlocal term $\mathcal{M}$. \textbf{(a) and (c):} Near the horizon. \textbf{(b) and (d):} Far from the horizon.}
    \label{fig: matrix-elements1}
\end{figure}

Secondly, for finite time interactions the Unruh vacuum no longer approximates well the Vaidya vacuum as the detectors move far away from the horizon. In Figure~\ref{fig: concurrence1}(b), we see that all four vacua distinguish themselves and all have different entangling power, and in particular we note that Vaidya vacuum is an interpolation of Unruh and Boulware vacuum: that is, the Vaidya vacuum is well-approximated by Boulware vacuum as measured by faraway observers while it is well-approximated by Unruh vacuum near the horizon. In Figure~\ref{fig: matrix-elements1} we separate the local noise contribution due to   detector A's excitation   and non-local contribution $\mathcal{M}$. In this particular example, the entangling power of the Vaidya vacuum is larger than the Unruh vacuum further from the horizon because the Boulware vacuum has a larger non-local term and smaller local noise. Again we observe that both local noise and non-local terms associated with the Vaidya vacuum interpolate between the Unruh and Boulware vacua; this suggests that the excellent approximation of the Vaidya vacuum by either Unruh or Boulware vacuum is generic and not unique to the  entanglement dynamics of the two detectors. We will see that this is true also for mutual information.

\begin{figure}[tp]
    \centering
    \includegraphics[scale=0.54]{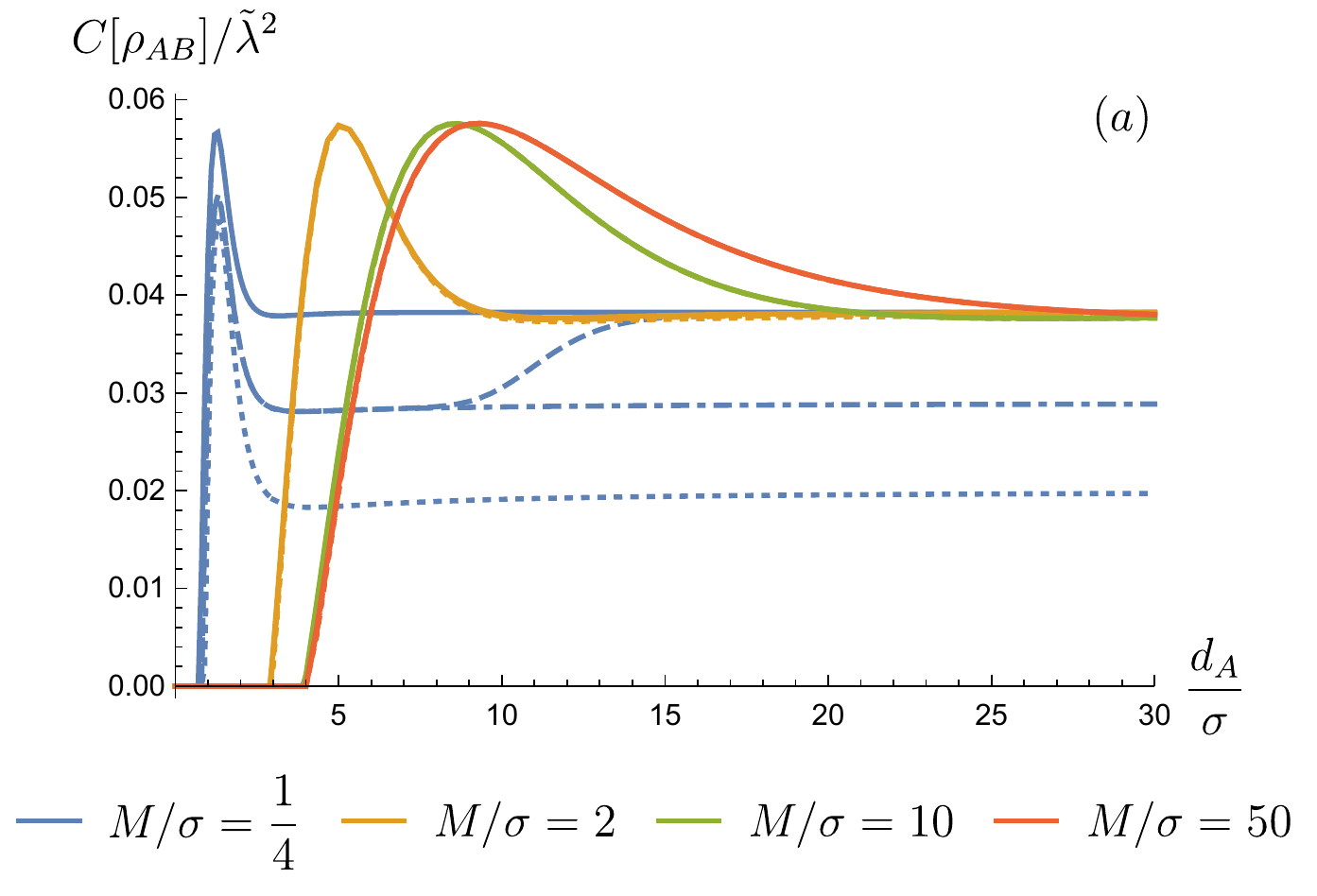}
    \includegraphics[scale=0.54]{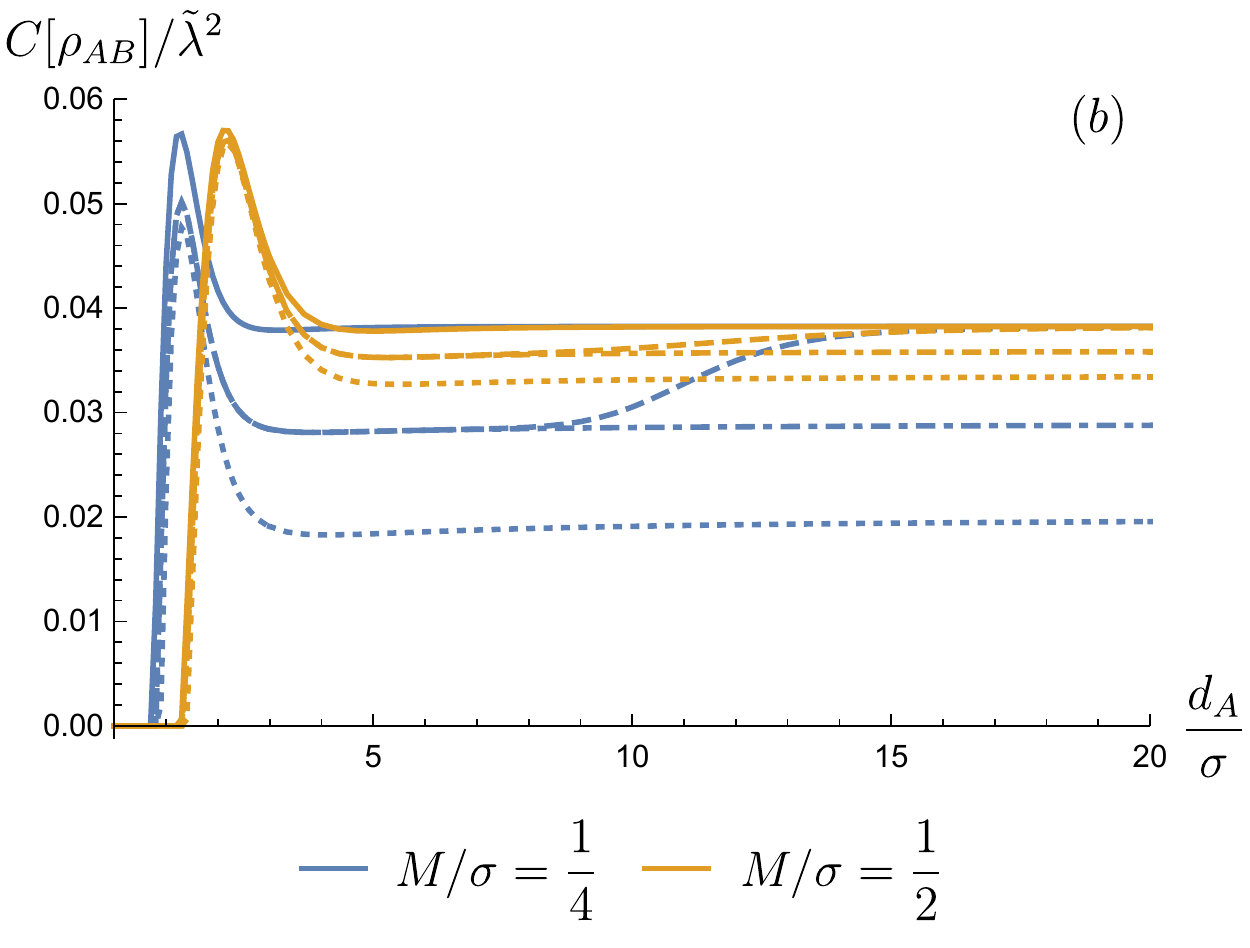}
    \caption{The concurrence as a function of proper distance of Alice's detector away from the horizon (in units of $\sigma$) for different black hole masses (in units of $\sigma$). In both figures, solid is Boulware, dotted is Hartle-Hawking-Israel,  dot-dash is Unruh, and dashed is Vaidya.  When the curves are indistinguishable, a solid curve is drawn. Here $\tilde\lambda = \lambda\sigma^{\frac{1-n}{2}}=\lambda$ is dimensionless coupling constant and we set $\Omega\sigma = 2$, with the two detectors   separated by proper distance $d(r_A,r_B)=2\sigma$. The detectors are turned on at $\tau_0=12 \sigma$. \textbf{(a)} the entanglement death zone is saturated for large masses, at $d_A\sim 4\sigma$ for $M/\sigma\geq 10$, and at large masses the concurrence tends to the zero-temperature Boulware limit. \textbf{(b)} Shrinking of the difference in entangling power between the four vacua as mass increases.
    }
    \label{fig: concurrence2-mass}
\end{figure}

 In Figure~\ref{fig: concurrence2-mass} we depict how concurrence varies with black hole mass. From Figure~\eqref{fig: concurrence2-mass}(a), we can make two observations. First, we see that as black hole mass increases, the entanglement death zone increases until at some point it is saturated for large enough mass. In our example, all vacua for $M/\sigma\geq 10$ have the same death zone, given by $d_A\sim 4\sigma$. Second, for large masses, the differences between the different vacua shrinks very quickly: in our example, for $M/\sigma\geq 2$ the four vacua (marked by different line style\footnote{Solid line: Boulware, dot-dashed line: Unruh, dotted: HHI, dashed: Vaidya.}) are practically indistinguishable from one another in the plot. Figure~\ref{fig: concurrence2-mass} shows how a small increase in mass (in units of $\sigma$) already shrinks the difference considerably, hence increasing the mass reduces the difference in entangling power of the four vacua. Note that for large masses, the curves for the four vacua overlap and approach the Boulware limit at large distances. This behaviour has a natural interpretation: it can be understood from the fact that as mass increases, the Hawking temperature decreases and hence in the limit of very large mass, the concurrence approaches that of zero-temperature vacuum in the sense of KMS condition \cite{Takagi1986noise}, i.e. the Boulware vacuum (see Section~\ref{sec: thermality} for brief discussion on thermality). Such small-mass distinctions are also present for the BTZ black hole \cite{henderson2018harvesting}.

\subsection{Harvesting mutual information}

\begin{figure}[tp]
    \centering
    \includegraphics[scale=0.54]{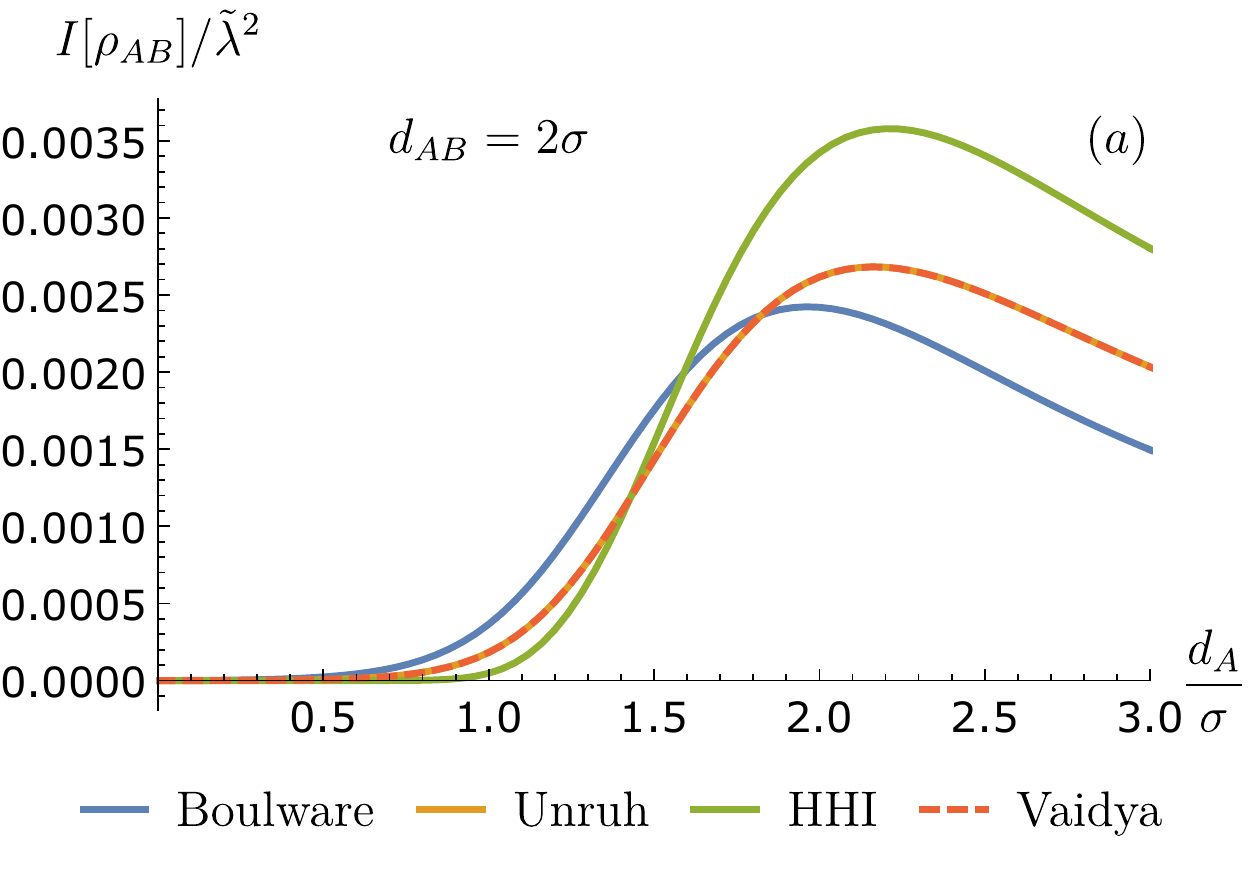}
    \includegraphics[scale=0.54]{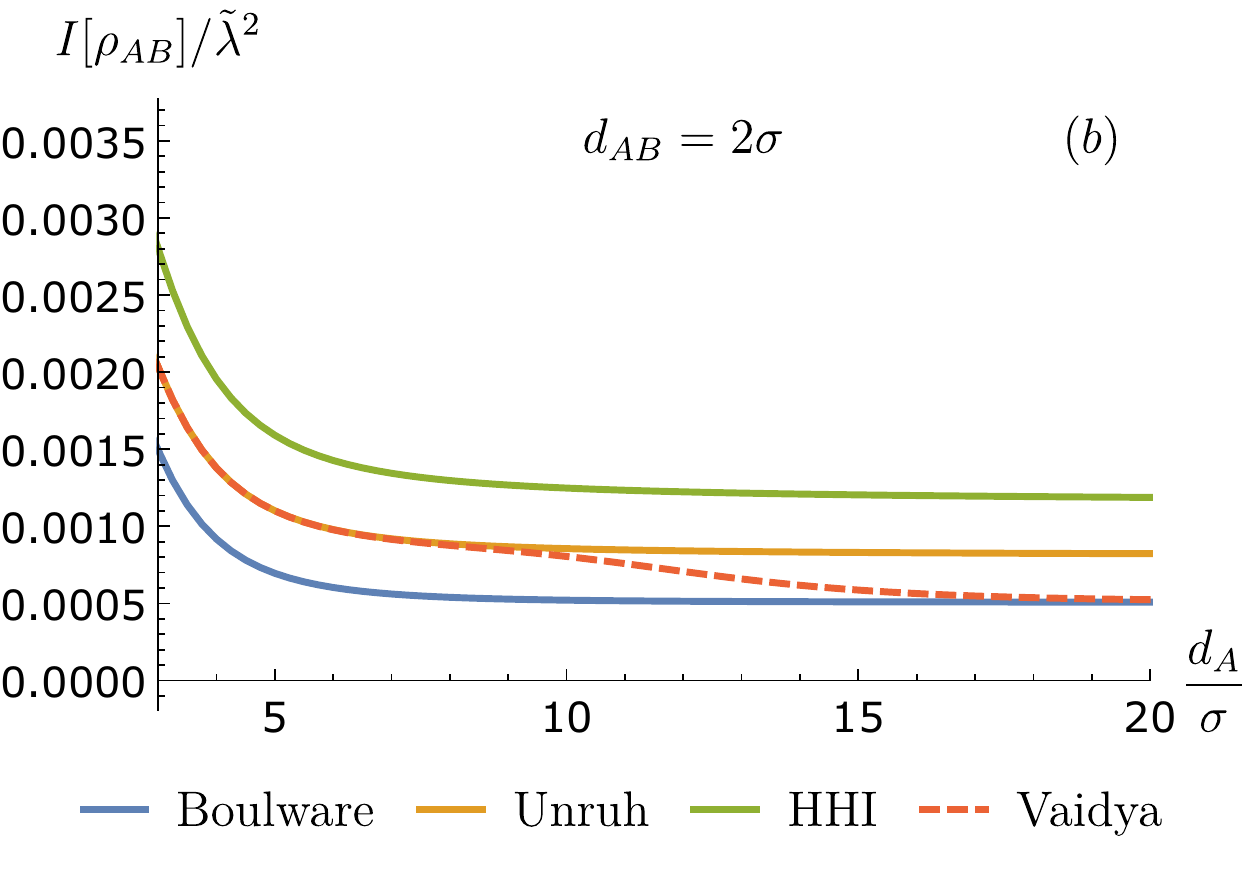}    
    \includegraphics[scale=0.54]{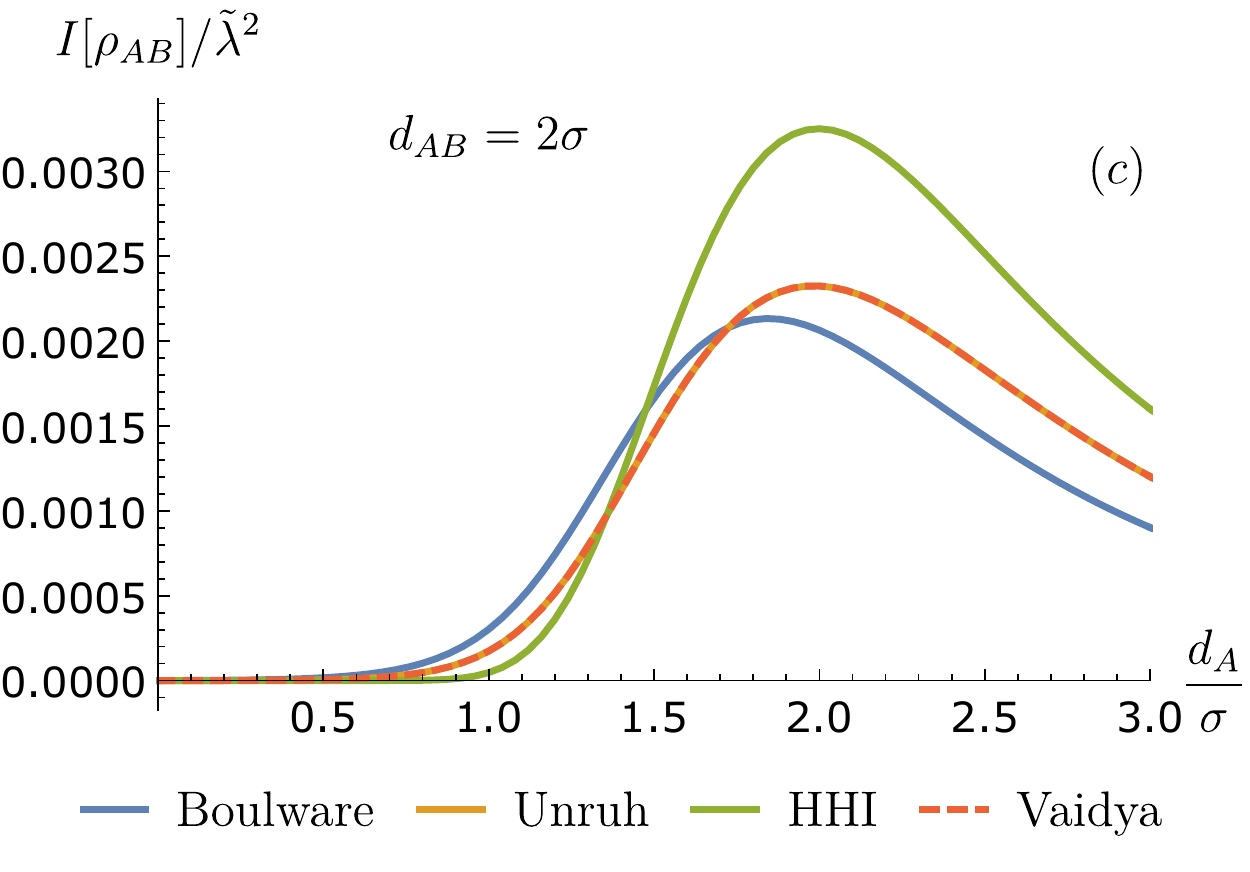}
    \includegraphics[scale=0.54]{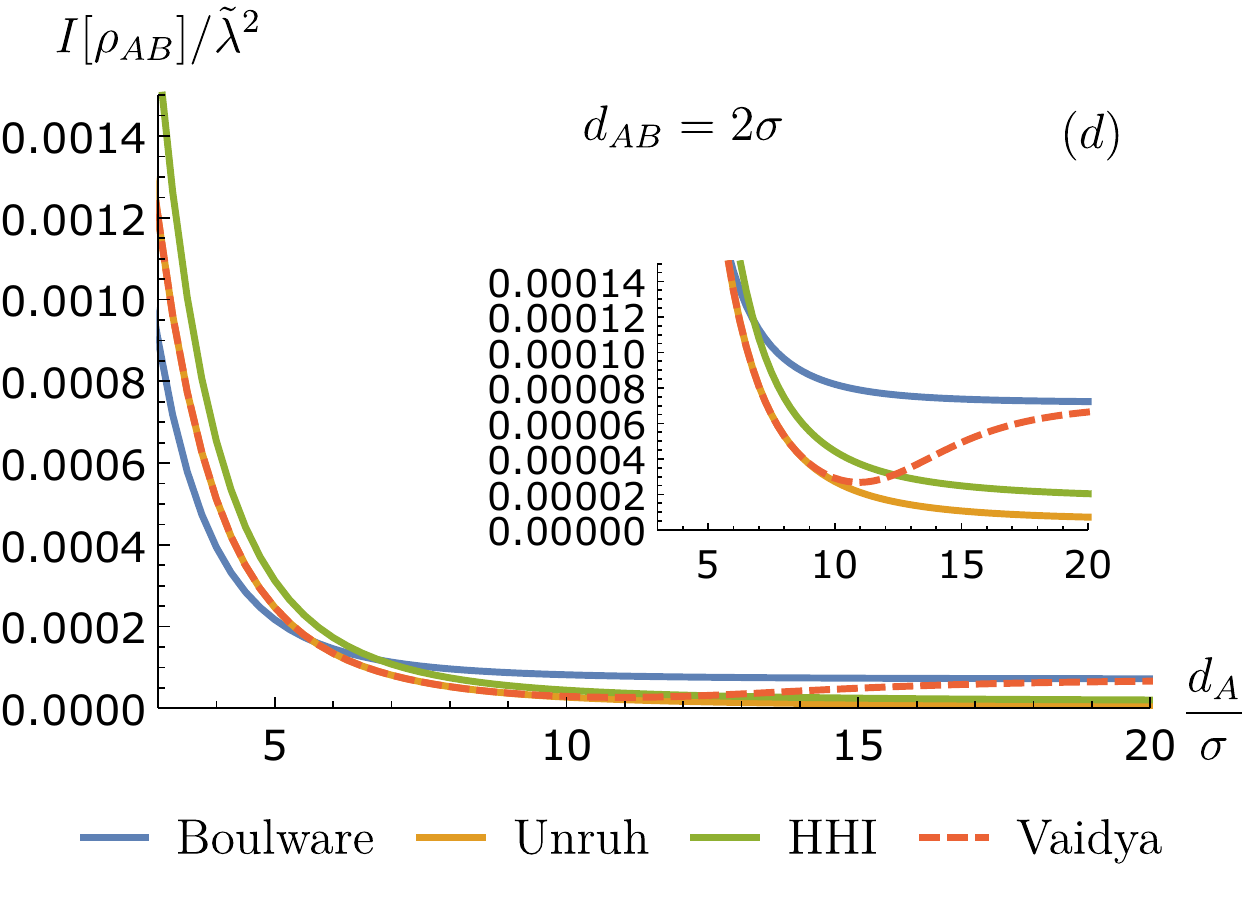}
    \caption{Mutual information as a function of proper distance of detector A away from the horizon (in units of $\sigma$) for various choice of vacua. Here $\tilde\lambda = \lambda\sigma^{\frac{1-n}{2}}=\lambda$ is dimensionless coupling constant. We set $\Omega\sigma = 2$, $M/ \sigma = \frac{1}{2}$. The detectors are turned on at $\tau_0=12 \sigma$ so that the Gaussian switching peak is very far from the shell. \textbf{(a)} the nonlocal term $\mathcal{M}$. \textbf{(a) and (b):} $d(r_A,r_B) = 2\sigma$. \textbf{(c) and (d):} $d(r_A,r_B) = 3\sigma$.}
    \label{fig: mutual-information1}
\end{figure}

So far we have looked at entanglement dynamics of the two detectors in a black hole background for various state. We now consider another correlation measure: mutual information.

\begin{figure}[tp]
    \centering
    \includegraphics[scale=0.55]{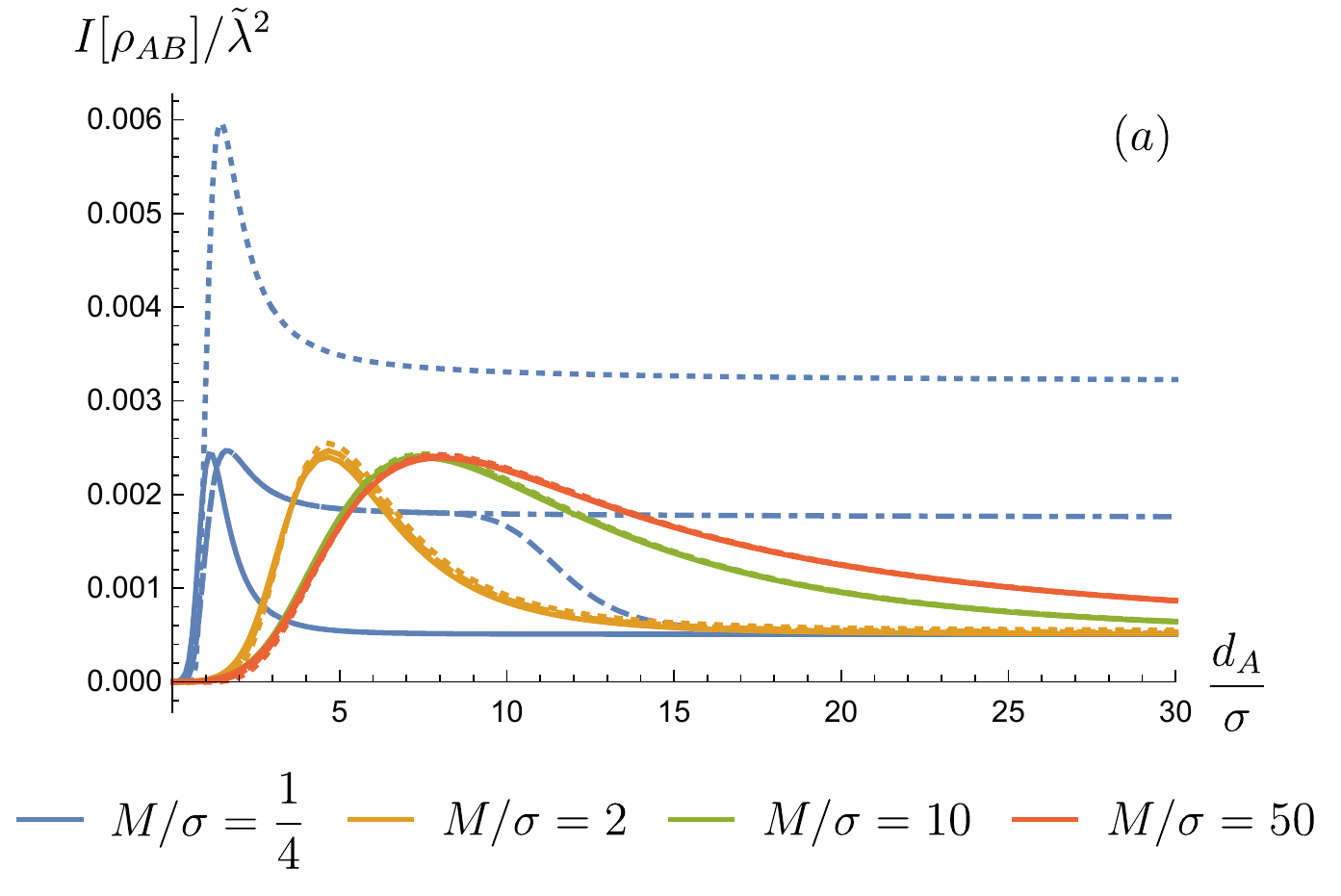}
    \includegraphics[scale=0.55]{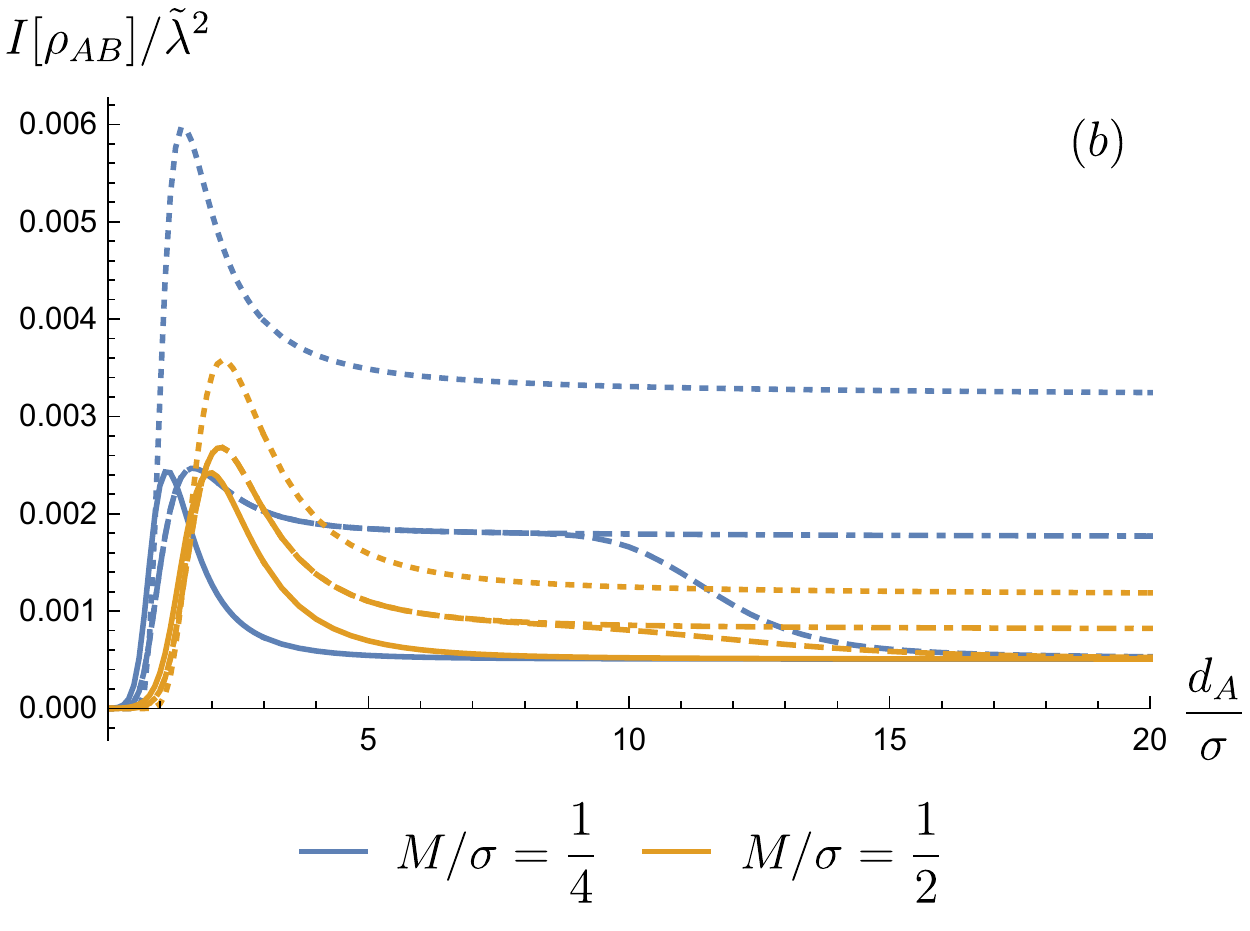}
    \caption{Mutual information as a function of proper distance of detector A away from the horizon (in units of $\sigma$) for various choice of vacua and for different black hole masses. In both figures, solid is Boulware, dotted is Hartle-Hawking-Israel,  dot-dash is Unruh, and dashed is Vaidya.  When the curves are indistinguishable, a solid curve is drawn.  Here $\tilde\lambda = \lambda\sigma^{\frac{1-n}{2}}=\lambda$ is dimensionless coupling constant. We set $\Omega\sigma = 2$ with detector separation $d(r_A,r_B)=2\sigma$, and the detectors are turned on at $\tau_0=12 \sigma$ so that the Gaussian switching peak is very far from the shell. Note that for smaller mass black hole, the `bifurcation' between Unruh and Vaidya vacuum occurs at larger radius.}
    \label{fig: mutual-information2-mass}
\end{figure}

In Figures~\ref{fig: mutual-information1} and~\ref{fig: mutual-information2-mass} we plot mutual information as a function of distance from the horizon for the same parameter choices as Figure~\ref{fig: concurrence1}. There are four observations that can be made here. First, from the right-hand diagrams in
figure~\ref{fig: mutual-information1}, again we see that the the Vaidya vacuum interpolates between  the Unruh and Boulware vacua insofar as mutual information is concerned. Second, we observe that far from the horizon, how much mutual information can be extracted from the Vaidya vacuum relative to the Unruh vacuum depends on detector separation. Comparing the right-hand diagrams in Figure~\ref{fig: mutual-information1}, larger detector separation tends to make the Vaidya vacuum better at `imparting' mutual information to the detectors than smaller separation, unlike concurrence where the Vaidya vacuum persistently `outperforms' the Unruh vacuum at large distances. 

 Third, the dependence of mutual correlation harvesting on black hole mass is similar to the concurrence, as we show in Figure~\ref{fig: mutual-information2-mass} (compare this with Figure~\ref{fig: concurrence2-mass}). We again see that the differences in how much mutual information can be extracted from the four vacua shrinks quickly with increasing mass, and that for large masses the mutual information harvesting approaches the Boulware limit. Mutual information also seems to exhibit some saturation near the horizon: in our example, $M/\sigma\geq 10$ seems to share similar mutual information for $d_A\sim 7.5\sigma$, and starts to show differences farther away.

 The third and the most important observation of this section concerns the behaviour of mutual information dynamics near the horizon. From Figures~\ref{fig: mutual-information1}(a,c) we see that, independently of the mass, mutual information vanishes quickly as detectors approach the horizon. Furthermore, near the horizon (for  every $d_A>0$) the amount of mutual information that can be harvested can be nonzero, even with zero entanglement (cf. Figure~\ref{fig: concurrence1}).   Thus very close to the horizon the detectors contain mostly non-entangling correlations. However, the mutual information quickly vanishes as the detectors approach the horizon regardless of detector separation. Figures~\ref{fig: mutual-information1}(a,c) show that even though mutual information can be extracted outside the horizon, the black hole horizon effectively extinguishes  \textit{all} correlations --- classical or quantum, at least for static detectors. 
 
 This result suggests an operational interpretation for detectors in static trajectory (constant $r$) that a black hole horizon is a null surface that breaks correlations: that is, two static detectors are increasingly unable to harvest \textit{any} correlations from the vacuum as they get closer to the horizon, and this is independent of which vacuum one uses for the description of the field's ground state. Our results therefore strengthen what was found in \cite{henderson2018harvesting}, in that not only does the horizon inhibit entanglement harvesting, in fact it also suppresses harvesting of \textit{any correlations}, regardless of the choice of black hole vacuum state. This is most likely related to the fact that detectors cannot maintain static trajectory at $r=2M$, thus it is very interesting to see how different the outcome would be when the detectors approach the horizon in free-falling motion; we leave this for future investigations.

\subsection{Communication between detectors: how timelike are the correlations?}

Another natural question to ask in the harvesting protocol is how much of the extracted correlations come from mutual signalling between the two detectors. So far in the literature, little attention has been paid to communication between the two detectors when it comes to harvesting correlations in curved spacetimes. The ability of the detectors to communicate important because two uncorrelated quantum systems can be correlated or entangled via signalling or their mutual interactions\footnote{In order to generate entanglement, one would need nonlocal operations in general \cite{nielsen2000quantum}. For example any LOCC (local operations and classical communications) cannot generate entanglement from uncorrelated state or increase the entanglement rank.}.

 In the UDW model, although two detectors interact locally with a common quantum field, due to relativistic causality the two detectors' ability to signal depends on their spacetime separations. In flat space, the notion of spacelike and timelike separation is straightforward and can be given in terms of the coordinate separation $Y^\mu\coloneqq  \sx_{A}^\mu-\sx_B^\mu$. In other words, if $Y^\mu Y_\mu \leq 0$ then the two points are causally connected (see e.g. \cite{pozas2015harvesting}). When non-compact switching or smearing is involved (such as via Gaussian functions), then one can define two detectors to be spacelike separated whenever the strong support of the switching or smearing functions of one detector is within another detector's causal complement. Relativistic causality of the underlying quantum field in flat space then demands that if the two points are spacelike separated ($Y^\mu Y_\mu>0$), then the any local observables constructed out of the field operators vanish:
\begin{align}
    [\hat O(\sx_A),\hat O(\sx_B)] = 0\,.
    \label{eq: microcausality}
\end{align}
Consequently, communication between detectors can be measured in terms of `signalling estimators' constructed out of the field commutators \cite{Causality2015Eduardo}.

It is worth noting that the microcausality condition \eqref{eq: microcausality} is quite simple to compute even when the background spacetime is curved if the Fourier mode decomposition of the underlying quantum field is known. This is interesting because in curved spacetimes it is generically very difficult in practice to characterize spacelike separation even classically: in presence of curvature, one has to show that the two points cannot be connected by any causal curve \cite{wald2010general}. This is especially prohibitive in practice for our detectors because we have to check how the entire Gaussian strong supports of the detectors are contained in each other's causal complement. Here we have a situation where quantum theory simplifies our task of quantifying communication between two detectors at two causally disjoint spacetime regions.

\begin{figure}[tp]
    \centering
    \includegraphics[scale=0.54]{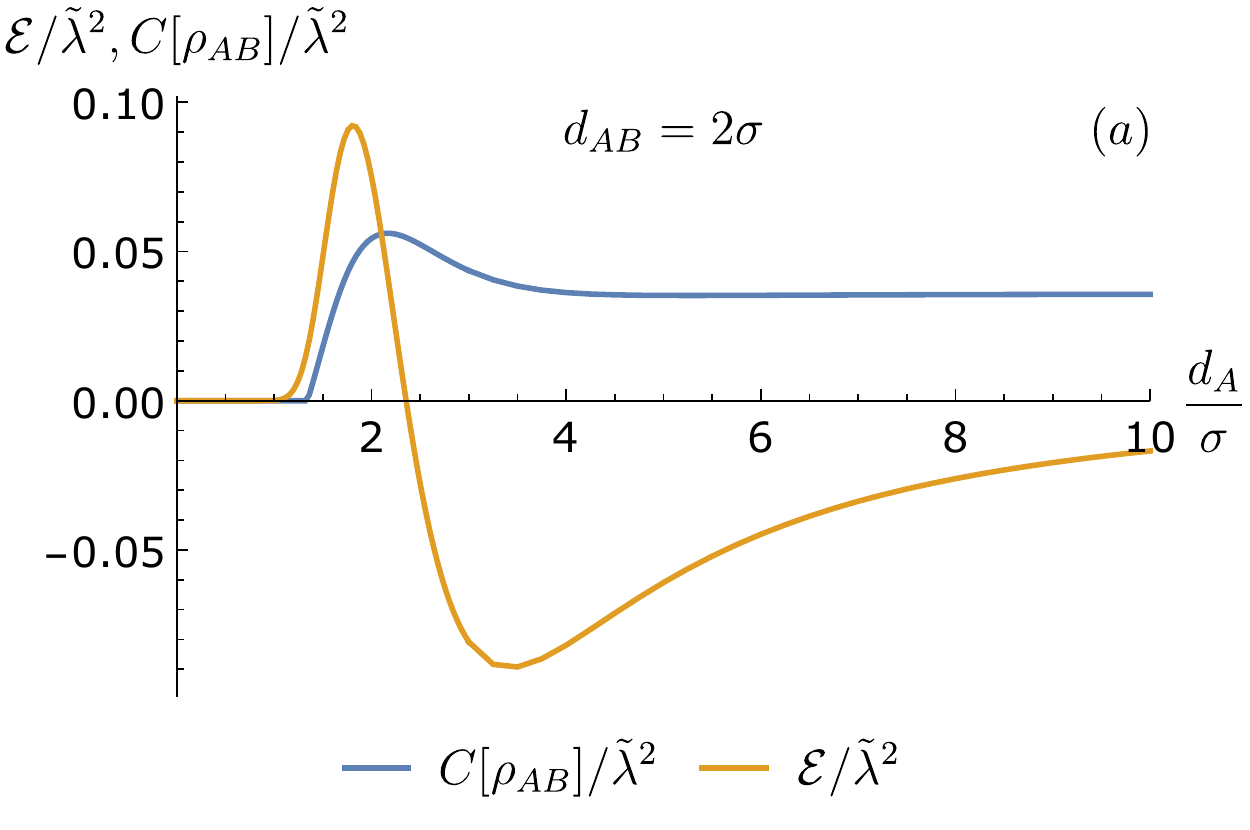}
    \includegraphics[scale=0.54]{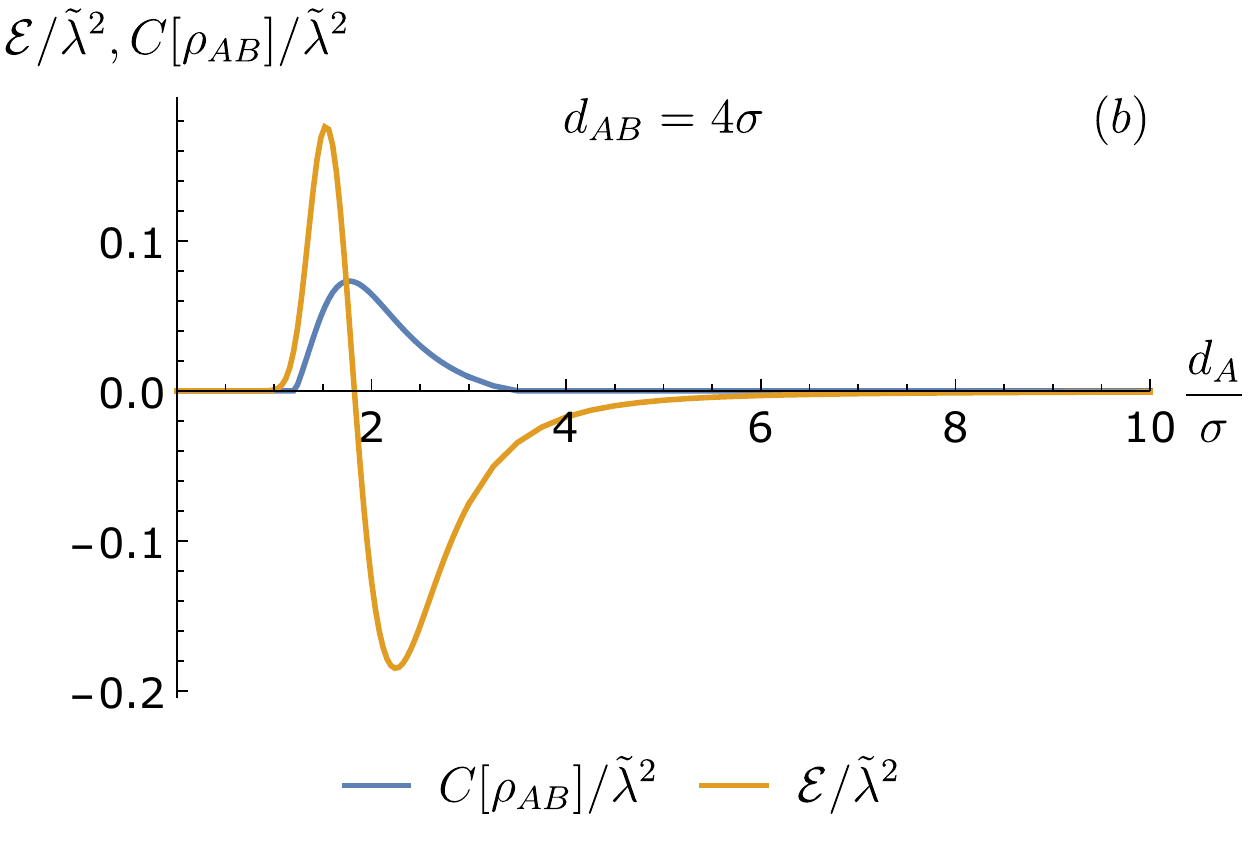}
    \includegraphics[scale=0.54]{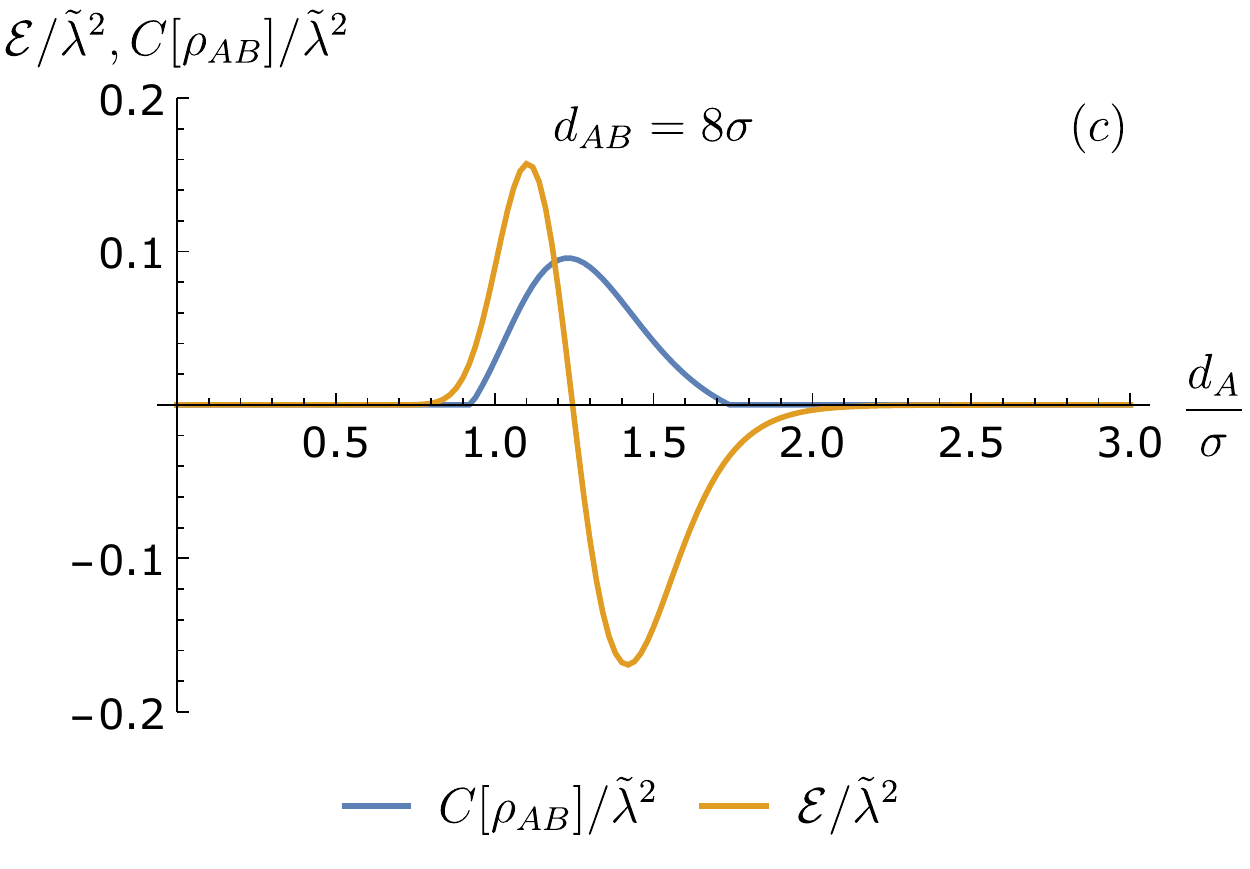}
    \includegraphics[scale=0.54]{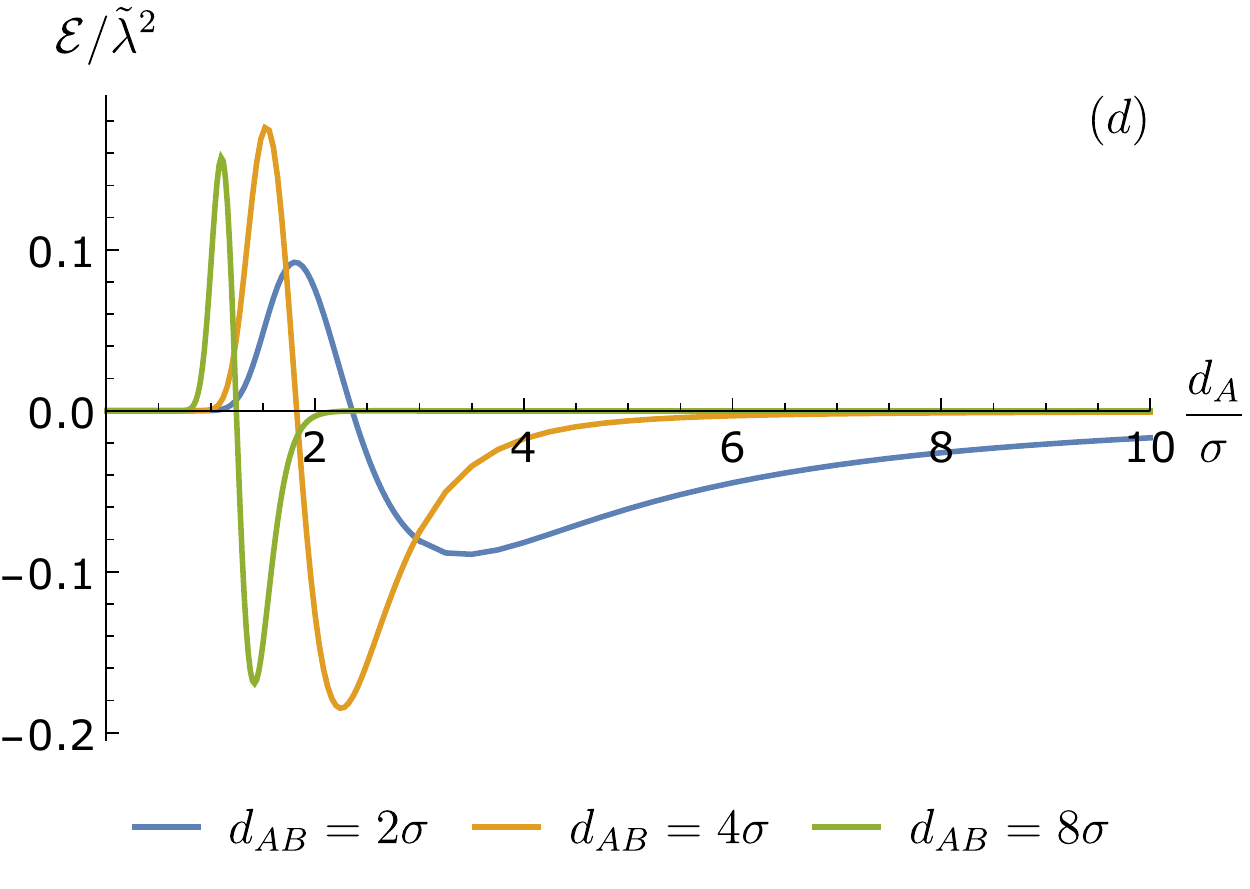}
    \caption{Concurrence $C[\rho_{AB}]$ and signalling estimator $\mathcal{E}$ as a function of proper distance of Alice's detector from the horizon for various detector separations $d_{AB}$. Here $\Omega\sigma=2$, $M/\sigma=1/2$. The concurrence is computed using the Unruh vacuum case as a reference. \textbf{(a)} $d_{AB} = 2\sigma$/ \textbf{(b)} $d_{AB} = 4\sigma$. \textbf{(c)} $d_{AB} = 8\sigma$. \textbf{(d)} the signalling estimators for various $d_{AB}$. Note that the dominant part of $\mathcal{E}$ is increasingly concentrated to a small region where entanglement can be extracted by the detectors as they become more spacelike separated.}
    \label{fig: causality-estimate}
\end{figure}

Let us construct a signalling estimator inspired by the construction in \cite{Causality2015Eduardo}: we define
\begin{align}
    \mathcal{E} \coloneqq -\frac{1}{2}\lambda_A\lambda_B\text{Im}\left(\int \dd\tau_A\dd\tau_B\,\chi_A(\tau_A)\chi_B(\tau_B)\braket{ 0_\alpha|[\pd_{\tau_A}\hat\phi(\sx_A(\tau_A)),\pd_{\tau_B}\hat\phi(\sx_B(\tau_B))]|0_\alpha }\right)\,,
\end{align}
where $\alpha=B,U,H,V$ label different vacua and we take the imaginary part since $\mathcal{E}$ is purely imaginary. The factor $-1/2$ is arbitrarily chosen so that in Figure~\ref{fig: causality-estimate} the magnitude of $\mathcal{E}$ is comparable to the concurrence, non-negative near the horizon and aids visualization. Note that we have used the proper time derivative $\pd_\tau\hat\phi(\sx(\tau))$ instead of the field operator $\hat\phi(\sx(\tau))$ because the commutator can be easily computed from the derivative Wightman function: as a distribution, this is given by
\begin{align}
    \braket{0_\alpha| [\pd_{\tau_A}\hat\phi(\sx_A(\tau_A)),\pd_{\tau_B}\hat\phi(\sx_B(\tau_B))] |0_\alpha} = \mathcal{A}_\alpha(\sx_A(\tau_A),\sx_B(\tau_B)) - \mathcal{A}_\alpha(\sx_B(\tau_B),\sx_A(\tau_A))\,.
\end{align}
Since the field commutator is state-independent we can drop the label $\alpha$, and also we assumed that the two detectors are identical. Therefore, the estimator can be simplified into
\begin{align}
    \mathcal{E} = -\frac{1}{2}\lambda^2\text{Im}\left(\int \dd\tau_A\dd\tau_B\,\chi(\tau_A)\chi(\tau_B)\bra{0} [\pd_{\tau_A}\hat\phi(\sx_A(\tau_A)),\pd_{\tau_B}\hat\phi(\sx_B(\tau_B))]\ket{0}\right)\,.
    \label{eq: signalling-estimator}
\end{align}
Certainly one could construct other estimators using operators associated with the field, but due to Eq.~\eqref{eq: microcausality} all observables constructed out of the field operators will behave similarly for spacelike separated regions. The estimators will only give different behaviour for different field observables when the detectors are timelike separated\footnote{By this we mean that if $\mathcal S_j$ is the strong support of the Gaussian switching of the detector $j$, then timelike separated here means $\mathcal S_A$ is contained within within causal past/future of $\mathcal S_B$.}.

We superimpose the concurrence for various detector separations with the signalling estimator $\mathcal{E}$ as shown in Figure~\ref{fig: causality-estimate}(a)-(c), using Unruh vacuum as a reference state for concurrence. We can make two important observations here. First, when the detectors become increasingly spacelike, the signalling estimator $\mathcal{E}$ is strongly confined to where the concurrence is nonzero. Therefore, at large distances the estimator vanishes for spacelike separation. This is especially manifest in Figure~\ref{fig: causality-estimate} where $d_{AB}=8\sigma$ would correspond to spacelike separated detectors if the spacetime were flat. In Figure~\ref{fig: causality-estimate}(d) we plot the estimators together for different spacelike separation and we see that the dominant part gets more concentrated to smaller spatial regions.

The second observation is that the estimator suggests that as we bring the two detectors very close to the black hole, the detectors very quickly become spacelike. The estimators fall off very quickly near the horizon. This is very remarkable because it shows that curvature modifies communication in highly non-trivial way: the signalling estimator $\mathcal{E}$ is effectively zero near the horizon, followed by an intermediate region the signalling is very enhanced (large $|\mathcal{E}|$), before eventually falling off again as one moves towards spatial infinity\footnote{ We have also checked the estimator when compact switching in \cite{Cong2020horizon} with approximately equal area as the Gaussian switching is used instead and the signalling estimator remains very similar. This provides an indication that despite the non-compact property of the Gaussian switching, the calculations done in this paper work as required.}. This observation demonstrates that in entanglement harvesting protocol, curvature and communication between detectors are very much related: curvature modifies causal relations between the two detectors as we move the them across different distances from the horizon.

Overall, while we do not explore the (extremely vast) parameter space of our setup, the signalling analysis highlights a mechanism through which entanglement harvesting between two detectors at the black hole exterior occurs: the efficiency of the protocol depends strongly on the ability of detectors to signal between them. The very specific issue of quantum communication in (3+1)D Schwarzschild black hole where angular variables of the metric have important role has been very recently investigated using state-of-the-art calculations in \cite{Jonsson2020:2002.05482v2}.

\subsection{Vaidya vacuum: near/far from horizon and early/late time limits}
\label{sec: late-time}

Our results thus far suggest that the Vaidya vacuum is an interpolation of the Unruh and Boulware vacua as we move from the horizon towards infinity. In order to better understand these, let us study the late-time and large distance limit of the respective Wightman functions. We stress that our notion of `late time' is not the same as \cite{Aubry2018Vaidya}: late-time means the detectors are turned on for \textit{finite} duration ($\sigma<\infty$) but the peak of the switching function is large ($\tau_0\to\infty$).

First, note that by taking the limit $r\gg 2M$ at \textit{fixed} coordinate time $t$ (or proper time $\tau$), the pullback of the Wightman function for the Boulware state approaches the Minkowski value:
\begin{align}
    \mathcal{A}_B\rr{\sx_A(\tau),\sx_B(\tau')} 
    &\sim -\frac{1}{4\pi}\rr{ \frac{1}{(r_A-r_B-(\tau-\tau'-\ii\epsilon))^2}+\frac{1}{(r_A-r_B+(\tau-\tau'-\ii\epsilon))^2}}\notag\\
    &\equiv \mathcal{A}_M(\sx_A(\tau),\sx_B(\tau'))\,,
\end{align}
where $\mathcal{A}_M(\sx_A(\tau),\sx_B(\tau'))$ is the derivative coupling Wightman function for \textit{Minkowski} vacuum (i.e. derivative version of Eq.~\eqref{eq: Wightman-flat-0} and \cite{Aubry2014derivative}). Second, for the HHI vacuum the Wightman function would approach that of a thermal bath in Minkowski space  with temperature $T_H = (8\pi M)^{-1}$ (cf. \cite{Weldon2000thermal}):
\begin{align}
    \mathcal{A}_H\rr{\sx_A(\tau),\sx_B(\tau')}\sim  -\frac{1}{4\pi}\frac{\text{csch}^2\left(\frac{r_A-r_B-(\tau-\tau'-\ii\epsilon)}{8 M}\right)+\text{csch}^2\left(\frac{r_A-r_B+(\tau-\tau'-\ii\epsilon)}{8 M}\right)}{64 M^2}\,.
\end{align}
For the Unruh vacuum, the Wightman function would approach the ``average'' of Minkowski vacuum and thermal bath:
\begin{align}
    \mathcal{A}_U\rr{\sx_A(\tau),\sx_B(\tau')}\sim  -\frac{1}{4\pi}\left[\frac{1}{(r_A-r_B+(\tau-\tau'-\ii\epsilon))^2}+ \frac{\text{csch}^2\left(\frac{r_A-r_B-(\tau-\tau'-\ii\epsilon)}{8 M}\right)}{64 M^2}\right]\,.
\end{align}
This averaging makes sense because the Wightman function is constructed by removing ``half'' of the HHI vacuum's radiation --- the ingoing flux (see Section~\ref{sec: thermality} for further discussion).

Now, let us try to make sense of the early time and far from  horizon limits. For the Vaidya vacuum, we recall from Eq.~\eqref{eq: Vaidya-U} that the Wightman function involves variable $\bar{u} = -4M\rr{1+\mathsf{W}(-U/e)}$, where $\mathsf{W}(z)$ is the Lambert-W function. For \textit{fixed} coordinate time $t$ (or proper time $\tau$) and large radial coordinate $r$, which corresponds to large $-U$, the asymptotic behaviour of the principal branch of the Lambert-W function  is \cite{NIST:DLMF}
\begin{align}
    \mathsf{W}(-U/e) \sim \log\rr{-\frac{U}{e}} = -\frac{t-r}{4M} - 1\,,
\end{align}
and hence $\bar{u}\sim t-r = u$, where $u$ is a null coordinate in Minkowski space. Therefore, we conclude that in the large $-U$ limit the Vaidya vacuum is well-approximated by the Boulware vacuum (and \textit{hence} also by \textit{Minkowski} vacuum of flat space). This happens when either (1) detectors are very far from the horizon ($r$ is very large), or (2) $\tau_0$ is very small (hence $t(\tau)$ along the strong support is small), i.e. detectors turned on very early but still within Schwarzschid exterior Region I shown in Figure~\ref{fig: Penrose2}. In particular, this calculation shows that for fixed switching peak $\tau_0$, once the detectors are sufficiently far away, the detectors cannot tell whether a black hole will form or not because they have the same joint density matrix as if the vacuum were Minkowski, even if they lie within Region I of Figure~\ref{fig: Penrose2}.

Let us now make sense of the late-time and near-horizon limit. When the detectors are switched on very late (very large $\tau_0$), the behaviour  increasingly approaches the Unruh limit. To see this, note that for any large but fixed radius $r$, one can always make $U$ very small by taking $t$ (or $\tau$) very large. This happens when we make the Gaussian switching peak $\tau_0$ very large\footnote{This also occurs when we make $\sigma$ large, though we need to make sure the strong support of the switching lies within the Schwarzschild exterior (which may involve increasing $\tau_0$).} (hence $t(\tau)$ along the strong support is large). In this case, one looks for the other branch of the Lambert-W function and the asymptotic behavior for small $-U$ is \cite{NIST:DLMF}
\begin{align}
    \mathsf{W}(-U/e)\sim -\log\rr{\frac{e}{U}} = -1 + \log U \approx -1 + U\,,
\end{align}
hence $\bar{u}\sim -4MU = \bar{U}$. This is precisely the null coordinate used for the definition of the Unruh vacuum (see Eq.~\eqref{eq: wightman-eddington} and Eq.~\eqref{eq: derivative-unruh})
Therefore, we conclude that the Vaidya vacuum is well-approximated by the Unruh vacuum when $-U$ is very small, i.e. either (1) when the detectors are very close to the horizon, or (2) when the detectors are switched on at very late times (even for finite, short interaction timescale $\sigma$).

\begin{figure}[tp]
    \centering
    \includegraphics[scale=0.57]{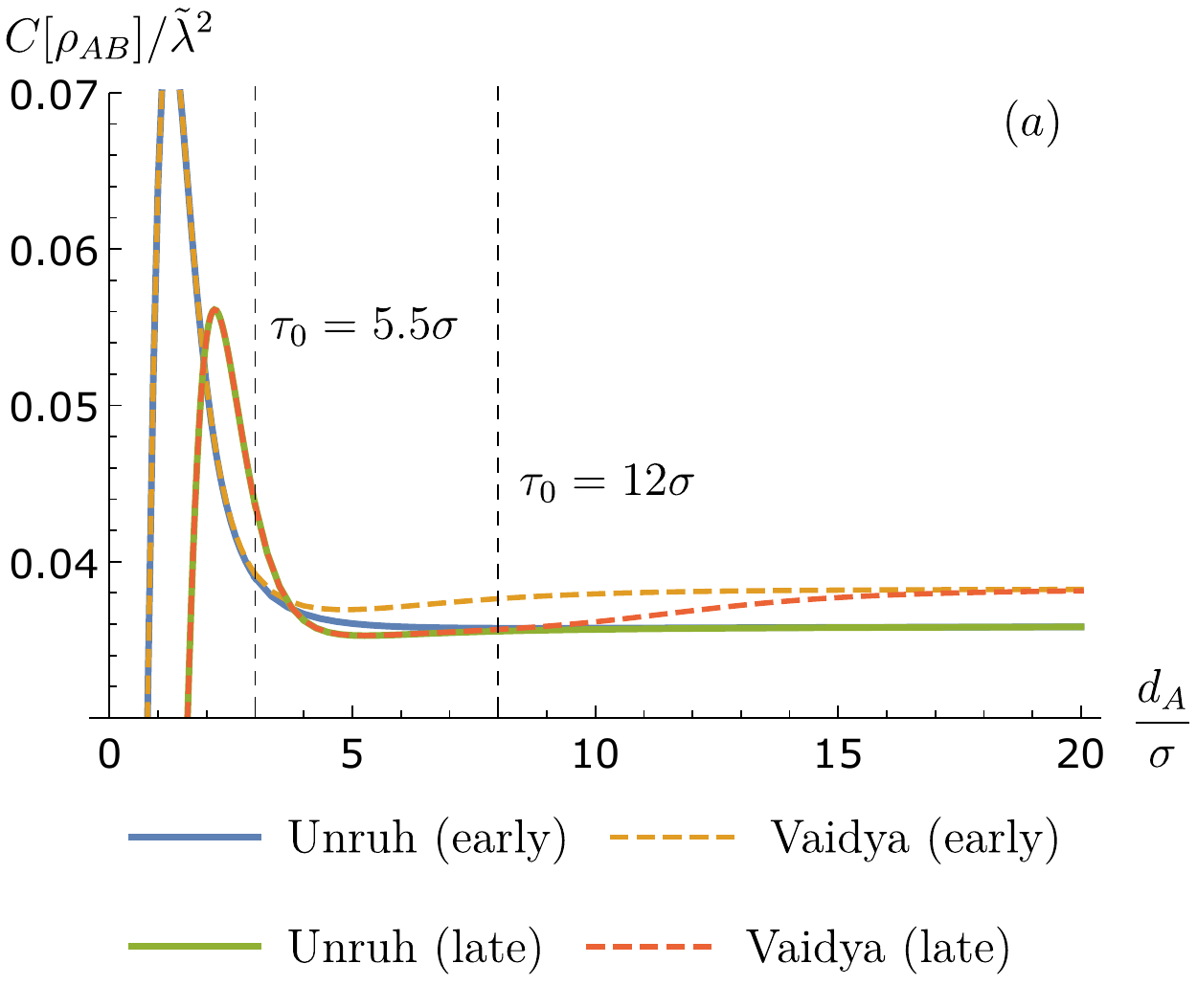}
    \includegraphics[scale=0.57]{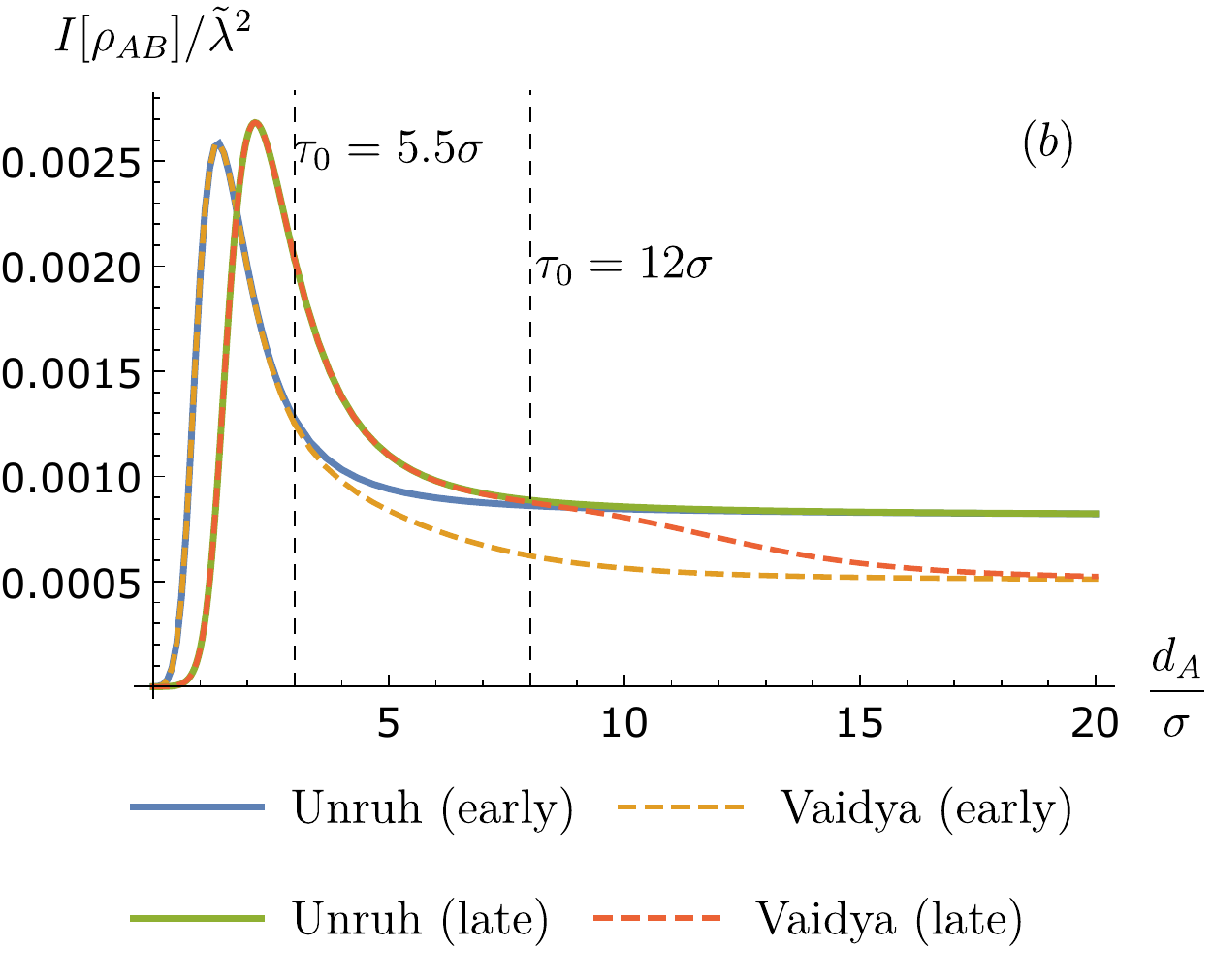}
    \caption{Comparison of (a) concurrence and (b) mutual information for Unruh and Vaidya vacua when the switching function is peaked at $\tau_0=5.5\sigma$ (early)
    and $\tau_0=12\sigma$ (late). Here $\tilde\lambda = \lambda\sigma^{\frac{1-n}{2}}=\lambda$ is dimensionless coupling constant. We set $\Omega\sigma = 2$, $M \sigma = \frac{1}{2}$ and $d(r_A,r_B)=3\sigma$. The `bifurcation point' at which the two states begin to show differences in concurrence is now at about $d_A \approx 3\sigma$ when $\tau_0=5.5\sigma$, as compared to $d_A\approx 8\sigma$ when $\tau_0=12\sigma$, 
    indicated by the vertical dashed lines.   }
    \label{fig: time-shift}
\end{figure}

The early/late time limit affecting the Boulware/Unruh approximation of Vaidya vacuum can be visualized in Figure~\ref{fig: time-shift}. We find that the primary  factor governing the point at which the approximation breaks down is \textit{when} the detector is switched on, i.e. the switching peak $\tau_0$. The earlier the switching time is, the Unruh/Vaidya difference becomes manifest nearer to the horizon. In Figure~\ref{fig: time-shift}, we see that this `bifurcation' point now begins when detector $A$ is at proper distance of $d_A\approx 3\sigma$ away from the horizon\footnote{The value of $\tau_0\approx 5.5\sigma$ in Figure~\ref{fig: time-shift} we chose is approximately the smallest for which the strong support of the Gaussian is entirely contained in Region I, and we do not push this earlier to avoid artifacts of shell-crossing where detector A's switching becomes highly non-Gaussian due to discontinuity of  the redshift factor across the shell.}, as compared to $d_A\approx 8\sigma$ when $\tau_0=12\sigma$. In other words, for finite-time interaction, how far away from the horizon the Unruh vacuum well-approximates   the Vaidya vacuum depends on how early/late the detectors are turned on relative to the null collapse time. This is precisely what we obtained earlier from the asymptotic analysis of the Wightman functions.

\subsection{Detailed balance condition and non-equilibrium states}
\label{sec: thermality}

In this section we briefly comment on the thermality (or lack thereof) of the black hole  vacua we have considered, using the notion of detailed balance.  

The Kubo-Martin-Schwinger (KMS) condition prescribes the necessary condition for a quantum state to be thermal with respect to some timelike Killing vector $\xi$ with KMS temperature $\mathsf{T}=\beta^{-1}$ \cite{Kubo1957thermality,Martin-Schwinger1959thermality}. A particularly operational formulation is given by the \textit{ detailed balance condition} \cite{Takagi1986noise}, which says that in the adiabatic limit (long, carefully switched on interaction), we have
\begin{align}
     \frac{\lim_{\sigma\to\infty} \mathcal{F}(\Omega,\sigma)}{\lim_{\sigma\to\infty} \mathcal{F}(-\Omega,\sigma)} = e^{-\beta_{\loc} \Omega}\,,
     \label{eq: KMS-balance}
\end{align}
where $\mathsf{T}_{\loc} = \beta^{-1}_\loc$ is the local temperature as measured an observer in curved spacetime. In our scenario, the local temperature will be given by the \textit{Tolman temperature} $\mathsf{T}_\loc = \mathsf{T}_H \sqrt{-g^{tt}} = (8\pi M\sqrt{1-2M/R})^{^{-1}}$ \cite{Tolman1930weight-heat,TolmanEhrenfest1930temperature}, where $\mathsf{T}_H=(8\pi M)^{-1}$ is the Hawking temperature of the Schwarzschild black hole.

The function $\mathcal{F}(\Omega,\sigma)$ is usually called the \textit{response function} (or transition rate \cite{Aubry2014derivative,hodgkinson2013particle}), defined as the excitation probability $\Pr(\Omega,\sigma)$ per unit time:
\begin{align}
    \mathcal{F}(\Omega,\sigma) &\coloneqq \frac{\Pr(\Omega,\sigma)}{\lambda^2\sigma}\,.
\end{align}
Recall that the expression for the excitation probability of a single detector is given precisely by the matrix element $\mathcal{L}_{jj}$, which takes the form (dropping the index $j$ since we only consider one detector, cf. Eq.\eqref{eq: local-noise}) 
\begin{align}
    \Pr(\Omega,\sigma) & = \lambda^2\int \dd\tau\,\dd\tau'\,\chi(\tau)\chi(\tau')e^{-\ii\Omega(\tau-\tau')}\mathcal{A}_\alpha(\sx(\tau),\sx(\tau'))\,,
\end{align}
where $\sigma$ is the interaction timescale encoded in $\chi(t)$. Note that by symmetry, $P(-\Omega,\sigma)$ corresponds to the de-excitation probability of a detector from its excited state to its ground state.

Formally, the detailed balance condition \eqref{eq: KMS-balance} is formulated in terms of the adiabatic limit of the response function $\mathcal{F}$. We are interested in the physical limit where the interaction timescale $\sigma$ is slowly increased to mimic longer and longer interaction time, and our choice of Gaussian switching guarantees that the interaction is smooth enough. Therefore, the ratio of the responses can instead be computed in terms of the \textit{excitation-to-deexcitation (EDR) ratio} 
\begin{align}
    \mathcal{R}(\Omega,\sigma) \coloneqq \frac{\mathcal{F}(\Omega,\sigma)}{\mathcal{F}(-\Omega,\sigma)} = \frac{\Pr(\Omega,\sigma)}{\Pr(-\Omega,\sigma)}\,.
\end{align}
The detailed balance condition \eqref{eq: KMS-balance} is therefore given by
\begin{align}
    \lim_{\sigma\to\infty} \mathcal{R}(\Omega,\sigma) = e^{-\beta_{\loc} \Omega}\,.
    \label{DB2}
\end{align}
Recently the EDR ratio has been used to study  interesting new phenomena, such as the Unruh effect without thermality \cite{Carballo2019Unruh-without-thermal}, anti-Unruh effects \cite{Brenna2016anti-unruh,Garay2016anti-unruh}, and a novel anti-Hawking effect \cite{Henderson2019anti-hawking}. 

\begin{figure}[tp]
    \centering
    \includegraphics[scale=0.57]{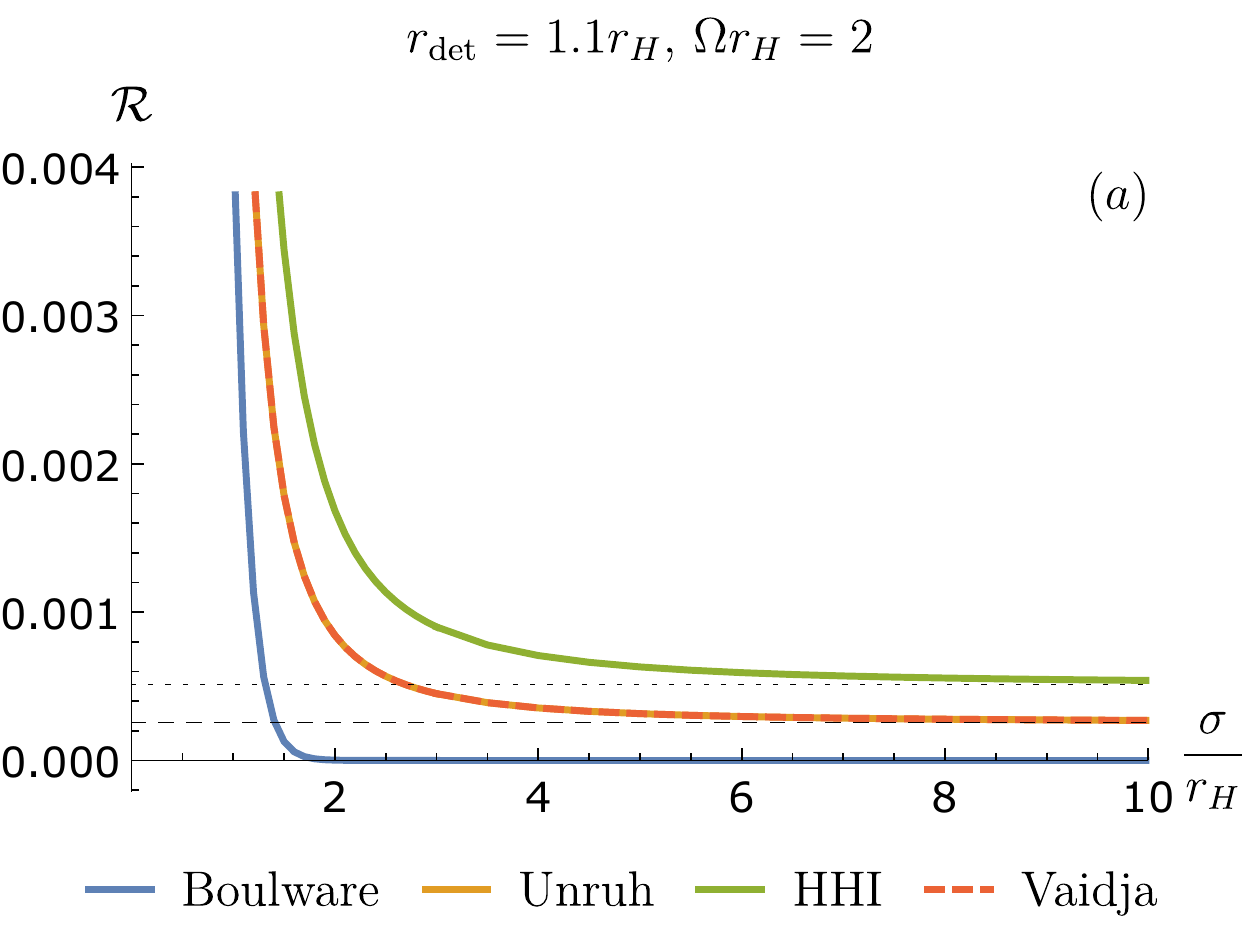}
    \includegraphics[scale=0.57]{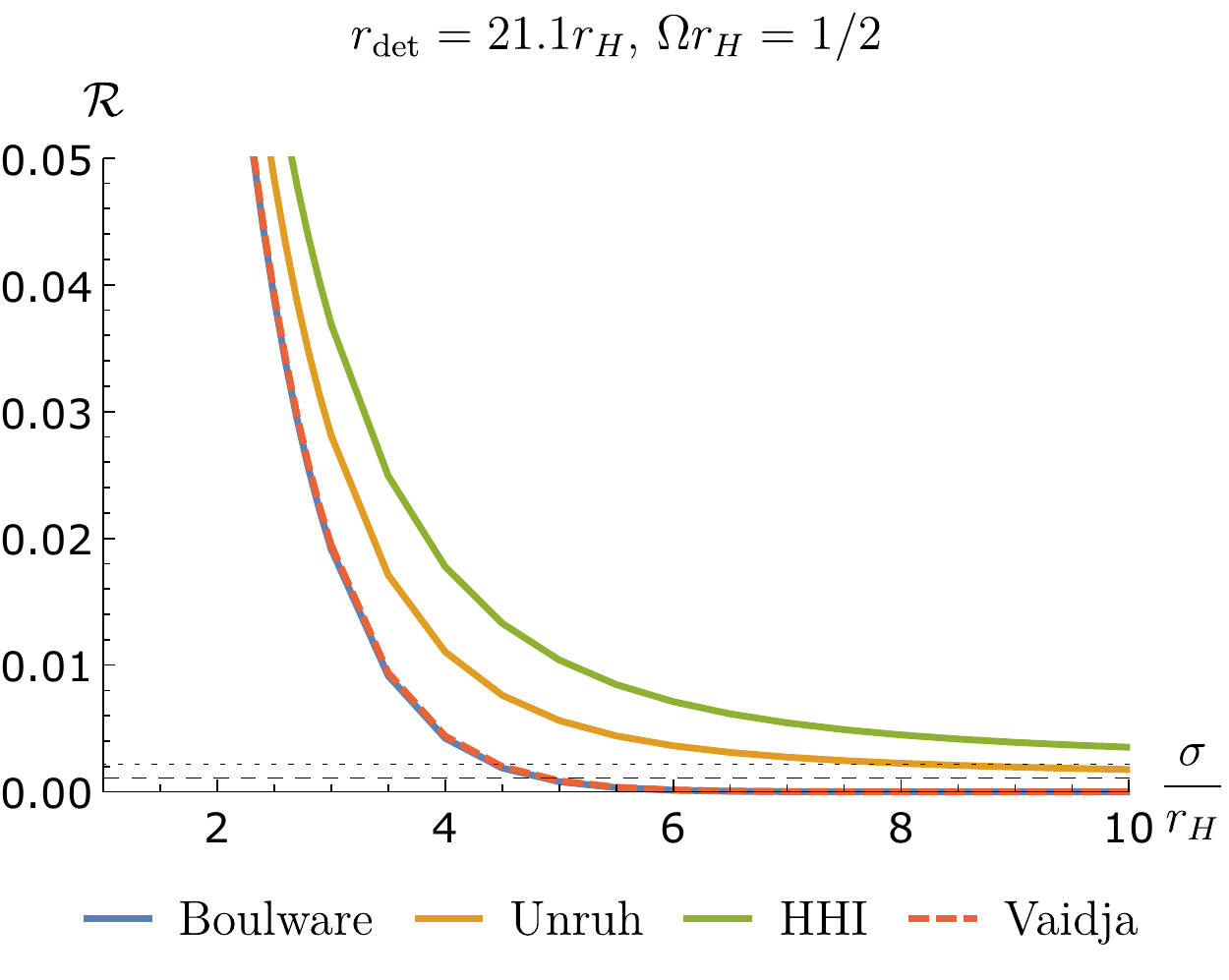}
    \caption{The excitation-to-deexcitation (EDR) ratio as a function of interaction time $T$ in units of $r_H=2M$ for various vacua. In (a) we set $\Omega r_H = 2$ with the detector is located at $r=2.2M = 1.1r_H.$  
    The horizontal dotted line corresponds to the EDR value obtained from KMS detailed balance $e^{-\beta_\loc\Omega} = e^{-\frac{8 \pi }{\sqrt{11}}}\approx 0.000512$, which is approached by HHI vacuum.  The horizontal dashed line corresponds to the non-thermal EDR ratio associated with the Unruh vacuum in Eq.~\eqref{eq: EDR-Unruh} $(2e^{\beta_{\text{loc}}\Omega}-1)^{-1} =(2e^{\frac{8 \pi }{\sqrt{11}}}-1)^{-1}  \approx 0.000256$.{ In
 (b) we depict the EDR ratios for $\Omega r_H = 1/2 $ with the detector is located at $e=21.1 r_H$.  }  
    }
    \label{fig: KMS1}
\end{figure}

In Figure~\ref{fig: KMS1}(a) we show how the EDR ratio $\mathcal{R}$ increases with interaction timescale $\sigma$ near the horizon, where we consider a static detector at $r=2.2M$. As expected, the HHI vacuum approaches the detailed balance prediction (horizontal dotted line) since it corresponds to black hole in thermal equilibrium with a thermal bath. Likewise,  the Boulware vacuum approaches the expected value of zero since it corresponds to a  zero-temperature state with $\beta\to\infty $ and hence $e^{-\beta_{\text{loc}}\Omega}\to 0$. By construction, we know that Unruh and Vaidya vacua correspond to non-equilibrium states since they are supposed to simulate radiation flux to infinity;  here we again see that the Unruh and Vaidya vacua agree very well in terms of their EDR near the horizon. This provides further evidence that at the level of single detector responses, the Unruh vacuum approximates very well the vacuum of a collapsing star spacetime near the horizon.

Inspection of Figure~\ref{fig: KMS1}(a) reveals that in the late time limit, {the response functions of} both  the Unruh and Vaidya vacua approach a constant value that is \textit{below} the detailed balance prediction \eqref{DB2} reflecting the non-thermal nature of these states. We can say more about the long time limit of EDR for these two states. From the Wightman functions (cf. Section~\ref{sec: geometry}), one can see that the Unruh vacuum can be thought of as an average of Boulware and HHI
states since the positive frequency modes are defined by $u=v-2r_*$ and $\bar U$. It then follows that in the long time limit, the EDR for the Unruh vacuum reads
\begin{align}
    \lim_{\sigma\to\infty} \frac{\Pr(\Omega,\sigma)_U}{\Pr(-\Omega,\sigma)_U} 
    &\to  \lim_{\sigma\to\infty}  \frac{\Pr(\Omega,\sigma)_B+\Pr(\Omega,\sigma)_H}{\Pr(-\Omega,\sigma)_B+\Pr(-\Omega,\sigma)_H} \notag\\
    &=\lim_{\sigma\to\infty}  \frac{\Pr(\Omega,\sigma)_H}{\Pr(-\Omega)_B+\Pr(-\Omega,\sigma)_H} \label{eq: KMS-estimate1}\\
    &\leq \lim_{\sigma\to\infty} \frac{\Pr(\Omega,\sigma)_H}{\Pr(-\Omega,\sigma)_H} = e^{-\beta_{\text{loc}}\Omega}\Bigr|_{\text{HHI}}\,,
\end{align}
which is always below the EDR for the HHI vacuum. The limit in Eq.~\eqref{eq: KMS-estimate1} can be computed exactly if we use transition rate. The long-time limit of the transition rates for the vacua are (denoted $\mathcal{\dot F}$ in \cite{Aubry2014derivative})
\begin{subequations}
\begin{align}
    \mathcal{\dot F}_B(\Omega) &= \Omega\Theta(-\Omega)\,,\\
    \mathcal{\dot F}_U(\Omega) &= \frac{\Omega}{2}\Theta(-\Omega)+\frac{\Omega}{2}\frac{1}{e^{\beta_{\text{loc}}\Omega}-1}\,,\\
    \mathcal{\dot F}_H(\Omega) &= \frac{1}{e^{\beta_{\text{loc}}\Omega}-1}\,.
\end{align}
\end{subequations}
Observe that the long-time transition rate for Unruh vacuum is precisely the average of the response of the Boulware and HHI vacua, and for $\Omega>0$ Unruh vacuum has Planckian spectrum but with the density of states halved from the HHI vacuum.   

Substituting into Eq.~\eqref{eq: KMS-estimate1}, we obtain the long time limit of the EDR ratio for the Unruh vacuum:
\begin{align}
    \mathcal{R}_U(\Omega) \coloneqq  \lim_{\sigma\to\infty} \frac{\Pr(\Omega,\sigma)_U}{\Pr(-\Omega,\sigma)_U} 
    &= \frac{ \mathcal{\dot F}_U(\Omega)}{ \mathcal{\dot F}_U(-\Omega)} = \frac{1}{2e^{\beta_{\text{loc}}\Omega}-1}<e^{-\beta_{\text{loc}}\Omega}\,.
    \label{eq: EDR-Unruh}
\end{align}
Indeed, $\mathcal{R}_U(\Omega)$ is precisely the asymptotic value of the EDR ratio that both the Unruh and Vaidya vacua approach in Figure~\ref{fig: KMS1}(a). What this tells us is that although the Unruh vacuum is not an equilibrium state, one can obtain the effective local temperature associated with only the outgoing flux of this state (the right-moving modes) by computing 
\begin{align}
    {\mathsf{T}}_{\text{loc}} \equiv \beta_{\text{loc}}^{-1} \coloneqq \Omega\rr{\log \frac{1+\mathcal{R}_U(\Omega)^{-1}}{2}}^{-1}\,,
\end{align}
Note that since $\mathsf{T}_\text{loc}$ depends on $\Omega$, strictly speaking this is not the detailed balance condition (which requires that it is independent of $\Omega$), since after all Unruh vacuum does not describe thermal equilibrium like Hartle-Hawking vacuum.


In Figure~\ref{fig: KMS1}(b), however, we see that the EDR ratio for the Vaidya vacuum approaches that of the Boulware vacuum once it is far from the horizon (in this example we take $r=21.1r_H$) for the same choice of switching peak $\tau_0$ as Figure~\ref{fig: KMS1}(a). This can be anticipated from earlier results in the entanglement harvesting calculation, where we saw that the transition probability approaches the Boulware vacuum far from the horizon. This is interesting because a faraway observer in Region I of Vaidya spacetime computing the EDR ratio would conclude that the vacuum state is KMS with zero temperature, associated with a Boulware vacuum. Therefore, a single detector faraway interacting for finite times will \textit{not} be able to infer the thermal behaviour of the radiation outflux to infinity for the Vaidya vacuum\footnote{One may wonder whether this is merely an artifact of the Gaussian switching `crossing the shell', since for fixed $\tau_0$ and $r$, large $\sigma$ will eventually lead to more parts of the switching function crossing the shell. It can be checked that for our choice of $\tau_0$ and $r$ this is not the case using Eq.~\eqref{eq: shell-crossing-limit}.} once it is far enough from the black hole (or equivalently, switched on early enough; see the asymptotic analysis in Section~\ref{sec: late-time}). 

Finally, we note that for finite time interactions, the EDR ratio for the Vaidya vacuum approaches that of the Unruh vacuum as one increases $\tau_0$. More concretely, if at fixed radius and fixed $\sigma$, we pick $\tau_0$ such that the EDR ratio agrees with the Boulware vacuum (as we did in Figure~\ref{fig: KMS1}(b)) increasing $\tau_0$ will bring the EDR curve towards the Unruh one.  This complements our earlier asymptotic analysis  in Section~\ref{sec: late-time}. From this perspective, the Unruh vacuum can be properly thought of as a physical, `late-time' limit of the Vaidya vacuum (corresponding to $\tau_0\to\infty$ limit), and this limit is approached very quickly. This is in contrast to the notion of the late time limit in \cite{Aubry2014derivative}, which has more to do with when the detector is switched off after it is turned on, thus more closely related to \textit{how long} the interaction is switched on.

\section{Conclusion}

We have studied the phenomenon of harvesting of correlations by two detectors from the vacuum state of a massless scalar field in background Vaidya spacetime, and compare the results with those associated to the three preferred vacua (Boulware, Unruh, Hartle-Hawking-Israel vacua) in Schwarzschild spacetime. We use the derivative coupling particle detector model where the Wightman functions have similar short-distance behaviour as the Wightman functions in the (3+1)-dimensional counterpart, as well as to resolve the infrared ambiguities associated to massless scalar fields in $(1+1)$-dimensional spacetimes. We perform these studies using a straightforward implementation of numerical contour integration, outlined in Appendix~\ref{appendix: numerical-contour}.

Let us summarize our results. First, we showed that from the perspective of harvesting of correlations between two detectors, near the horizon the Unruh vacuum agrees very well with the Vaidya vacuum \textit{even for finite-time interactions}, complementing the long-interaction result from \cite{Aubry2014derivative}. Second, all four vacua have different capacities for creating correlations between the detectors, with the Vaidya vacuum's capacity interpolating between that of the Unruh vacuum near the horizon and the Boulware vacuum far from the horizon. Third, our examination of mutual information indicated that the black hole horizon inhibits \textit{any} correlations, not just entanglement, complementing the results found in \cite{henderson2018harvesting}. Finally, efficiency of the harvesting protocol depends strongly on the signalling ability of the two detectors,  which is highly non-trivial in the presence of curvature. We have also studied the asymptotic behaviour of the Vaidya vacuum analytically to understand {how it approximates   the Boulware/Minkowski vacuum in the early time/large distance limit, and approximates  the  Unruh vacuum in the} late time/near-horizon limit. Our asymptotic analysis clarifies the distinction between the late-time and long-time limits in transition rate calculations \cite{Aubry2014derivative,Aubry2018Vaidya}.

A natural extension to our results would be to analyse the correlations between two detectors, one of which free-falling through the horizon during the interactions. Since our results on the exterior region shows that the horizon inhibits all forms of correlations from being extracted from the vacuum, how free-falling detectors break correlations between them is an operational question related to information problem in black hole thermodynamics. Another related question concerns the effect of null shockwaves from the perspective of supertranslation \cite{Compere2019, Kolekar2017accelmemory,Kolekar2018rindlerhair}: it would be interesting to study the correlations between two detectors in presence of supertranslations. Finally, it would also be interesting to see how the four vacua considered in this paper affect communication efficiencies such as channel capacities between detectors \cite{Jonsson:2020npo}. We leave these questions for future work.

\section*{Acknowledgment}

The authors thank Eduardo Mart\'in-Mart\'inez and Robie A. Hennigar for useful discussions on the physics, and Nicholas Funai for illuminating discussion on some aspects of complex analysis needed for the numerical method employed in this work. E. T. acknowledges support from Mike-Ophelia Lazaridis Fellowship during this work. This work was supported in part by the Natural Sciences and Engineering Research Council of Canada.

\appendix

\section{Numerical contour integration}
\label{appendix: numerical-contour}

Here we present our method of performing contour integration using \textit{Mathematica} that we employed in this paper. The basic idea was first demonstrated in the context of moving mirror spacetimes \cite{Tjoa2019MSc}.  We recapitulate this approach, making a  slight improvement.  We believe this method is worth outlining because it seems to be useful for many purposes beyond relativistic quantum information settings,   since it is essentially the problem of evaluating a double integral over a distribution.

Let us illustrate the technique by computing the transition probability of an Unruh-DeWitt detector comoving with the quantization frame in (3+1)-dimensional Minkowski space. The detector-field interaction is prescribed by the usual amplitude coupling $\hat H_I = \lambda\chi(t)\hat\mu(t)\hat\phi(t,\bx)$, where $(t,\bx)$ denotes the coordinates of the detector. For convenience we will set the detector to be at the origin, so that $\bx = \mathbf{0}$. It is easy to see that the Wightman function associated with the Minkowski vacuum, $W_{M}(t,t')\equiv W_M(t,\bm{0},t',\bm{0})$, reduces to the simple expression\footnote{Observe that this is the same as the Wightman function for derivative coupling in (1+1)-dimensional Minkowski space, up to the constant prefactor.} \cite{birrell1984quantum,smith2016topology}
\begin{align}
    W_{{M}}(t,t') = -\frac{1}{4\pi^2}\frac{1}{(t-t'-\ii\epsilon)^2}\,.
\end{align}
We have kept the $\ii\epsilon$ prescription here since it is the common way of describing the distributional nature of the vacuum Wightman functions\footnote{In the simple case above, we could also remove the $\ii\epsilon$ in exchange of using Dirac delta function and principal value integral via Sokhotsky's formula \cite{mukhanov2007introduction}
\begin{align}
    \lim_{\epsilon\to 0^+} \frac{1}{x\mp\ii\epsilon} = \text{P.V.}\rr{\frac{1}{x}}\pm \ii\pi \delta(x)\,,
\end{align}
where P.V. denotes principal value and the limit is understood in the distributional sense.}. Let us set the switching to be the Gaussian switching $\chi(t) = e^{-t^2/(2\sigma^2)}$ for convenience, since we will later make comparison with some results in the literature that uses this switching function. Note that this is different from the switching considered in the main body of the paper (cf. Eq.~\eqref{eq: switching}) by substitution $\sigma\to \sqrt{2}\sigma$. The transition probability of a pointlike detector prepared in the ground state  to leading order in perturbation theory is
\begin{align}
    \Pr(\Omega,\sigma) &= \lambda^2\int \dd t\,\dd t'\,e^{-t^2/(2\sigma^2)}e^{-{t'}^2/(2\sigma^2)}e^{-\ii\Omega(t-t')}W_M(t,t')\,,
\end{align}
where $\Omega>0$. We can write this as
\begin{align}
    \Pr(\Omega,\sigma) &= \lim_{\epsilon\to 0^+}\lambda^2 \mathcal{J}(\Omega,\sigma,\epsilon)\,,\\
    \mathcal{J}(\Omega,\sigma,\epsilon) &\coloneqq -\frac{1}{4\pi^2}\int \dd t\,\dd t'\,e^{-\ii\Omega(t-t')}\frac{e^{-t^2/(2\sigma^2)}e^{-{t'}^2/(2\sigma^2)}}{(t-t'-\ii\epsilon)^2}\,.
\label{A5}    
\end{align}
For this particular case, we are fortunate because the closed-form expression for Eq.~\eqref{A5} is known, which we can use to check our calculations. This is given by \cite{smith2016topology}
\begin{align}
    J_0 \coloneqq \frac{e^{-\sigma ^2 \Omega ^2}-\sqrt{\pi } \sigma  \Omega\;   \text{erfc}(\sigma  \Omega )}{4 \pi }\,. 
    \label{eq: exact-minkowski-response}
\end{align}
We remark that there is another closed form expression derived differently in \cite{Sriramkumar1994finitetime} that only works correctly for $\Omega>0$, whereas Eq.~\eqref{eq: exact-minkowski-response} is valid for all $\Omega\in \R$. 

Let us now compare this with numerical computation\footnote{All numerical computations in this paper are done using \textit{Mathematica} 12.0 \cite{Mathematica}.}. We will refer to an integral $J_j$ as \textbf{Method \textit{j}}, and we will call $J_0$ \textbf{Method 0}.

\subsection{Method 1: direct $\ii\epsilon$ integration}

We denote $J_1$ to be  {the integral \eqref{A5}}
evaluated by brute force, picking a small enough $\epsilon$ during integration. We will evaluate this for $\Omega\sigma=1$ for concreteness.  In this case Method 0 gives
\begin{align}\label{J0W1}
    J_0\bigr|_{\Omega\sigma=1} = \frac{e^{-1}-\sqrt{\pi } \text{erfc}(1)}{4 \pi } \approx 0.00708827\,.
\end{align}

The values of $J_1$ can be computed using various settings and optimizations. In our case, reasonable results are obtained using $\text{MinRecursion}\to 3, \text{MaxRecursion}\to 20, \text{AccuracyGoal}\to \infty, \text{PrecisionGoal}\to 10$. We also cut the integral at   strong support, i.e. $t,t'\in (-5\sigma,5\sigma)$ for better convergence; one can check that the results are generally worse if one chooses to numerically integrate over $\R$. The results are shown in Table~\ref{tab: double_ieps}.
\begin{table}[htp]
    \centering
    \begin{tabular}{c|c}
        $\epsilon/\sigma$  &   $J_1$ \\
        \hline
        $10^{-1}$   &   $0.00670272 + 2.67648\times 10^{-9}\ii$    \\
        $10^{-2}$   &   $0.00704838 - 6.50088\times 10^{-16} \ii$    \\
        $10^{-3}$   &   $-0.732931 - 9.03524\times 10^{-7}\ii$   \\
        $10^{-4}$   &   $-6.84952 + 1.52192 \ii$  \\
        $10^{-5}$   &   $-27.3218 - 0.246437 \ii$
    \end{tabular}
    \caption{Values of $J_1$ using Method 1 (direct $\ii\epsilon$ integration) as $\epsilon$ varies.}
    \label{tab: double_ieps}
\end{table}

Observe that $\epsilon\sim 10^{-2}\sigma$ 
reasonably approximates \eqref{J0W1}, but the rest of the values do not work. To our knowledge, any other settings within this scheme do not help much, and we believe that while in principle there should be a way to make this method work, it would require a great deal of effort and understanding of the back-end numerical analysis to make this worthwhile in terms of both the computational time and numerical stability. We stress that the sorts of computations done in \cite{Aubry2014derivative} or \cite{henderson2019entangling} have one particular advantage: they can be recast into one-dimensional integrals that can be dealt with much better numerically. For example, the $\text{Method}\to\text{``DoubleExponentialOscillatory''}$ used in \cite{henderson2018harvesting} is not available for higher-dimensional integrals.

\subsection{Method 2: numerical contour integration}

\begin{figure}[tp]
    \centering
    \includegraphics[scale=0.3]{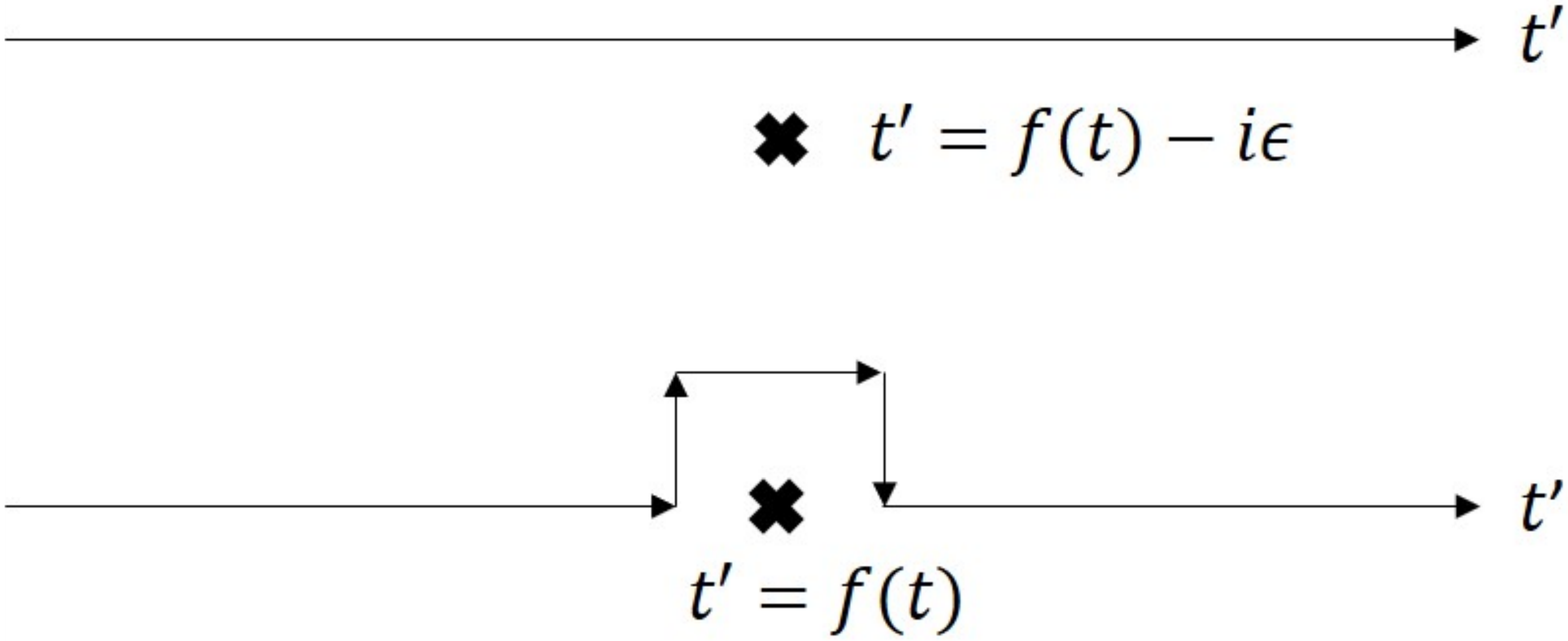}
    \caption{The choice of contour about $t'=f(t)$ in contrast to $\ii\epsilon$ prescription. For our example, we have $f(t) = t$.}
    \label{fig: contour-choice}
\end{figure}

The idea is basically to perform the following integral:
\begin{align}
    J_2 \coloneqq -\frac{1}{4\pi^2}\int_{\R}\dd t\int_{C(\epsilon)}\dd t'\,e^{-\ii\Omega(t-t')}\frac{e^{-t^2/(2\sigma^2)}e^{-{t'}^2/(2\sigma^2)}}{(t-t')^2}\,,
\end{align}
where $C(\epsilon)$ is a contour deformed to the upper complex plane around the pole $t'=t$. The contour is shown in Figure~\ref{fig: contour-choice} and shown  contrasted to the $\ii\epsilon $ prescription. If we choose to instead perform the integral over $t$ first, then the contour is deformed to the lower complex plane around the pole $t=t'$. We let $\epsilon$ here to be the distance from the pole (in units of $\sigma$): that is, we integrate $t'$ from $-\infty$ to $t-\epsilon$, then from $t-\epsilon$ to $t-\epsilon+\ii$, then to $t+\epsilon+\ii$, followed by $t+\epsilon$ and finally from $t\to\infty$. That is, we set $\epsilon$ to be the distance from the pole along the $t'$ axis\footnote{A minor point: the unit of $\epsilon$ depends on the pole. Typically one views $\epsilon$ as a UV regulator (hence typically in natural units it has units of length), but mathematically it is really just a prescription for describing the distributional nature of the distribution at hand. Thus it can be dimensionless, depending on where it appears.}. Again we integrate over strong support $(-5\sigma,5\sigma)$ as in general integration over $\R$ is of lower quality. The results are shown in Table~\ref{tab: num-contour}. 

\begin{table}[htp]
    \centering
    \begin{tabular}{c|c}
        $\epsilon/\sigma$  &   $J_2$ \\
        \hline
        $10^{-0}$   &   $0.00708827 - 6.59116\times 10^{-19} \ii$ \\
        $10^{-1}$   &   $0.00708827 - 8.73414\times 10^{-19} \ii$ \\
        $10^{-2}$   &   $0.00708827 - 1.69173\times 10^{-18} \ii$ \\
        $10^{-3}$   &   $0.00708827 + 1.09202\times 10^{-9} \ii$ \\
        $10^{-4}$   &   $0.00708827 + 1.09423\times 10^{-9} \ii$ \\
        $10^{-5}$   &   $0.00708712 + 1.17660\times 10^{-9} \ii$ \\
    \end{tabular}
    \caption{Values of $J_2$ using Method 2 (numerical contour) as $\epsilon$ varies.}
    \label{tab: num-contour}
\end{table}

Notice that the results are  a much better approximation
to \eqref{J0W1}
than Method 1. Furthermore, to achieve the quality shown above, the only setting needed was $\text{MinRecursion}\to 3$ and nothing more. This could be improved with more optimization. It is quite remarkable that this method works very well with minimal settings   whereas the usual $\ii\epsilon$ approach of Method 1 fails terribly. Method 2 only starts to deviate very little when we get too close to the pole ($\epsilon\sim 10^{-5}\sigma$) due to numerical resolution.

We make four observations here. First, the fact that $J_2$ is numerically constant across a broad range of values of $\epsilon$ is a manifestation of a basic principle in complex analysis, namely the \textit{deformation theorem}. The theorem states that within a holomorphic region we can deform the contour of an integral without changing the value of the integral, which follows from Cauchy's integral theorem. Since the pole is along the $t'$ axis, any $\epsilon$ will give the same result since there is no other pole in the upper complex plane. Therefore Method 2 provides a very nice way of checking numerical stability: if the integral is no longer constant as we vary $\epsilon$ across  a
broad but 
reasonable range, (recall from Table~\ref{tab: num-contour} that we would not want $\epsilon$ to be too small numerically), then perhaps one needs to check if something has gone wrong or the method itself no longer works stably. Second, because of the deformation theorem, in practice the contour shown in Fig~\ref{fig: contour-choice} is flexible: we chose this contour because we felt this to be the simplest for illustration. Third, notice that in computing $\mathcal{J}(\Omega,\sigma,\epsilon)$,  \textit{Jordan's lemma} cannot be used due to the Gaussian switching function  -- hence we do  not have the benefit of using the residue technique numerically. Finally, what we have performed here is effectively a two-dimensional contour integration, where the poles are continuous (one pole on the $t'$ axis for every $t$) on the $(t,t')$ plane.

We pause to remark that actually there are two more methods that work well for Minkowski vacuum calculations, which are used in \cite{hodgkinson2013particle,Satz2006howoften,Satz2007transitionrate, Aubry2014derivative}. One of them in fact can be written in a form free of the UV regulator $\epsilon$, thus it is either correct or incorrect. A brief investigation  \cite{Tjoa2019MSc} indicated that for Minkowski vacua these two are competitive methods and behave very well. However they failed in the presence of a (possibly dynamical) Dirichlet boundary condition (such as a moving mirror  \cite{cong2019entanglement}) because the mirror introduces new poles that reduce the utility of these methods. Nevertheless the resultant  calculations remained valid because the transition rate is given by a one-dimensional integral in time. It was shown that numerical contour integration remains superior in the context of moving mirrors \cite{Tjoa2019MSc}.

The results  in this section and the observations above testify to the especial appeal of numerical contour integration in practical calculation.

\subsection{Better contour for entanglement harvesting: Vaidya spacetime}

For calculations carried out in this paper, the choice of contour in Figure~\ref{fig: contour-choice} is not good enough. The problem is that the contour we picked relies on finding the location of the poles, i.e. we are solving for $t'=f(t)$. Even for derivative Wightman functions for the Unruh and HHI vacua in Eq.~\eqref{eq: derivative-unruh} and \eqref{eq: derivative-HHI}, $f$ is in general not a linear function of $t$ and depends on the black hole mass $M$. Even more importantly, for entanglement harvesting the nonlocal term $\mathcal{M}$ depends on two different radial coordinates with different gravitational redshifts, so $f$ is highly non-trivial. 

An even bigger problem is caused by time ordering in $\mathcal{M}$: one would have to constantly track whether within the strong support the poles are included or not when time ordering is applied. For derivative Wightman functions in Vaidya spacetime, this is worsened by the two additional equally complicated terms. It is not hard to check that the contour prescription we did earlier, where the deformation is somewhat close to the poles (say $\epsilon\sim 10^{-1}\sigma$) did not quite work, let alone direct integration via $\ii\epsilon$.

\begin{figure}[tp]
    \centering
    \includegraphics[scale=0.7]{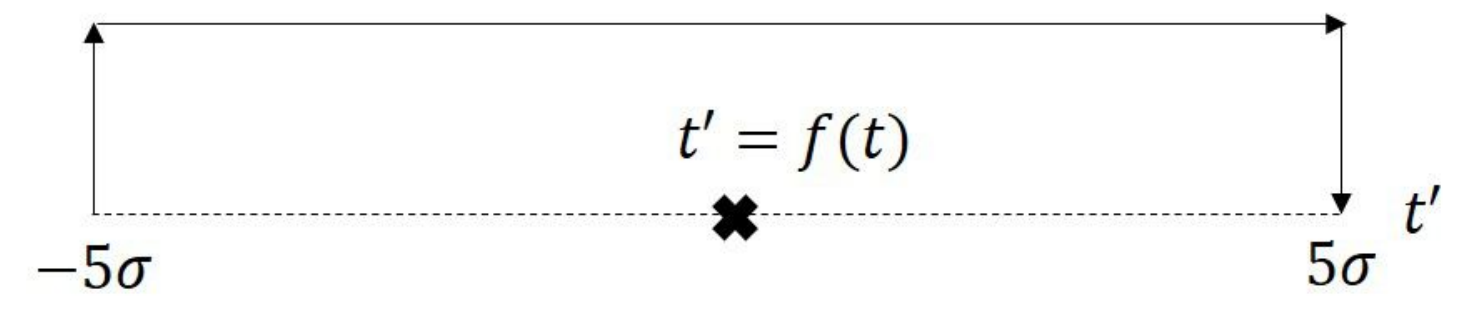}
    \caption{The modified contour used for harvesting calculations in this paper. The contour effectively does away with finding the poles for derivative Wightman functions.}
    \label{fig: contour-choice2}
\end{figure}

The deformation theorem \textit{and} Gaussian suppression coming from the switching function provide us with a new contour that we can use\footnote{The authors thank Nicholas Funai for useful discussions on the complex analysis aspects of this numerical technique.}. The idea is that   if the strong support of the Gaussian switching is $(-b\sigma,b\sigma)$, then we set $\epsilon = b\sigma$. In other words, we adopt the contour in Figure~\ref{fig: contour-choice2}. This contour has the advantage that we effectively do away with finding the poles: the poles are either within the within the strip or outside the strip, and a single contour covers all possible positions of the pole independent  of the complexity of the function $f$ that describes the location of the pole as a function of $t$. It also solves the issue of time ordering, by replacing the upper limit of $t'$ from $5\sigma$ to $t$.

As it turns out, with relatively minimal settings (such as $\text{MinRecursion}\sim 3-6$), the numerical computation works very well even for Vaidya spacetime. We also chose for simplicity to deform the contour by one imaginary unit $+\ii$ to the upper complex plane. In fact, this numerical calculation is stable enough for computation of long-time observables such as the EDR ratio for KMS detailed balance condition. We have also used infinite precision such as using fractions $1/2$ instead of $0.5$ whenever possible, and we set the global precision setting to 50 digits using \text{PreRead} command.

We  note that with more complicated problems (and depending on the issue at hand), we expect that different optimizations and variations may be needed on top of what we have done here. The main point is that numerical contour integration provides a sufficiently robust and \textit{straightforward} implementation without having to construct a separate numerical scheme from scratch, and the fact that optimization is possible at all (unlike the direct integration by $\ii\epsilon$ prescription). We also did not attempt to optimize computational time; this is perhaps left for future investigations.

\bibliography{Vaidyabib}

\providecommand{\href}[2]{#2}\begingroup\raggedright\begin{thebibliography}{10}

\bibitem{summers1985bell}
S.~J. Summers and R.~Werner, \emph{{The vacuum violates Bell's inequalities}},
  \href{https://doi.org/https://doi.org/10.1016/0375-9601(85)90093-3}{\emph{Phys.
  Lett. A} {\bfseries 110} (1985) 257 }.

\bibitem{summers1987bell}
S.~J. Summers and R.~Werner, \emph{Bell’s inequalities and quantum field
  theory. i. general setting}, \href{https://doi.org/10.1063/1.527733}{\emph{J.
  Math. Phys.} {\bfseries 28} (1987) 2440}.

\bibitem{hawking1975particle}
S.~W. Hawking, \emph{Particle creation by black holes},
  \href{https://doi.org/https://link.springer.com/article/10.1007/BF01608497}{\emph{Comm.
  Math. Phys.} {\bfseries 43} (1975) 199}.

\bibitem{almheiri2013black}
A.~Almheiri, D.~Marolf, J.~Polchinski and J.~Sully, \emph{Black holes:
  complementarity or firewalls?},
  \href{https://doi.org/https://doi.org/10.1007/JHEP02(2013)062}{\emph{J. High
  Energ. Phys.} {\bfseries 2013} (2013) 62}.

\bibitem{marolf2017black}
D.~Marolf, \emph{The black hole information problem: past, present, and
  future}, \href{https://doi.org/10.1088/1361-6633/aa77cc}{\emph{Rep. Prog.
  Phys.} {\bfseries 80} (2017) 092001}.

\bibitem{haco2018black}
S.~Haco, S.~W. Hawking, M.~J. Perry and A.~Strominger, \emph{Black hole entropy
  and soft hair},
  \href{https://doi.org/https://doi.org/10.1007/JHEP12(2018)098}{\emph{J. High
  Energ. Phys.} {\bfseries 2018} (2018) 98}.

\bibitem{nielsen2000quantum}
M.~Nielsen and I.~Chuang, \emph{Quantum Computation and Quantum Information},
  Cambridge Series on Information and the Natural Sciences. Cambridge
  University Press, 2000.

\bibitem{chitambar2019quantum}
E.~Chitambar and G.~Gour, \emph{Quantum resource theories},
  \href{https://doi.org/10.1103/RevModPhys.91.025001}{\emph{Rev. Mod. Phys.}
  {\bfseries 91} (2019) 025001}.

\bibitem{Brandao2015reversibleQRT}
F.~G. S.~L. Brand\~ao and G.~Gour, \emph{Reversible framework for quantum
  resource theories},
  \href{https://doi.org/10.1103/PhysRevLett.115.070503}{\emph{Phys. Rev. Lett.}
  {\bfseries 115} (2015) 070503}.

\bibitem{Valentini1991nonlocalcorr}
A.~Valentini, \emph{Non-local correlations in quantum electrodynamics},
  \href{https://doi.org/https://doi.org/10.1016/0375-9601(91)90952-5}{\emph{Phys.
  Lett. A} {\bfseries 153} (1991) 321 }.

\bibitem{reznik2003entanglement}
B.~Reznik, \emph{Entanglement from the vacuum},
  \href{https://doi.org/https://doi.org/10.1023/A:1022875910744}{\emph{Found.
  Phys.} {\bfseries 33} (2003) 167}.

\bibitem{reznik2005violating}
B.~Reznik, A.~Retzker and J.~Silman, \emph{{Violating Bell's inequalities in
  vacuum}}, \href{https://doi.org/10.1103/PhysRevA.71.042104}{\emph{Phys. Rev.
  A} {\bfseries 71} (2005) 042104}.

\bibitem{salton2015acceleration}
G.~Salton, R.~B. Mann and N.~C. Menicucci, \emph{Acceleration-assisted
  entanglement harvesting and rangefinding},
  \href{https://doi.org/10.1088/1367-2630/17/3/035001}{\emph{New J. Phys.}
  {\bfseries 17} (2015) 035001}.

\bibitem{pozas2015harvesting}
A.~Pozas-Kerstjens and E.~Mart\'{\i}n-Mart\'{\i}nez, \emph{Harvesting
  correlations from the quantum vacuum},
  \href{https://doi.org/10.1103/PhysRevD.92.064042}{\emph{Phys. Rev. D}
  {\bfseries 92} (2015) 064042}.

\bibitem{Steeg2009}
G.~V. Steeg and N.~C. Menicucci, \emph{Entangling power of an expanding
  universe}, \href{https://doi.org/10.1103/PhysRevD.79.044027}{\emph{Phys. Rev.
  D} {\bfseries 79} (2009) 044027}.

\bibitem{kukita2017harvesting}
S.~Kukita and Y.~Nambu, \emph{{Harvesting Large Scale Entanglement in de Sitter
  Space with Multiple Detectors}},
  \href{https://doi.org/10.3390/e19090449}{\emph{Entropy} {\bfseries 19} (2017)
  449}.

\bibitem{henderson2018harvesting}
L.~J. Henderson, R.~A. Hennigar, R.~B. Mann, A.~R.~H. Smith and J.~Zhang,
  \emph{Harvesting entanglement from the black hole vacuum},
  \href{https://doi.org/10.1088/1361-6382/aae27e}{\emph{Class. Quantum Gravity}
  {\bfseries 35} (2018) }.

\bibitem{ng2018AdS}
K.~K. Ng, R.~B. Mann and E.~Mart\'{\i}n-Mart\'{\i}nez, \emph{{Unruh-DeWitt
  detectors and entanglement: The anti--de Sitter space}},
  \href{https://doi.org/10.1103/PhysRevD.98.125005}{\emph{Phys. Rev. D}
  {\bfseries 98} (2018) 125005}.

\bibitem{smith2016topology}
E.~Mart\'{\i}n-Mart\'{\i}nez, A.~R.~H. Smith and D.~R. Terno, \emph{Spacetime
  structure and vacuum entanglement},
  \href{https://doi.org/10.1103/PhysRevD.93.044001}{\emph{Phys. Rev. D}
  {\bfseries 93} (2016) 044001}.

\bibitem{henderson2019entangling}
L.~J. Henderson, R.~A. Hennigar, R.~B. Mann, A.~R. Smith and J.~Zhang,
  \emph{{Entangling detectors in anti-de Sitter space}},
  \href{https://doi.org/https://doi.org/10.1007/JHEP05(2019)178}{\emph{J. High
  Energ. Phys.} {\bfseries 2019} (2019) 178}.

\bibitem{cong2019entanglement}
W.~Cong, E.~Tjoa and R.~B. Mann, \emph{Entanglement harvesting with moving
  mirrors},
  \href{https://doi.org/https://doi.org/10.1007/JHEP06(2019)021}{\emph{J. High
  Energ. Phys} {\bfseries 2019} (2019) 21}.

\bibitem{Henderson2020temporal}
L.~J. Henderson, A.~Belenchia, E.~Castro-Ruiz, C.~Budroni, M.~Zych, C.~Brukner
  et~al., \emph{{Quantum Temporal Superposition: the case of QFT}},
  \href{https://arxiv.org/abs/arxiv:2002.06208}{{\ttfamily arxiv:2002.06208}}.

\bibitem{Foo2020:2003.12774v3}
J.~Foo, S.~Onoe and M.~Zych, \emph{{ Unruh-deWitt detectors in quantum
  superpositions of trajectories}},
  \href{https://arxiv.org/abs/Arxiv:2003.12774v3}{{\ttfamily
  Arxiv:2003.12774v3}}.

\bibitem{Foo2020:2005.03914v1}
J.~Foo, S.~Onoe, R.~B. Mann and M.~Zych, \emph{Thermality, causality and the
  quantum-controlled unruh-dewitt detector},
  \href{https://arxiv.org/abs/Arxiv:2005.03914v1}{{\ttfamily
  Arxiv:2005.03914v1}}.

\bibitem{olson2011entanglement}
S.~J. Olson and T.~C. Ralph, \emph{Entanglement between the future and the past
  in the quantum vacuum},
  \href{https://doi.org/10.1103/PhysRevLett.106.110404}{\emph{Phys. Rev. Lett.}
  {\bfseries 106} (2011) 110404}.

\bibitem{olson2012extraction}
S.~J. Olson and T.~C. Ralph, \emph{Extraction of timelike entanglement from the
  quantum vacuum},
  \href{https://doi.org/10.1103/PhysRevA.85.012306}{\emph{Phys. Rev. A}
  {\bfseries 85} (2012) 012306}.

\bibitem{sabin2010dynamics}
C.~Sab\'{\i}n, J.~J. Garc\'{\i}a-Ripoll, E.~Solano and J.~Le\'on,
  \emph{Dynamics of entanglement via propagating microwave photons},
  \href{https://doi.org/10.1103/PhysRevB.81.184501}{\emph{Phys. Rev. B}
  {\bfseries 81} (2010) 184501}.

\bibitem{sabin2012extracting}
C.~Sab\'{\i}n, B.~Peropadre, M.~del Rey and E.~Mart\'{\i}n-Mart\'{\i}nez,
  \emph{{Extracting past-future vacuum correlations using circuit QED}},
  \href{https://doi.org/10.1103/PhysRevLett.109.033602}{\emph{Phys. Rev. Lett.}
  {\bfseries 109} (2012) 033602}.

\bibitem{EMM2013farming}
E.~Mart\'{\i}n-Mart\'{\i}nez, E.~G. Brown, W.~Donnelly and A.~Kempf,
  \emph{Sustainable entanglement production from a quantum field},
  \href{https://doi.org/10.1103/PhysRevA.88.052310}{\emph{Phys. Rev. A}
  {\bfseries 88} (2013) 052310}.

\bibitem{ardenghi2018entanglement}
J.~S. Ardenghi, \emph{Entanglement harvesting in double-layer graphene by
  vacuum fluctuations in a microcavity},
  \href{https://doi.org/10.1103/PhysRevD.98.045006}{\emph{Phys. Rev. D}
  {\bfseries 98} (2018) 045006}.

\bibitem{beny2018energy}
C.~B{\'e}ny, C.~T. Chubb, T.~Farrelly and T.~J. Osborne, \emph{Energy cost of
  entanglement extraction in complex quantum systems},
  \href{https://doi.org/https://doi.org/10.1038/s41467-018-06153-w}{\emph{Nat.
  Commun.} {\bfseries 9} (2018) 1}.

\bibitem{Unruh1979evaporation}
W.~G. Unruh, \emph{Notes on black-hole evaporation},
  \href{https://doi.org/10.1103/PhysRevD.14.870}{\emph{Phys. Rev. D} {\bfseries
  14} (1976) 870}.

\bibitem{DeWitt1979}
B.~S. {Dewitt}, \emph{{Quantum gravity: the new synthesis}},  in \emph{General
  Relativity: An Einstein centenary survey}, S.~W. {Hawking} and W.~{Israel},
  eds., pp.~680--745, 1979.

\bibitem{pozas2016entanglement}
A.~Pozas-Kerstjens and E.~Mart\'{\i}n-Mart\'{\i}nez, \emph{Entanglement
  harvesting from the electromagnetic vacuum with hydrogenlike atoms},
  \href{https://doi.org/10.1103/PhysRevD.94.064074}{\emph{Phys. Rev. D}
  {\bfseries 94} (2016) 064074}.

\bibitem{Sachs:2018trp}
A.~M. Sachs, R.~B. Mann and E.~Martín-Martínez, \emph{{Entanglement
  harvesting from multiple massless scalar fields and divergences in
  Unruh-DeWitt detector models}},
  \href{https://arxiv.org/abs/1808.05980}{{\ttfamily 1808.05980}}.

\bibitem{Aubry2014derivative}
B.~A. Ju{\'{a}}rez-Aubry and J.~Louko, \emph{{Onset and decay of the 1 + 1
  Hawking-Unruh effect: what the derivative-coupling detector saw}},
  \href{https://doi.org/10.1088/0264-9381/31/24/245007}{\emph{Class. and
  Quantum Gravity} {\bfseries 31} (2014) 245007}.

\bibitem{Hodgkinson2012corotatingBTZ}
L.~Hodgkinson and J.~Louko, \emph{{Static, stationary, and inertial
  Unruh-DeWitt detectors on the BTZ black hole}},
  \href{https://doi.org/10.1103/PhysRevD.86.064031}{\emph{Phys. Rev. D}
  {\bfseries 86} (2012) 064031}.

\bibitem{Xu:2020pbj}
Q.~Xu, S.~A. Ahmad and A.~R.~H. Smith, \emph{Gravitational waves affect vacuum
  entanglement},  \href{https://arxiv.org/abs/Arxiv:2006.11301v1}{{\ttfamily
  Arxiv:2006.11301v1}}.

\bibitem{Aubry2018Vaidya}
B.~A. Ju{\'a}rez-Aubry and J.~Louko, \emph{{Quantum fields during black hole
  formation: how good an approximation is the Unruh state?}},
  \href{https://doi.org/10.1007/JHEP05(2018)140}{\emph{Journal of High Energy
  Physics} {\bfseries 2018} (2018) 140}.

\bibitem{hodgkinson2013particle}
L.~Hodgkinson, \emph{Particle detectors in curved spacetime quantum field
  theory},
  \href{https://doi.org/https://arxiv.org/abs/1309.7281}{\emph{arXiv:1309.7281}
  (2013) }.

\bibitem{NIST:DLMF}
``{\it NIST Digital Library of Mathematical Functions}.''
  http://dlmf.nist.gov/, Release 1.0.27 of 2020-06-15.

\bibitem{birrell1984quantum}
N.~Birrell, N.~Birrell and P.~Davies, \emph{Quantum Fields in Curved Space},
  Cambridge Monographs on Mathematical Physics. Cambridge University Press,
  1984.

\bibitem{EMM2014zeromode}
E.~Mart\'{\i}n-Mart\'{\i}nez and J.~Louko, \emph{Particle detectors and the
  zero mode of a quantum field},
  \href{https://doi.org/10.1103/PhysRevD.90.024015}{\emph{Phys. Rev. D}
  {\bfseries 90} (2014) 024015}.

\bibitem{Marolf2016}
D.~Marolf and A.~C. Wall, \emph{State-dependent divergences in the entanglement
  entropy}, \href{https://doi.org/10.1007/JHEP10(2016)109}{\emph{Journal of
  High Energy Physics} {\bfseries 2016} (2016) 109}.

\bibitem{Tales2020GRQO}
E.~Mart\'{\i}n-Mart\'{\i}nez, T.~R. Perche and B.~de~S.~L.~Torres,
  \emph{General relativistic quantum optics: Finite-size particle detector
  models in curved spacetimes},
  \href{https://doi.org/10.1103/PhysRevD.101.045017}{\emph{Phys. Rev. D}
  {\bfseries 101} (2020) 045017}.

\bibitem{Bruno2020time-ordering}
E.~Mart\'in-Mart\'inez, T.~R. Perche and B.~de~S.~L.~Torres, \emph{Broken
  covariance of particle detector models in relativistic quantum information},
  \href{https://arxiv.org/abs/Arxiv:2006.12514v1}{{\ttfamily
  Arxiv:2006.12514v1}}.

\bibitem{Wotters1998entanglementmeasure}
W.~K. Wootters, \emph{Entanglement of formation of an arbitrary state of two
  qubits}, \href{https://doi.org/10.1103/PhysRevLett.80.2245}{\emph{Phys. Rev.
  Lett.} {\bfseries 80} (1998) 2245}.

\bibitem{Vidal2002negativity}
G.~Vidal and R.~F. Werner, \emph{Computable measure of entanglement},
  \href{https://doi.org/10.1103/PhysRevA.65.032314}{\emph{Phys. Rev. A}
  {\bfseries 65} (2002) 032314}.

\bibitem{simidzija2018harvesting}
P.~Simidzija and E.~Mart\'{\i}n-Mart\'{\i}nez, \emph{Harvesting correlations
  from thermal and squeezed coherent states},
  \href{https://doi.org/10.1103/PhysRevD.98.085007}{\emph{Phys. Rev. D}
  {\bfseries 98} (2018) 085007}.

\bibitem{Zurek2001discord}
H.~Ollivier and W.~H. Zurek, \emph{Quantum discord: A measure of the
  quantumness of correlations},
  \href{https://doi.org/10.1103/PhysRevLett.88.017901}{\emph{Phys. Rev. Lett.}
  {\bfseries 88} (2001) 017901}.

\bibitem{Henderson2001correlations}
L.~Henderson and V.~Vedral, \emph{Classical, quantum and total correlations},
  \href{https://doi.org/10.1088/0305-4470/34/35/315}{\emph{Journal of Physics
  A: Mathematical and General} {\bfseries 34} (2001) 6899}.

\bibitem{Cong2020horizon}
W.~Cong, C.~Qian, M.~R.~R. Good and R.~B. Mann, \emph{Effects of horizons on
  entanglement harvesting},
  \href{https://arxiv.org/abs/Arxiv:2006.01720v1}{{\ttfamily
  Arxiv:2006.01720v1}}.

\bibitem{Ng2014Schwarzschild}
K.~K. Ng, L.~Hodgkinson, J.~Louko, R.~B. Mann and E.~Mart\'{\i}n-Mart\'{\i}nez,
  \emph{{Unruh-DeWitt detector response along static and circular-geodesic
  trajectories for Schwarzschild--anti-de Sitter black holes}},
  \href{https://doi.org/10.1103/PhysRevD.90.064003}{\emph{Phys. Rev. D}
  {\bfseries 90} (2014) 064003}.

\bibitem{Jonsson2020:2002.05482v2}
R.~H. Jonsson, D.~Q. Aruquipa, M.~Casals, A.~Kempf and E.~Martín-Martínez,
  \emph{Communication through quantum fields near a black hole},
  \href{https://arxiv.org/abs/Arxiv:2002.05482v2}{{\ttfamily
  Arxiv:2002.05482v2}}.

\bibitem{Satz2006howoften}
J.~Louko and A.~Satz, \emph{{ How often does the Unruh-DeWitt detector click?
  Regularisation by a spatial profile }}, {\emph{Classical and Quantum Gravity}
  {\bfseries 23} (2006) 6321}
  [\href{https://arxiv.org/abs/Arxiv:gr-qc/0606067v3}{{\ttfamily
  Arxiv:gr-qc/0606067v3}}].

\bibitem{Satz2007transitionrate}
J.~Louko and A.~Satz, \emph{{ Transition rate of the Unruh-DeWitt detector in
  curved spacetime }}, {\emph{Classical and Quantum Gravity} {\bfseries 25}
  (2007) 055012} [\href{https://arxiv.org/abs/Arxiv:0710.5671v3}{{\ttfamily
  Arxiv:0710.5671v3}}].

\bibitem{Takagi1986noise}
S.~Takagi, \emph{{Vacuum Noise and Stress Induced by Uniform Acceleration,
  Hawking-Unruh Effect in Rindler Manifold of Arbitrary Dimension}},
  \href{https://doi.org/10.1143/PTP.88.1}{\emph{Progress of Theoretical Physics
  Supplement} {\bfseries 88} (1986) 1}.

\bibitem{Causality2015Eduardo}
E.~Mart\'{\i}n-Mart\'{\i}nez, \emph{{Causality issues of particle detector
  models in QFT and quantum optics}},
  \href{https://doi.org/10.1103/PhysRevD.92.104019}{\emph{Phys. Rev. D}
  {\bfseries 92} (2015) 104019}.

\bibitem{wald2010general}
R.~Wald, \emph{General Relativity}. University of Chicago Press, 2010.

\bibitem{Weldon2000thermal}
H.~A. Weldon, \emph{{Thermal Green functions in coordinate space for massless
  particles of any spin}},
  \href{https://doi.org/10.1103/PhysRevD.62.056010}{\emph{Phys. Rev. D}
  {\bfseries 62} (2000) 056010}.

\bibitem{Kubo1957thermality}
R.~Kubo, \emph{Statistical-mechanical theory of irreversible processes. i.
  general theory and simple applications to magnetic and conduction problems},
  \href{https://doi.org/10.1143/JPSJ.12.570}{\emph{Journal of the Physical
  Society of Japan} {\bfseries 12} (1957) 570}
  [\href{https://arxiv.org/abs/https://doi.org/10.1143/JPSJ.12.570}{{\ttfamily
  https://doi.org/10.1143/JPSJ.12.570}}].

\bibitem{Martin-Schwinger1959thermality}
P.~C. Martin and J.~Schwinger, \emph{Theory of many-particle systems. i},
  \href{https://doi.org/10.1103/PhysRev.115.1342}{\emph{Phys. Rev.} {\bfseries
  115} (1959) 1342}.

\bibitem{Tolman1930weight-heat}
R.~C. Tolman, \emph{On the weight of heat and thermal equilibrium in general
  relativity}, \href{https://doi.org/10.1103/PhysRev.35.904}{\emph{Phys. Rev.}
  {\bfseries 35} (1930) 904}.

\bibitem{TolmanEhrenfest1930temperature}
R.~C. Tolman and P.~Ehrenfest, \emph{Temperature equilibrium in a static
  gravitational field},
  \href{https://doi.org/10.1103/PhysRev.36.1791}{\emph{Phys. Rev.} {\bfseries
  36} (1930) 1791}.

\bibitem{Carballo2019Unruh-without-thermal}
R.~Carballo-Rubio, L.~J. Garay, E.~Mart\'{\i}n-Mart\'{\i}nez and J.~de~Ram\'on,
  \emph{Unruh effect without thermality},
  \href{https://doi.org/10.1103/PhysRevLett.123.041601}{\emph{Phys. Rev. Lett.}
  {\bfseries 123} (2019) 041601}.

\bibitem{Brenna2016anti-unruh}
W.~Brenna, R.~B. Mann and E.~Martín-Martínez, \emph{{Anti-Unruh phenomena}},
  \href{https://doi.org/https://doi.org/10.1016/j.physletb.2016.04.002}{\emph{Physics
  Letters B} {\bfseries 757} (2016) 307 }.

\bibitem{Garay2016anti-unruh}
L.~J. Garay, E.~Mart\'{\i}n-Mart\'{\i}nez and J.~de~Ram\'on,
  \emph{{Thermalization of particle detectors: The Unruh effect and its
  reverse}}, \href{https://doi.org/10.1103/PhysRevD.94.104048}{\emph{Phys. Rev.
  D} {\bfseries 94} (2016) 104048}.

\bibitem{Henderson2019anti-hawking}
L.~J. Henderson, R.~A. Hennigar, R.~B. Mann, A.~R.~H. Smith and J.~Zhang,
  \emph{{ The BTZ black hole exhibits anti-Hawking phenomena }},
  \href{https://arxiv.org/abs/Arxiv:1911.02977v1}{{\ttfamily
  Arxiv:1911.02977v1}}.

\bibitem{Compere2019}
G.~Comp\`ere, J.~Long and M.~Riegler, \emph{{Invariance of Unruh and Hawking
  radiation under matter-induced supertranslations}},
  \href{https://doi.org/10.1007/JHEP05(2019)053}{\emph{Journal of High Energy
  Physics} {\bfseries 2019} (2019) 53}.

\bibitem{Kolekar2017accelmemory}
S.~Kolekar and J.~Louko, \emph{Gravitational memory for uniformly accelerated
  observers}, \href{https://doi.org/10.1103/PhysRevD.96.024054}{\emph{Phys.
  Rev. D} {\bfseries 96} (2017) 024054}.

\bibitem{Kolekar2018rindlerhair}
S.~Kolekar and J.~Louko, \emph{Quantum memory for rindler supertranslations},
  \href{https://doi.org/10.1103/PhysRevD.97.085012}{\emph{Phys. Rev. D}
  {\bfseries 97} (2018) 085012}.

\bibitem{Jonsson:2020npo}
R.~H. Jonsson, D.~Q. Aruquipa, M.~Casals, A.~Kempf and E.~Martín-Martínez,
  \emph{{Communication through quantum fields near a black hole}},
  \href{https://arxiv.org/abs/2002.05482}{{\ttfamily 2002.05482}}.

\bibitem{Tjoa2019MSc}
E.~Tjoa, \emph{{Aspects of Quantum Field Theory with Boundary Conditions
  \emph{(}MSc. Thesis\emph{)}}}. UWSpace, 2019.

\bibitem{mukhanov2007introduction}
V.~Mukhanov and S.~Winitzki, \emph{Introduction to Quantum Effects in Gravity}.
  Cambridge University Press, 2007.

\bibitem{Sriramkumar1994finitetime}
L.~Sriramkumar and T.~Padmanabhan, \emph{Response of finite-time particle
  detectors in non-inertial frames and curved spacetime}, {\emph{Classical and
  Quantum Gravity} {\bfseries 13} (1994) 2061}
  [\href{https://arxiv.org/abs/Arxiv:gr-qc/9408037v1}{{\ttfamily
  Arxiv:gr-qc/9408037v1}}].

\bibitem{Mathematica}
W.~R. Inc., ``Mathematica, {V}ersion 12.0.''

\end{thebibliography}\endgroup
\bibliographystyle{JHEP}
\end{document}